\documentclass[%
reprint,
superscriptaddress,
amsmath,amssymb,
aps,
pra,
]{revtex4-2}
\usepackage[utf8]{inputenc}
\UseRawInputEncoding
\usepackage[english]{babel}
\usepackage{blindtext}
\usepackage{bbm}
\usepackage{graphicx}   %include figure files
\graphicspath{{figs/}}   %add figure path
\usepackage{dcolumn}% Align table columns on decimal point
\usepackage{bm}% bold math
\usepackage[hidelinks]{hyperref}% add hypertext capabilities
\usepackage{svg}
\usepackage{wasysym}
\usepackage{textcomp}

\usepackage{epstopdf}
\usepackage{braket}
\usepackage{upgreek}
\usepackage{amsmath}
\usepackage{siunitx}
\usepackage{changes}
\usepackage{natbib}
\usepackage[title]{appendix}

\usepackage{multirow}
\usepackage{import}
\usepackage{calc}
\usepackage{xcolor}

\newcommand{\secref}[1]{\hyperref[#1]{{Section~\ref{#1}}}}
\newcommand{\chapref}[1]{\hyperref[#1]{{Chapter~\ref{#1}}}}
\newcommand{\suppref}[1]{\hyperref[#1]{{App.~\ref{#1}}}}
\newcommand{\figref}[1]{\hyperref[#1]{{Fig.~\ref*{#1}}}}
\newcommand{\Figref}[1]{\hyperref[#1]{{Figure~\ref*{#1}}}}
\newcommand{\figrefadd}[2]{\hyperref[#1]{{Fig.~\ref*{#1}#2}}}
\newcommand{\Figrefadd}[2]{\hyperref[#1]{{Figure~\ref*{#1}#2}}}
\newcommand{\tabref}[1]{\hyperref[#1]{{Table~\ref*{#1}}}}
\renewcommand{\eqref}[1]{\hyperref[#1]{{Eq.~\ref*{#1}}}}

\newcommand{\revisex}[1]{{}}

\definecolor{myBrown}{HTML}{947139}
\definecolor{myDarkgreen}{HTML}{6E692A}
\definecolor{myGreen}{HTML}{00A76D}
\definecolor{myLightgreen}{HTML}{A6CE42}
\definecolor{myYellow}{HTML}{FFF200}
\definecolor{myRed}{HTML}{ED1C1A}
\definecolor{myOrange}{HTML}{F7931D}
\definecolor{myLightblue}{HTML}{00C0F3}
\definecolor{myPink}{HTML}{F6979F}
\definecolor{myBlue}{HTML}{0071BC}
\definecolor{myGold}{HTML}{FFCB04}
\definecolor{myPurple}{HTML}{A066AA}
\definecolor{myDarkgrey}{HTML}{6D6E70}
\definecolor{myLightgrey}{HTML}{9D9FA1}

\begin{document}

\title{Low crosstalk modular flip-chip architecture for coupled superconducting qubits}

\author{Soeren Ihssen}
\affiliation{IQMT,~Karlsruhe~Institute~of~Technology,~76131~Karlsruhe,~Germany}

\author{Simon Geisert}
\thanks{First two authors contributed equally.}
\affiliation{IQMT,~Karlsruhe~Institute~of~Technology,~76131~Karlsruhe,~Germany}

\author{Gabriel Jauma}
\affiliation{Institute of Fundamental Physics IFF-CSIC, Calle Serrano 113b, 28006 Madrid, Spain}
\affiliation{Applied Physics Department, Salamanca University, Salamanca 37008, Spain}

\author{Patrick Winkel}
\affiliation{IQMT,~Karlsruhe~Institute~of~Technology,~76131~Karlsruhe,~Germany}
\affiliation{Departments~of~Applied~Physics~and~Physics,~Yale University,~New Haven,~CT 06520,~USA}
\affiliation{Yale~Quantum~Institute,~Yale~University,~New~Haven,~CT 06520,~USA}

\author{Martin Spiecker}
\affiliation{IQMT,~Karlsruhe~Institute~of~Technology,~76131~Karlsruhe,~Germany}

\author{Nicolas Zapata}
\affiliation{IQMT,~Karlsruhe~Institute~of~Technology,~76131~Karlsruhe,~Germany}

\author{Nicolas Gosling}
\affiliation{IQMT,~Karlsruhe~Institute~of~Technology,~76131~Karlsruhe,~Germany}

\author{Patrick Paluch}
\affiliation{IQMT,~Karlsruhe~Institute~of~Technology,~76131~Karlsruhe,~Germany}
\affiliation{PHI,~Karlsruhe~Institute~of~Technology,~76131~Karlsruhe,~Germany}

\author{Manuel Pino}
\affiliation{Nanotechnology Group, USAL-Nanolab, Salamanca University, Salamanca 37008, Spain}
\affiliation{Institute of Fundamental Physics IFF-CSIC, Calle Serrano 113b, 28006 Madrid, Spain}

\author{Thomas Reisinger}
\affiliation{IQMT,~Karlsruhe~Institute~of~Technology,~76131~Karlsruhe,~Germany}

\author{Wolfgang Wernsdorfer}
\affiliation{IQMT,~Karlsruhe~Institute~of~Technology,~76131~Karlsruhe,~Germany}
\affiliation{PHI,~Karlsruhe~Institute~of~Technology,~76131~Karlsruhe,~Germany}

\author{Juan Jose Garcia-Ripoll}
\affiliation{Institute of Fundamental Physics IFF-CSIC, Calle Serrano 113b, 28006 Madrid, Spain}

\author{Ioan M. Pop}
\email{ioan.pop@kit.edu}
\affiliation{IQMT,~Karlsruhe~Institute~of~Technology,~76131~Karlsruhe,~Germany}
\affiliation{PHI,~Karlsruhe~Institute~of~Technology,~76131~Karlsruhe,~Germany}
\affiliation{Physics~Institute~1,~Stuttgart~University,~70569~Stuttgart,~Germany}

\date{\today}

\begin{abstract}
     We present a flip-chip architecture for an array of coupled superconducting qubits, in which circuit components reside inside individual microwave enclosures. In contrast to other flip-chip approaches, the qubit chips in our architecture are electrically floating, which guarantees a simple, fully modular assembly of capacitively coupled circuit components such as qubit, control, and coupling structures, as well as reduced crosstalk between the components. We validate the concept with a chain of three nearest neighbor coupled generalized flux qubits in which the center qubit acts as a frequency-tunable coupler. Using this coupler, we demonstrate a transverse coupling on/off $\text{ratio} \approx 50$, $\text{zz-crosstalk}\,\approx 0.7\,\text{kHz}$ between resonant qubits and isolation between the qubit $\text{enclosures}\,>60\,\text{dB}$. 
\end{abstract}

\maketitle

% Background and research problem
Circuit quantum electrodynamics~\cite{Blais2021} --- using  
superconducting qubits, resonators and waveguides as building blocks --- is one of the leading platforms for quantum hardware development. Following an approach similar to the semiconductor industry, the on-chip integration of an increasing number of such components is, in principle, a matter of sophisticated engineering and it has led to the realization of impressively complex quantum processors~\cite{Arute__Google_Processor__2019, Wu__Quantum_Processor__2021, Krinner2022, Marques2022, GoogleQuantAI2023, Kim2023, GoogleQuantAI2024}. However, as quantum processors grow in size and complexity, new physical phenomena emerge, such as correlated quasiparticle and phonon bursts~\cite{Cardani2021, McEwen__Resolving_catastrophic_error_bursts__2021}, charge offsets~\cite{Wilen2021} and two-level-system reconfigurations due to ionizing radiation~\cite{Thorbeck2023, Yelton2024}. Moreover, microwave crosstalk emerges as one of the main limitations in some of the most sophisticated architectures~\cite{GoogleQuantAI2024}. Last, but not least, it becomes increasingly likely that frequency crowding or a single defective component limits the performance of an entire chip. In view of these challenges, it is essential for scale-up strategies to minimize microwave and mechanical (i.e. phonon) crosstalk and maximize modularity.

% List conventional solutions and present modular architectures as an alternative solution
On-chip strategies to scale-up quantum processors include from a technological perspective relatively involved techniques such as air-bridges~\cite{Janzen__Aluminum_air_bridges__2022}, spring-loaded pogo pins~\cite{Bronn__High_coherence_plane_breaking_packaging_for_superconducting_qubits__2018} or deep silicon vias~\cite{Mallek__Fabrication_of_superconducting_through_silicon_vias__2021}. While these monolithic devices can achieve impressive microwave performance in terms of isolation and spurious mode suppression, they are susceptible to correlated phonon bursts and lack modularity due to cold-welded connections between layers. 
A complementary approach is to assemble a processor from chip-level building blocks, allowing for individual component fabrication, as well as flexible and reconfigurable arrangements. Several promising techniques are currently being pursued~\cite{Mollenhauer__high_efficiency_plug_and_play_superconducting_qubit__2024,Chou__A_superconducting_dual_rail_cavity_qubit_with_erasure_detected_logical_measurements__2024,Copetudo2024Feb, Spring2022}, including flip-chip~\cite{Kosen__Two_Tmon_FlipChip_Device__2022, Conner__Superconducting_qubits_in_a_flip_chip_architecture__2021} and chiplet~\cite{Field__Modular_Superconducting_Qubit_Architecture_with_a_Multi_chip_Tunable_Coupler__2024, Bild2023Apr, Yost__Solid_state_qubits_integrated_with_superconducting_through_silicon_vias__2020} architectures.

% Our alternative approach as a combination of isolated 3D enclosures and 2D circuitry
Here we present an alternative modular approach, where we combine several features of 2D and 3D architectures to form an array of coupled, but crosstalk-resilient superconducting qubits. Our goal is to facilitate a highly modular assembly, with microwave-isolated sub-components insensitive to correlated errors caused by the propagation of phonon waves. The key feature in our design is the separation of the qubits on dedicated, electrically floating chips within individual microwave enclosures.
%Each qubit is coupled without galvanic connections to its readout and control electronics, as well as to its neighbors.
To validate the architecture, we implemented a device consisting of three enclosures, each hosting a generalized flux qubit, inductively coupled to a dedicated readout mode (QR system) similar to Ref.~\cite{Geisert__GFQ__2024}. We use the central qubit as a flux-tunable coupler and demonstrate tunable two-qubit interactions. This concept can be extended by adding microwave enclosures, both in-plane and out-of-plane, to increase the number of adjacent qubits.

\begin{figure*}
    \resizebox{\textwidth}{!}{ 
        \def\svgwidth{\textwidth}
        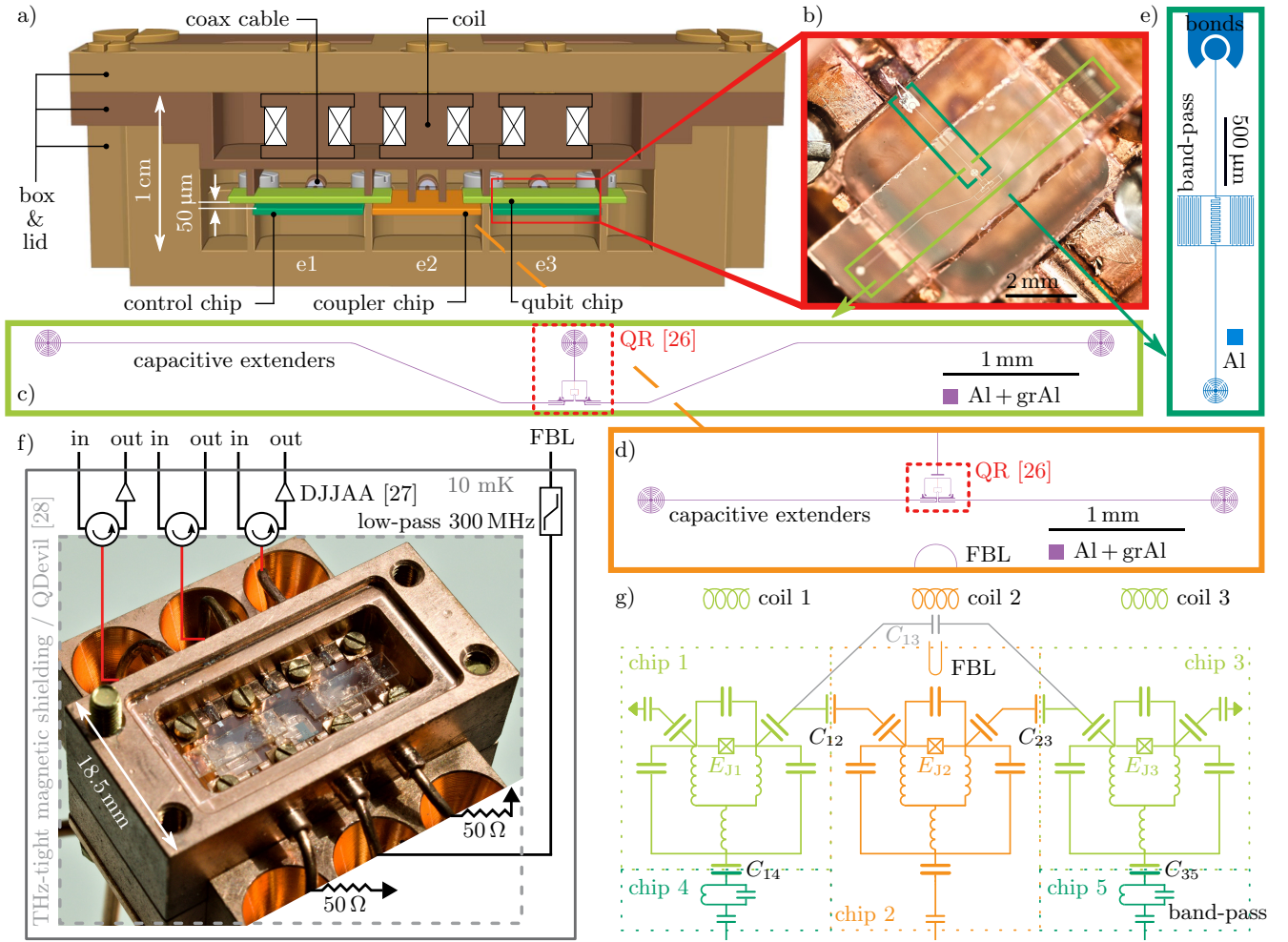 
    }
    \caption{ \textbf{Modular flip-chip architecture based on individual qubit and coupler enclosures} \textbf{a)} Cross-section of the sample box: two of the three enclosures (e1 \& e3) contain control (dark green) and qubit (light green) chips, one enclosure (e2) houses a qubit chip used as coupler (orange). Each enclosure can be accessed by two coaxial cables perpendicular to the cross-section plane, one on each side. The static magnetic field in each enclosure is controlled by coils integrated in the top lid, which are highlighted via the black X symbols. \textbf{b)} Optical image of the qubit-chip above the control-chip in e1, without a coupler chip present. \textbf{c-e)} Design layouts of the qubit, coupler and control chips, respectively. The qubit(Q)-resonator(R) system consists of a generalized flux qubit inductively coupled to a readout resonator, as described in Ref.~\cite{Geisert__GFQ__2024}. The QR system is indicated by the red frames in panels c and d. Two capacitive extenders are capacitively coupled to the qubit junction electrodes to enable capacitive coupling to the adjacent chips via the skeletal shaped pads. The middle pad in panel c is used to couple the QR system to the input-output control chip shown in panel e. The band-pass filter implemented by the meandered inductor and finger capacitor visible in the center of the chip in panel d is used to reduce Purcell decay~\cite{Reed2010, Sete2015}. Through pulses on the flux bias line (FBL) visible in panel d, the frequency of the coupler qubit is tunable. The colour legend indicates the material used for each circuit element: blue for aluminum (Al) and purple for Al covered with granular aluminium (grAl). \textbf{f)} Optical image of the fully equipped sample box and schematics of the reflection measurement setup in a magnetically shielded environment at 10\,mK. The FBL is connected to a bulk commercial low-pass filter with a cutoff frequency of 300\,MHz. \textbf{g)} Circuit diagram of the coupled qubit array. Each qubit chip (chip 1 \& 3) contains a QR system~\cite{Geisert__GFQ__2024} with a corresponding band-pass filter (cf. panel d) on its control chip (chip 4 \& 5, respectively). The QR used as coupler is located on chip 2. The coupling capacitances $C_{12}$, $C_{23}$, $C_{14}$ and $C_{35}$ bridge the gaps between individual chips, enabling a modular flip-chip architecture. The capacitance $C_{13}$ is not implemented on the chips and represents the direct capacitive coupling between the outer qubits.}
    \label{fig:sample_box}
\end{figure*}

\begin{figure*}
        \includegraphics[width=0.85\textwidth]{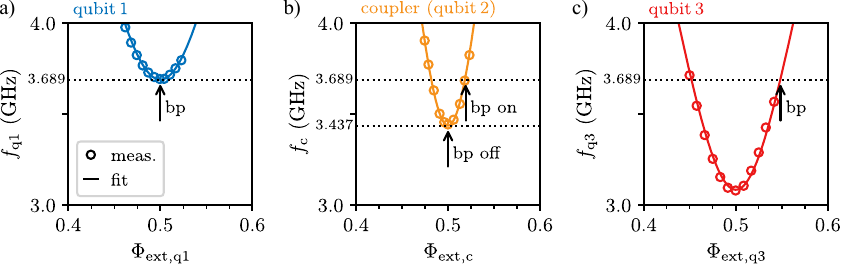}
        \caption{ \textbf{Overview of qubit and coupler spectra with bias points (bp).} The measured (circles) and fitted (lines) spectra of \textbf{a)} qubit 1 (q1, blue) in e1, \textbf{b)} the coupler (c, orange)---i.e. qubit 2---in e2 and \textbf{c)} qubit 3 (q3, red) in e3 vs. the external flux $\Phi_\text{ext}$ through the respective qubit/coupler loops are shown. During all experiments both qubits (q1\,\&\,q3) are operated at a frequency of $f_\text{q1}\approx f_\text{q3}\approx3.689$\,GHz (see the corresponding bp in panel a and c), corresponding to the half-flux point of q1 ($\Phi_\text{ext,\,q1}=\Phi_0/2$). To switch the coupler off, coil\,2 is used to park the coupler at its half flux point ($\Phi_\text{ext,\,c}=\Phi_0/2$), corresponding to bp off at a frequency of 3.437\,GHz. To switch the coupler on, a DC pulse is played on the FBL, aligning $f_\text{c}$ at bp on, resonant with $f_\text{q1}$ and $f_\text{q3}$.
        }
        \label{fig:static_coupling}
\end{figure*}

% Design: describe experimental setup e.g. Fig 1
The three enclosures of our prototype, each hosting a QR system on a dedicated chip, are shown in \figref{fig:sample_box}a. They have dimensions of $6.1\times 6.1\times 6.0\,\text{mm}^3$, pushing the enclosures lowest frequency eigenmodes above 16\,GHz. The 6.5\,mm spacing between neighboring qubits facilitates their assembly without the need for sophisticated packaging tools and strongly reduces crosstalk.
%qubit chip on top of control chip
In the outer enclosures on the left and right (e1 \& e3 respectively, see \figref{fig:sample_box}b) we mount an upside-down flipped qubit chip on top of a control chip, while the middle enclosure (e2) contains only a single coupler chip. The qubit chip dimensions ($2.85\times10.0\,\text{mm}^2$) exceed the size of their enclosures, so that they overlap with the adjacent enclosures. We attach the top qubit chips with a small amount of vacuum grease to pedestals inside the enclosures. We fix the bottom chips with titanium or copper screws to the sample box and wire bond to the coaxial ports. The pedestals define the $d=50 \pm 10$ $\upmu$m gap between the top and bottom chips, avoiding cold-welded bump bonds or on-chip spacers.
%qubit chip description
The qubit chips contain a QR systems, as described in Ref.~\cite{Geisert__GFQ__2024}, to which we add capacitive extenders that reach into the adjacent enclosures (see \figref{fig:sample_box}b,c). The pads at the end of the extenders and in the center of the chips are used for capacitive coupling across the gap between the chips, which is discussed in more detail in \suppref{sec:ANSYS_simulations}. We chose a skeletal pattern for the pads to avoid flux trapping.
%coupler chip description
The coupler chip in e2 also houses a QR system connected via capacitive extenders to pads (see \figref{fig:sample_box}d), aligned with the corresponding neighboring qubits' coupling pads. Details of the fabrication can be found in \suppref{sec:fabrication}.

\begin{figure*}
        \includegraphics[width=1.0\textwidth]{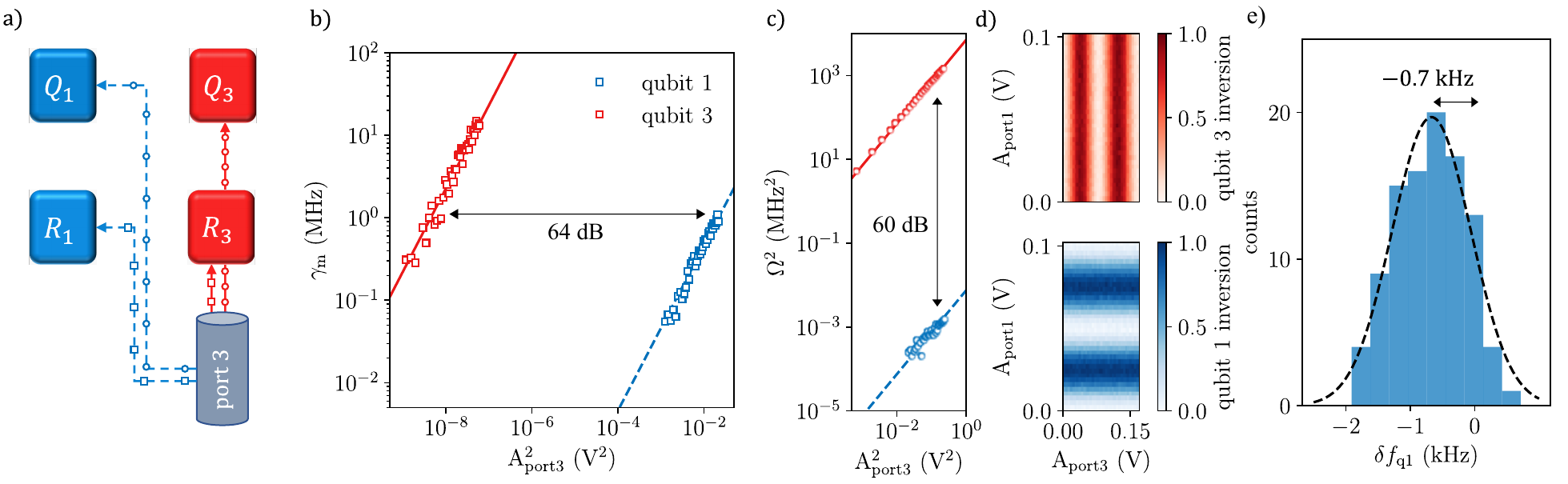}
        \caption{ \textbf{Isolation and crosstalk with coupler switched off.} \textbf{a)} Sketch of the microwave crosstalk measurement scheme: qubit 3 and resonator 3 are driven via microwave signals applied at port 3 (solid red lines). Port-to-qubit and port-to-resonator crosstalk affects qubit 1 and resonator 1 when driving on resonance through port 3 (dashed blue lines). \textbf{b)} Measurement-induced dephasing rate $\gamma_{\text{m}}$ of qubit 1 (blue) and qubit 3 (red), vs. power at port 3. The dephasing rate is extracted from damped Ramsey fringes and fitted to a dispersive model, compare \suppref{sec:power_calibration}. We measure an isolation between enclosures 1 and 3 of $64\pm 0.5$\,dB. The black arrow is a qualitative guide to the eye. \textbf{c)} Rabi frequencies $\Omega$ of qubit 1 (blue) and qubit 3 (red) extracted from Rabi oscillations induced by driving at port 3. The measured port-to-qubit isolation is 60 dB. \textbf{d)} Simultaneous driving of qubits 1 and 3 biased on resonance ($f_\text{q1} = f_\text{q3} = 3.689 \, \text{GHz}$) with the coupler switched off ($f_\text{c} = 3.437 \, \text{GHz}$). The x- and y-axis depict the drive amplitudes at port 1 and port 3, respectively. The qubits can be manipulated independently while being on resonance. \textbf{e)} Measured $zz$-crosstalk between qubit 1 and qubit 3. We extract $\delta f_\text{q1}$ from the difference of Ramsey fringe frequency of qubit 1 when the qubit 3 population is inverted. From a Gaussian fit to the measured $\delta f_\text{q1}$ histogram we extract $\overline{\delta f_\text{q1}} = -0.7 \pm 0.5$\,kHz.}
        \label{fig:isolation_experiments}
\end{figure*} 

%control chip description
The control chips below each qubit chip in e1 and e3 accommodate a bond pad, capacitively coupled to a band-pass filter,  which is connected to a capacitive pad in the center of the qubit chip (see \figref{fig:sample_box}e). The band-pass filter reduces Purcell decay~\cite{Reed2010, Sete2015}, and enables the coupling of the readout mode to the coaxial input port (see \suppref{sec:ANSYS_simulations}).
The control structures for the coupler, which consist of a fast flux bias line (FBL) and a readout line as shown in \figref{fig:sample_box}d, are on the same chip. The FBL on the center coupler chip is wire-bonded to a coaxial cable, which is equipped with a 300\,MHz low-pass filter.

%entire device picture, simplified mes. schematics mu and cryogenic anchoring
The picture in~\figref{fig:sample_box}f shows an open sample box made of copper and it provides an overview of the entire assembly of five chips thermally anchored to the base plate of a dilution cryostat. We show an equivalent circuit diagram of the assembly in~\figref{fig:sample_box}g. Note that the three individual enclosures are defined by the lid, which is not visible in the picture. 
% going over to how to measure as a link to Fig. 2
The simplified schematics of the microwave readout and control lines is also shown in~\figref{fig:sample_box}f. We perform individual qubit readout by probing the readout resonators on the qubit chips in reflection. We use dimer Josephson junction array parametric amplifiers (DJJAAs)~\cite{Winkel__Amplifier__2020} for the readout of the two qubits in e1 \& e3. Qubit control pulses are also send through the readout lines. Simultaneous readout and single qubit $\pi$-pulse calibration measurement are shown in \suppref{sec:single_qubit}.

%Explain the qubit spectra: Off-situation and bias point %shown in~\figref{fig:static_coupling}.
We show the measured two-tone spectroscopy for the qubits and coupler close to their half-flux points (i.e. the magnetic flux sweet spots) in~\figref{fig:static_coupling}. Their frequencies can be tuned independently via the three magnetic field coils, after calibrating the static magnetic field crosstalks. To operate the device, we define the idle points to be the sweet spots of qubit 1 and coupler, which are at $f_\text{q1}=3.689$\,GHz and $f_\text{c}=3.437$\,GHz, respectively. Exploiting the fact that the qubits are isolated in distant enclosures, we chose to operate qubit 1 and 3 on resonance. To turn on coupling between qubit 1 and 3, we play fast flux pulses on the FBL to tune the coupler closer to the qubit 1 and 3 bias points.

% Isolation between enclosure 1 and 3
In view of the fact that we operate the qubits on resonance, it is paramount to know the isolation and crosstalk between the enclosures when the coupling is turned off. \figref{fig:isolation_experiments}a illustrates schematically the crosstalk present in our system. To identify the port-to-resonator isolation, we drive resonator 1 ($f_\text{r1}=6.508$\,GHz) and resonator 3 ($f_\text{r3}=5.226$\,GHz) through the readout port of e3. We measure readout-induced dephasing for both qubits via Ramsey interferometry vs. drive power and drive frequency and fit it to a dispersive model~\cite{Gambetta2006}, detailed in \suppref{sec:power_calibration}. We extract the isolation between enclosures from the ratio of power transmission coefficients. Fitting the measured data shown in \figref{fig:isolation_experiments}, we obtain an isolation of $64 \pm 0.5$ dB.

% Simultaneous Rabi Oscillations
To test qubit control crosstalk we drive via port 3 and measure Rabi oscillations vs. drive amplitude on both qubits for a $1 \upmu$s long pulse. We fit the linear dependence of the measured Rabi frequencies $\Omega_{i}$ with applied drive amplitude, as shown in \figref{fig:isolation_experiments}c. Following references~\cite{Spring2022} and~\cite{Kosen2024}, we extract a port-to-qubit crosstalk, also known as qubit drive selectivity, of

\begin{equation}
    10\log_{10} \left( \frac{\Omega_{1}}{\Omega_{3}} \right)^2 \approx -60 \ \mathrm{dB},
\end{equation} which allows to drive simultaneous Rabi oscillations of qubit 1 and qubit 3 on resonance  ($f_\text{q1} = f_\text{q3}$) with less than 1\permil~crosstalk error.

% Longitudinal zz-Crosstalk
We quantify the longitudinal interaction between qubits 1 and 3 by measuring Ramsey fringes on qubit 1, with and without playing a $\pi$-pulse on qubit 3. The measured shifts in qubit 1 frequencies, $\delta f_\text{q1}$, are summarized in~\figref{fig:isolation_experiments}d. The resulting zz-crosstalk is described by a Gaussian distribution with an average $\overline{\delta f_\text{q1}} = -0.7$\,kHz and standard deviation $\sigma = 0.5$\,kHz.

Compared to state-of-the-art conventional flip-chip architectures based on coplanar waveguide architectures~\cite{Kosen2024}, the isolation presented here is several orders of magnitude larger. Compared to similar architectures based on 3D-integrated floating chips in enclosures~\cite{Spring2022}, the isolation is at least comparable or one order of magnitude larger on average.
This proves the effective screening of microwave signals across the enclosures, even while capacitive extenders between the enclosures are present and while the qubits are operated on resonance.

\begin{figure}[h!]
        \includegraphics[width=0.5\textwidth]{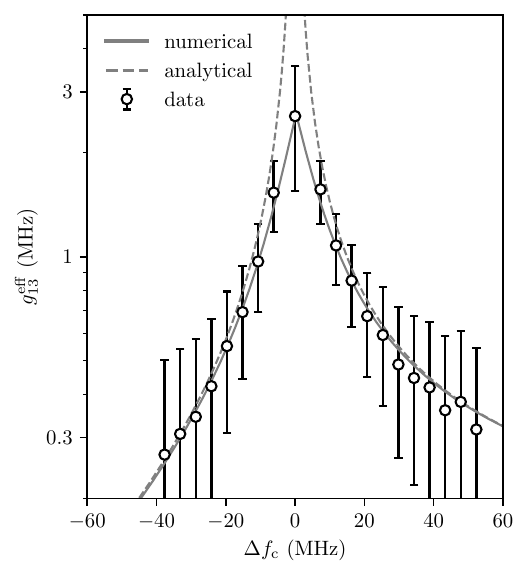}
        \caption{ \textbf{Effective qubit-qubit coupling strength $g_{13}^{\text{eff}}$ vs. coupler detuning $\Delta f_{\text{c}}$.} The data points show the measured coupling strength between q1 \& q3 that are operated on resonance ($f_\text{q1} = f_\text{q3} = 3.689\, \text{GHz}$) for different coupler detunings $\Delta f_\text{c}$. The errorbars represent the spectroscopic linewidth of the $f_\text{q1}$-transition used to measure the level splittings. The numerical (continuous) and analytical (dashed) theoretical curves correspond to an effective two-qubit model obtained using a Schrieffer-Wolff transformation, as detailed in~\suppref{sec:SWT}.
        \label{fig:coupling_experiments}
        }
        
        \label{fig:static_coupling_experiments}
\end{figure}

The effective transverse coupling strength $g_{13}^{\text{eff}}$ between qubit 1 and 3 can be tuned by adapting the coupler detuning $\Delta f_{\text{c}} = f_{\text{c}} - f_{\text{q1}}$. To illustrate this, we show the measured $g_{13}^{\text{eff}}$ in~\figref{fig:static_coupling_experiments} which are obtained from two-tone spectroscopic measurements of avoided level splitting between qubits 1 and 3 for different coupler detunings (cf.~\suppref{sec:avoided_crossings}). By applying a Schrieffer-Wolff transformation  (SWT) to the coupled three-qubit array, we obtain an effective two-qubit model \cite{hita2021three, consani2020effective}. The SWT can be calculated numerically without approximation as explained in Ref.~\cite{hita2021three}, and analytically using second-order perturbation, as detailed in \suppref{sec:SWT}. 
The continuous and dashed lines in~\figref{fig:static_coupling_experiments} show fits of the SWT model to the measured $g_{13}^{\text{eff}}$ using two parameters: the effective qubit-coupler $C_{12} = C_{23} \approx 0.2\,$fF and direct qubit-qubit $C_{13} \approx 5$\,aF coupling capacitances, which are given by the combination of capacitive extenders and shunting capacitances via ground (cf.~\figref{fig:sample_box}g).
The individual qubit and coupler parameters are obtained from separate fits of their flux-dependent spectra~\cite{Geisert__GFQ__2024} and summarized in \suppref{sec:qubit_parameters}. As expected, perturbation theory fails to predict the level splitting close to resonance, whereas the numerical results capture the tunability of the effective qubit-qubit coupling correctly. 

The coupling strength $g^{\text{max}}_{13} = 2.5$\,MHz is maximal on resonance, confirming the measured qubit population swap frequency $f_{\text{SWAP}} \approx 10\,\text{MHz} = 4g_{13}^{\text{max}}$, as detailed in~\suppref{sec:pop_transfer}. At the coupler sweet spot, which corresponds to a detuning of $\Delta f_{\text{c}} = -252$\,MHz, we measure a residual $g_{13}^{\text{eff}} = 50$\,kHz, corresponding to an ON/OFF ratio of 50 (see ~\suppref{sec:avoided_crossings},~\figref{fig:App_AC_extrema}). In future experiments, the ON/OFF ratio can be optimized by designing the coupling capacitors $C_{12}$ and $C_{23}$ in such a way that the direct charge interaction between qubit 1 and 3, represented by a parasitic $C_{13}$, cancels with the one mediated by the coupler, similarly to Ref.~\cite{Sung2021}.

In summary, we have demonstrated a linear array of three coupled flux qubits, each on a dedicated chip in a separate microwave enclosure, which ensures low microwave crosstalk between the outermost enclosures, below -60 dB. We leverage this isolation and operate the qubits on resonance, using the center qubit as flux-tunable coupler. As a result, the architecture provides strong isolation while offering tunable coupling at the same time. In addition, the system is fully modular and allows for the substitution and reassembly of individual circuit parts. 
In the future, this architecture can be used to implement two-qubit gates on resonant qubits by using the coupler off-resonantly~\cite{Chen2014, McKay2016, Weber2017, Sung2021, Moskalenko2022} or to investigate novel gate schemes including the coupler degree of freedom. The \textit{one qubit one enclosure} concept presented here is directly scalable in two dimensions (see \suppref{sec:5erBox}) and could potentially be miniaturized using micromachining techniques~\cite{Brecht2017, Rosenberg2020, Spring2022}.\\
\ \\
%
%\section*{Supplementary Material}
%See the supplementary material for more details on the capacitive coupling between the chips including corresponding finite-element simulations, the fabrication procedure, the readout and single qubit gate fidelity, the power calibration used to calculate the dephasing rate $\gamma_\mathrm{m}$, the avoided level crossings between the qubits from which we extract the effective qubit-qubit coupling $g_{13}^{\text{eff}}$, the Schrieffer-Wolff transformation used to predict $g_{13}^{\text{eff}}$, the qubit spectra and parameters, the population transfer between the qubits and the scaling of the architecture in two dimensions. \\
%

\section*{Acknowledgements}
 We are grateful to L. Radtke and S. Diewald for technical assistance.
 This project was financed by the European Commission via project FET-Open AVaQus, GA 899561. 
 M.S., P.P., N.G. and T.R.~acknowledge partial funding from the German Ministry of Education and Research (BMBF) within the project GEQCOS (FKZ:~13N15683). 
 N.Z.~acknowledges funding from the Deutsche Forschungsgemeinschaft (DFG – German Research Foundation) under project number 450396347 (GeHoldeQED).
 G.J.~acknowledges support from the “la Caixa” Foundation (ID 100010434) via fellowship code LCF/BQ/DR22/11950032.
 M.P.~acknowledges funding from European Union NextGenerationEU/PRTR project Consolidación Investigadora CNS2022-136025. 
 Facilities use was supported by the KIT Nanostructure Service Laboratory. 
 We acknowledge the measurement software framework qKit.

\bibliography{arxiv__aipsamp}% Produces the bibliography via BibTeX.

\providecommand{\noopsort}[1]{}\providecommand{\singleletter}[1]{#1}%
\begin{thebibliography}{45}%
\makeatletter
\providecommand \@ifxundefined [1]{%
 \@ifx{#1\undefined}
}%
\providecommand \@ifnum [1]{%
 \ifnum #1\expandafter \@firstoftwo
 \else \expandafter \@secondoftwo
 \fi
}%
\providecommand \@ifx [1]{%
 \ifx #1\expandafter \@firstoftwo
 \else \expandafter \@secondoftwo
 \fi
}%
\providecommand \natexlab [1]{#1}%
\providecommand \enquote  [1]{``#1''}%
\providecommand \bibnamefont  [1]{#1}%
\providecommand \bibfnamefont [1]{#1}%
\providecommand \citenamefont [1]{#1}%
\providecommand \href@noop [0]{\@secondoftwo}%
\providecommand \href [0]{\begingroup \@sanitize@url \@href}%
\providecommand \@href[1]{\@@startlink{#1}\@@href}%
\providecommand \@@href[1]{\endgroup#1\@@endlink}%
\providecommand \@sanitize@url [0]{\catcode `\\12\catcode `\$12\catcode
  `\&12\catcode `\#12\catcode `\^12\catcode `\_12\catcode `\%12\relax}%
\providecommand \@@startlink[1]{}%
\providecommand \@@endlink[0]{}%
\providecommand \url  [0]{\begingroup\@sanitize@url \@url }%
\providecommand \@url [1]{\endgroup\@href {#1}{\urlprefix }}%
\providecommand \urlprefix  [0]{URL }%
\providecommand \Eprint [0]{\href }%
\providecommand \doibase [0]{https://doi.org/}%
\providecommand \selectlanguage [0]{\@gobble}%
\providecommand \bibinfo  [0]{\@secondoftwo}%
\providecommand \bibfield  [0]{\@secondoftwo}%
\providecommand \translation [1]{[#1]}%
\providecommand \BibitemOpen [0]{}%
\providecommand \bibitemStop [0]{}%
\providecommand \bibitemNoStop [0]{.\EOS\space}%
\providecommand \EOS [0]{\spacefactor3000\relax}%
\providecommand \BibitemShut  [1]{\csname bibitem#1\endcsname}%
\let\auto@bib@innerbib\@empty
%</preamble>
\bibitem [{\citenamefont {Blais}\ \emph {et~al.}(2021)\citenamefont {Blais},
  \citenamefont {Grimsmo}, \citenamefont {Girvin},\ and\ \citenamefont
  {Wallraff}}]{Blais2021}%
  \BibitemOpen
  \bibfield  {author} {\bibinfo {author} {\bibfnamefont {A.}~\bibnamefont
  {Blais}}, \bibinfo {author} {\bibfnamefont {A.~L.}\ \bibnamefont {Grimsmo}},
  \bibinfo {author} {\bibfnamefont {S.~M.}\ \bibnamefont {Girvin}},\ and\
  \bibinfo {author} {\bibfnamefont {A.}~\bibnamefont {Wallraff}},\ }\bibfield
  {title} {\bibinfo {title} {{Circuit quantum electrodynamics}},\ }\href
  {https://doi.org/10.1103/RevModPhys.93.025005} {\bibfield  {journal}
  {\bibinfo  {journal} {Rev. Mod. Phys.}\ }\textbf {\bibinfo {volume} {93}},\
  \bibinfo {pages} {025005} (\bibinfo {year} {2021})}\BibitemShut {NoStop}%
\bibitem [{\citenamefont {Arute}\ \emph {et~al.}(2019)\citenamefont {Arute},
  \citenamefont {Arya}, \citenamefont {Babbush}, \citenamefont {Bacon},
  \citenamefont {Bardin}, \citenamefont {Barends}, \citenamefont {Biswas},
  \citenamefont {Boixo}, \citenamefont {Brandao}, \citenamefont {Buell},
  \citenamefont {Burkett}, \citenamefont {Chen}, \citenamefont {Chen},
  \citenamefont {Chiaro}, \citenamefont {Collins}, \citenamefont {Courtney},
  \citenamefont {Dunsworth}, \citenamefont {Farhi}, \citenamefont {Foxen},
  \citenamefont {Fowler}, \citenamefont {Gidney}, \citenamefont {Giustina},
  \citenamefont {Graff}, \citenamefont {Guerin}, \citenamefont {Habegger},
  \citenamefont {Harrigan}, \citenamefont {Hartmann}, \citenamefont {Ho},
  \citenamefont {Hoffmann}, \citenamefont {Huang}, \citenamefont {Humble},
  \citenamefont {Isakov}, \citenamefont {Jeffrey}, \citenamefont {Jiang},
  \citenamefont {Kafri}, \citenamefont {Kechedzhi}, \citenamefont {Kelly},
  \citenamefont {Klimov}, \citenamefont {Knysh}, \citenamefont {Korotkov},
  \citenamefont {Kostritsa}, \citenamefont {Landhuis}, \citenamefont
  {Lindmark}, \citenamefont {Lucero}, \citenamefont {Lyakh}, \citenamefont
  {Mandr{\`{a}}}, \citenamefont {McClean}, \citenamefont {McEwen},
  \citenamefont {Megrant}, \citenamefont {Mi}, \citenamefont {Michielsen},
  \citenamefont {Mohseni}, \citenamefont {Mutus}, \citenamefont {Naaman},
  \citenamefont {Neeley}, \citenamefont {Neill}, \citenamefont {Niu},
  \citenamefont {Ostby}, \citenamefont {Petukhov}, \citenamefont {Platt},
  \citenamefont {Quintana}, \citenamefont {Rieffel}, \citenamefont {Roushan},
  \citenamefont {Rubin}, \citenamefont {Sank}, \citenamefont {Satzinger},
  \citenamefont {Smelyanskiy}, \citenamefont {Sung}, \citenamefont
  {Trevithick}, \citenamefont {Vainsencher}, \citenamefont {Villalonga},
  \citenamefont {White}, \citenamefont {Yao}, \citenamefont {Yeh},
  \citenamefont {Zalcman}, \citenamefont {Neven},\ and\ \citenamefont
  {Martinis}}]{Arute__Google_Processor__2019}%
  \BibitemOpen
  \bibfield  {author} {\bibinfo {author} {\bibfnamefont {F.}~\bibnamefont
  {Arute}}, \bibinfo {author} {\bibfnamefont {K.}~\bibnamefont {Arya}},
  \bibinfo {author} {\bibfnamefont {R.}~\bibnamefont {Babbush}}, \bibinfo
  {author} {\bibfnamefont {D.}~\bibnamefont {Bacon}}, \bibinfo {author}
  {\bibfnamefont {J.~C.}\ \bibnamefont {Bardin}}, \bibinfo {author}
  {\bibfnamefont {R.}~\bibnamefont {Barends}}, \bibinfo {author} {\bibfnamefont
  {R.}~\bibnamefont {Biswas}}, \bibinfo {author} {\bibfnamefont
  {S.}~\bibnamefont {Boixo}}, \bibinfo {author} {\bibfnamefont {F.~G. S.~L.}\
  \bibnamefont {Brandao}}, \bibinfo {author} {\bibfnamefont {D.~A.}\
  \bibnamefont {Buell}}, \bibinfo {author} {\bibfnamefont {B.}~\bibnamefont
  {Burkett}}, \bibinfo {author} {\bibfnamefont {Y.}~\bibnamefont {Chen}},
  \bibinfo {author} {\bibfnamefont {Z.}~\bibnamefont {Chen}}, \bibinfo {author}
  {\bibfnamefont {B.}~\bibnamefont {Chiaro}}, \bibinfo {author} {\bibfnamefont
  {R.}~\bibnamefont {Collins}}, \bibinfo {author} {\bibfnamefont
  {W.}~\bibnamefont {Courtney}}, \bibinfo {author} {\bibfnamefont
  {A.}~\bibnamefont {Dunsworth}}, \bibinfo {author} {\bibfnamefont
  {E.}~\bibnamefont {Farhi}}, \bibinfo {author} {\bibfnamefont
  {B.}~\bibnamefont {Foxen}}, \bibinfo {author} {\bibfnamefont
  {A.}~\bibnamefont {Fowler}}, \bibinfo {author} {\bibfnamefont
  {C.}~\bibnamefont {Gidney}}, \bibinfo {author} {\bibfnamefont
  {M.}~\bibnamefont {Giustina}}, \bibinfo {author} {\bibfnamefont
  {R.}~\bibnamefont {Graff}}, \bibinfo {author} {\bibfnamefont
  {K.}~\bibnamefont {Guerin}}, \bibinfo {author} {\bibfnamefont
  {S.}~\bibnamefont {Habegger}}, \bibinfo {author} {\bibfnamefont {M.~P.}\
  \bibnamefont {Harrigan}}, \bibinfo {author} {\bibfnamefont {M.~J.}\
  \bibnamefont {Hartmann}}, \bibinfo {author} {\bibfnamefont {A.}~\bibnamefont
  {Ho}}, \bibinfo {author} {\bibfnamefont {M.}~\bibnamefont {Hoffmann}},
  \bibinfo {author} {\bibfnamefont {T.}~\bibnamefont {Huang}}, \bibinfo
  {author} {\bibfnamefont {T.~S.}\ \bibnamefont {Humble}}, \bibinfo {author}
  {\bibfnamefont {S.~V.}\ \bibnamefont {Isakov}}, \bibinfo {author}
  {\bibfnamefont {E.}~\bibnamefont {Jeffrey}}, \bibinfo {author} {\bibfnamefont
  {Z.}~\bibnamefont {Jiang}}, \bibinfo {author} {\bibfnamefont
  {D.}~\bibnamefont {Kafri}}, \bibinfo {author} {\bibfnamefont
  {K.}~\bibnamefont {Kechedzhi}}, \bibinfo {author} {\bibfnamefont
  {J.}~\bibnamefont {Kelly}}, \bibinfo {author} {\bibfnamefont {P.~V.}\
  \bibnamefont {Klimov}}, \bibinfo {author} {\bibfnamefont {S.}~\bibnamefont
  {Knysh}}, \bibinfo {author} {\bibfnamefont {A.}~\bibnamefont {Korotkov}},
  \bibinfo {author} {\bibfnamefont {F.}~\bibnamefont {Kostritsa}}, \bibinfo
  {author} {\bibfnamefont {D.}~\bibnamefont {Landhuis}}, \bibinfo {author}
  {\bibfnamefont {M.}~\bibnamefont {Lindmark}}, \bibinfo {author}
  {\bibfnamefont {E.}~\bibnamefont {Lucero}}, \bibinfo {author} {\bibfnamefont
  {D.}~\bibnamefont {Lyakh}}, \bibinfo {author} {\bibfnamefont
  {S.}~\bibnamefont {Mandr{\`{a}}}}, \bibinfo {author} {\bibfnamefont {J.~R.}\
  \bibnamefont {McClean}}, \bibinfo {author} {\bibfnamefont {M.}~\bibnamefont
  {McEwen}}, \bibinfo {author} {\bibfnamefont {A.}~\bibnamefont {Megrant}},
  \bibinfo {author} {\bibfnamefont {X.}~\bibnamefont {Mi}}, \bibinfo {author}
  {\bibfnamefont {K.}~\bibnamefont {Michielsen}}, \bibinfo {author}
  {\bibfnamefont {M.}~\bibnamefont {Mohseni}}, \bibinfo {author} {\bibfnamefont
  {J.}~\bibnamefont {Mutus}}, \bibinfo {author} {\bibfnamefont
  {O.}~\bibnamefont {Naaman}}, \bibinfo {author} {\bibfnamefont
  {M.}~\bibnamefont {Neeley}}, \bibinfo {author} {\bibfnamefont
  {C.}~\bibnamefont {Neill}}, \bibinfo {author} {\bibfnamefont {M.~Y.}\
  \bibnamefont {Niu}}, \bibinfo {author} {\bibfnamefont {E.}~\bibnamefont
  {Ostby}}, \bibinfo {author} {\bibfnamefont {A.}~\bibnamefont {Petukhov}},
  \bibinfo {author} {\bibfnamefont {J.~C.}\ \bibnamefont {Platt}}, \bibinfo
  {author} {\bibfnamefont {C.}~\bibnamefont {Quintana}}, \bibinfo {author}
  {\bibfnamefont {E.~G.}\ \bibnamefont {Rieffel}}, \bibinfo {author}
  {\bibfnamefont {P.}~\bibnamefont {Roushan}}, \bibinfo {author} {\bibfnamefont
  {N.~C.}\ \bibnamefont {Rubin}}, \bibinfo {author} {\bibfnamefont
  {D.}~\bibnamefont {Sank}}, \bibinfo {author} {\bibfnamefont {K.~J.}\
  \bibnamefont {Satzinger}}, \bibinfo {author} {\bibfnamefont {V.}~\bibnamefont
  {Smelyanskiy}}, \bibinfo {author} {\bibfnamefont {K.~J.}\ \bibnamefont
  {Sung}}, \bibinfo {author} {\bibfnamefont {M.~D.}\ \bibnamefont
  {Trevithick}}, \bibinfo {author} {\bibfnamefont {A.}~\bibnamefont
  {Vainsencher}}, \bibinfo {author} {\bibfnamefont {B.}~\bibnamefont
  {Villalonga}}, \bibinfo {author} {\bibfnamefont {T.}~\bibnamefont {White}},
  \bibinfo {author} {\bibfnamefont {Z.~J.}\ \bibnamefont {Yao}}, \bibinfo
  {author} {\bibfnamefont {P.}~\bibnamefont {Yeh}}, \bibinfo {author}
  {\bibfnamefont {A.}~\bibnamefont {Zalcman}}, \bibinfo {author} {\bibfnamefont
  {H.}~\bibnamefont {Neven}},\ and\ \bibinfo {author} {\bibfnamefont {J.~M.}\
  \bibnamefont {Martinis}},\ }\bibfield  {title} {\bibinfo {title} {Quantum
  supremacy using a programmable superconducting processor},\ }\href
  {https://doi.org/10.1038/s41586-019-1666-5} {\bibfield  {journal} {\bibinfo
  {journal} {Nature}\ }\textbf {\bibinfo {volume} {574}},\ \bibinfo {pages}
  {505} (\bibinfo {year} {2019})}\BibitemShut {NoStop}%
\bibitem [{\citenamefont {Wu}\ \emph {et~al.}(2021)\citenamefont {Wu},
  \citenamefont {Bao}, \citenamefont {Cao}, \citenamefont {Chen}, \citenamefont
  {Chen}, \citenamefont {Chen}, \citenamefont {Chung}, \citenamefont {Deng},
  \citenamefont {Du}, \citenamefont {Fan}, \citenamefont {Gong}, \citenamefont
  {Guo}, \citenamefont {Guo}, \citenamefont {Guo}, \citenamefont {Han},
  \citenamefont {Hong}, \citenamefont {Huang}, \citenamefont {Huo},
  \citenamefont {Li}, \citenamefont {Li}, \citenamefont {Li}, \citenamefont
  {Li}, \citenamefont {Liang}, \citenamefont {Lin}, \citenamefont {Lin},
  \citenamefont {Qian}, \citenamefont {Qiao}, \citenamefont {Rong},
  \citenamefont {Su}, \citenamefont {Sun}, \citenamefont {Wang}, \citenamefont
  {Wang}, \citenamefont {Wu}, \citenamefont {Xu}, \citenamefont {Yan},
  \citenamefont {Yang}, \citenamefont {Yang}, \citenamefont {Ye}, \citenamefont
  {Yin}, \citenamefont {Ying}, \citenamefont {Yu}, \citenamefont {Zha},
  \citenamefont {Zhang}, \citenamefont {Zhang}, \citenamefont {Zhang},
  \citenamefont {Zhang}, \citenamefont {Zhao}, \citenamefont {Zhao},
  \citenamefont {Zhou}, \citenamefont {Zhu}, \citenamefont {Lu}, \citenamefont
  {Peng}, \citenamefont {Zhu},\ and\ \citenamefont
  {Pan}}]{Wu__Quantum_Processor__2021}%
  \BibitemOpen
  \bibfield  {author} {\bibinfo {author} {\bibfnamefont {Y.}~\bibnamefont
  {Wu}}, \bibinfo {author} {\bibfnamefont {W.-S.}\ \bibnamefont {Bao}},
  \bibinfo {author} {\bibfnamefont {S.}~\bibnamefont {Cao}}, \bibinfo {author}
  {\bibfnamefont {F.}~\bibnamefont {Chen}}, \bibinfo {author} {\bibfnamefont
  {M.-C.}\ \bibnamefont {Chen}}, \bibinfo {author} {\bibfnamefont
  {X.}~\bibnamefont {Chen}}, \bibinfo {author} {\bibfnamefont {T.-H.}\
  \bibnamefont {Chung}}, \bibinfo {author} {\bibfnamefont {H.}~\bibnamefont
  {Deng}}, \bibinfo {author} {\bibfnamefont {Y.}~\bibnamefont {Du}}, \bibinfo
  {author} {\bibfnamefont {D.}~\bibnamefont {Fan}}, \bibinfo {author}
  {\bibfnamefont {M.}~\bibnamefont {Gong}}, \bibinfo {author} {\bibfnamefont
  {C.}~\bibnamefont {Guo}}, \bibinfo {author} {\bibfnamefont {C.}~\bibnamefont
  {Guo}}, \bibinfo {author} {\bibfnamefont {S.}~\bibnamefont {Guo}}, \bibinfo
  {author} {\bibfnamefont {L.}~\bibnamefont {Han}}, \bibinfo {author}
  {\bibfnamefont {L.}~\bibnamefont {Hong}}, \bibinfo {author} {\bibfnamefont
  {H.-L.}\ \bibnamefont {Huang}}, \bibinfo {author} {\bibfnamefont {Y.-H.}\
  \bibnamefont {Huo}}, \bibinfo {author} {\bibfnamefont {L.}~\bibnamefont
  {Li}}, \bibinfo {author} {\bibfnamefont {N.}~\bibnamefont {Li}}, \bibinfo
  {author} {\bibfnamefont {S.}~\bibnamefont {Li}}, \bibinfo {author}
  {\bibfnamefont {Y.}~\bibnamefont {Li}}, \bibinfo {author} {\bibfnamefont
  {F.}~\bibnamefont {Liang}}, \bibinfo {author} {\bibfnamefont
  {C.}~\bibnamefont {Lin}}, \bibinfo {author} {\bibfnamefont {J.}~\bibnamefont
  {Lin}}, \bibinfo {author} {\bibfnamefont {H.}~\bibnamefont {Qian}}, \bibinfo
  {author} {\bibfnamefont {D.}~\bibnamefont {Qiao}}, \bibinfo {author}
  {\bibfnamefont {H.}~\bibnamefont {Rong}}, \bibinfo {author} {\bibfnamefont
  {H.}~\bibnamefont {Su}}, \bibinfo {author} {\bibfnamefont {L.}~\bibnamefont
  {Sun}}, \bibinfo {author} {\bibfnamefont {L.}~\bibnamefont {Wang}}, \bibinfo
  {author} {\bibfnamefont {S.}~\bibnamefont {Wang}}, \bibinfo {author}
  {\bibfnamefont {D.}~\bibnamefont {Wu}}, \bibinfo {author} {\bibfnamefont
  {Y.}~\bibnamefont {Xu}}, \bibinfo {author} {\bibfnamefont {K.}~\bibnamefont
  {Yan}}, \bibinfo {author} {\bibfnamefont {W.}~\bibnamefont {Yang}}, \bibinfo
  {author} {\bibfnamefont {Y.}~\bibnamefont {Yang}}, \bibinfo {author}
  {\bibfnamefont {Y.}~\bibnamefont {Ye}}, \bibinfo {author} {\bibfnamefont
  {J.}~\bibnamefont {Yin}}, \bibinfo {author} {\bibfnamefont {C.}~\bibnamefont
  {Ying}}, \bibinfo {author} {\bibfnamefont {J.}~\bibnamefont {Yu}}, \bibinfo
  {author} {\bibfnamefont {C.}~\bibnamefont {Zha}}, \bibinfo {author}
  {\bibfnamefont {C.}~\bibnamefont {Zhang}}, \bibinfo {author} {\bibfnamefont
  {H.}~\bibnamefont {Zhang}}, \bibinfo {author} {\bibfnamefont
  {K.}~\bibnamefont {Zhang}}, \bibinfo {author} {\bibfnamefont
  {Y.}~\bibnamefont {Zhang}}, \bibinfo {author} {\bibfnamefont
  {H.}~\bibnamefont {Zhao}}, \bibinfo {author} {\bibfnamefont {Y.}~\bibnamefont
  {Zhao}}, \bibinfo {author} {\bibfnamefont {L.}~\bibnamefont {Zhou}}, \bibinfo
  {author} {\bibfnamefont {Q.}~\bibnamefont {Zhu}}, \bibinfo {author}
  {\bibfnamefont {C.-Y.}\ \bibnamefont {Lu}}, \bibinfo {author} {\bibfnamefont
  {C.-Z.}\ \bibnamefont {Peng}}, \bibinfo {author} {\bibfnamefont
  {X.}~\bibnamefont {Zhu}},\ and\ \bibinfo {author} {\bibfnamefont {J.-W.}\
  \bibnamefont {Pan}},\ }\bibfield  {title} {\bibinfo {title} {Strong quantum
  computational advantage using a superconducting quantum processor},\ }\href
  {https://doi.org/10.1103/PhysRevLett.127.180501} {\bibfield  {journal}
  {\bibinfo  {journal} {Phys. Rev. Lett.}\ }\textbf {\bibinfo {volume} {127}},\
  \bibinfo {pages} {180501} (\bibinfo {year} {2021})}\BibitemShut {NoStop}%
\bibitem [{\citenamefont {Krinner}\ \emph {et~al.}(2022)\citenamefont
  {Krinner}, \citenamefont {Lacroix}, \citenamefont {Remm}, \citenamefont
  {Di~Paolo}, \citenamefont {Genois}, \citenamefont {Leroux}, \citenamefont
  {Hellings}, \citenamefont {Lazar}, \citenamefont {Swiadek}, \citenamefont
  {Herrmann}, \citenamefont {Norris}, \citenamefont {Andersen}, \citenamefont
  {M{\ifmmode\ddot{u}\else\"{u}\fi}ller}, \citenamefont {Blais}, \citenamefont
  {Eichler},\ and\ \citenamefont {Wallraff}}]{Krinner2022}%
  \BibitemOpen
  \bibfield  {author} {\bibinfo {author} {\bibfnamefont {S.}~\bibnamefont
  {Krinner}}, \bibinfo {author} {\bibfnamefont {N.}~\bibnamefont {Lacroix}},
  \bibinfo {author} {\bibfnamefont {A.}~\bibnamefont {Remm}}, \bibinfo {author}
  {\bibfnamefont {A.}~\bibnamefont {Di~Paolo}}, \bibinfo {author}
  {\bibfnamefont {E.}~\bibnamefont {Genois}}, \bibinfo {author} {\bibfnamefont
  {C.}~\bibnamefont {Leroux}}, \bibinfo {author} {\bibfnamefont
  {C.}~\bibnamefont {Hellings}}, \bibinfo {author} {\bibfnamefont
  {S.}~\bibnamefont {Lazar}}, \bibinfo {author} {\bibfnamefont
  {F.}~\bibnamefont {Swiadek}}, \bibinfo {author} {\bibfnamefont
  {J.}~\bibnamefont {Herrmann}}, \bibinfo {author} {\bibfnamefont {G.~J.}\
  \bibnamefont {Norris}}, \bibinfo {author} {\bibfnamefont {C.~K.}\
  \bibnamefont {Andersen}}, \bibinfo {author} {\bibfnamefont {M.}~\bibnamefont
  {M{\ifmmode\ddot{u}\else\"{u}\fi}ller}}, \bibinfo {author} {\bibfnamefont
  {A.}~\bibnamefont {Blais}}, \bibinfo {author} {\bibfnamefont
  {C.}~\bibnamefont {Eichler}},\ and\ \bibinfo {author} {\bibfnamefont
  {A.}~\bibnamefont {Wallraff}},\ }\bibfield  {title} {\bibinfo {title}
  {{Realizing repeated quantum error correction in a distance-three surface
  code}},\ }\href {https://doi.org/10.1038/s41586-022-04566-8} {\bibfield
  {journal} {\bibinfo  {journal} {Nature}\ }\textbf {\bibinfo {volume} {605}},\
  \bibinfo {pages} {669} (\bibinfo {year} {2022})}\BibitemShut {NoStop}%
\bibitem [{\citenamefont {Marques}\ \emph {et~al.}(2022)\citenamefont
  {Marques}, \citenamefont {Varbanov}, \citenamefont {Moreira}, \citenamefont
  {Ali}, \citenamefont {Muthusubramanian}, \citenamefont {Zachariadis},
  \citenamefont {Battistel}, \citenamefont {Beekman}, \citenamefont {Haider},
  \citenamefont {Vlothuizen}, \citenamefont {Bruno}, \citenamefont {Terhal},\
  and\ \citenamefont {DiCarlo}}]{Marques2022}%
  \BibitemOpen
  \bibfield  {author} {\bibinfo {author} {\bibfnamefont {J.~F.}\ \bibnamefont
  {Marques}}, \bibinfo {author} {\bibfnamefont {B.~M.}\ \bibnamefont
  {Varbanov}}, \bibinfo {author} {\bibfnamefont {M.~S.}\ \bibnamefont
  {Moreira}}, \bibinfo {author} {\bibfnamefont {H.}~\bibnamefont {Ali}},
  \bibinfo {author} {\bibfnamefont {N.}~\bibnamefont {Muthusubramanian}},
  \bibinfo {author} {\bibfnamefont {C.}~\bibnamefont {Zachariadis}}, \bibinfo
  {author} {\bibfnamefont {F.}~\bibnamefont {Battistel}}, \bibinfo {author}
  {\bibfnamefont {M.}~\bibnamefont {Beekman}}, \bibinfo {author} {\bibfnamefont
  {N.}~\bibnamefont {Haider}}, \bibinfo {author} {\bibfnamefont
  {W.}~\bibnamefont {Vlothuizen}}, \bibinfo {author} {\bibfnamefont
  {A.}~\bibnamefont {Bruno}}, \bibinfo {author} {\bibfnamefont {B.~M.}\
  \bibnamefont {Terhal}},\ and\ \bibinfo {author} {\bibfnamefont
  {L.}~\bibnamefont {DiCarlo}},\ }\bibfield  {title} {\bibinfo {title}
  {{Logical-qubit operations in an error-detecting surface code}},\ }\href
  {https://doi.org/10.1038/s41567-021-01423-9} {\bibfield  {journal} {\bibinfo
  {journal} {Nat. Phys.}\ }\textbf {\bibinfo {volume} {18}},\ \bibinfo {pages}
  {80} (\bibinfo {year} {2022})}\BibitemShut {NoStop}%
\bibitem [{\citenamefont {AI}(2023)}]{GoogleQuantAI2023}%
  \BibitemOpen
  \bibfield  {author} {\bibinfo {author} {\bibfnamefont {G.~Q.}\ \bibnamefont
  {AI}},\ }\bibfield  {title} {\bibinfo {title} {{Suppressing quantum errors by
  scaling a surface code logical qubit}},\ }\href
  {https://doi.org/10.1038/s41586-022-05434-1} {\bibfield  {journal} {\bibinfo
  {journal} {Nature}\ }\textbf {\bibinfo {volume} {614}},\ \bibinfo {pages}
  {676} (\bibinfo {year} {2023})},\ \bibinfo {note} {[Online; accessed 6. Sep.
  2024]}\BibitemShut {NoStop}%
\bibitem [{\citenamefont {Kim}\ \emph {et~al.}(2023)\citenamefont {Kim},
  \citenamefont {Wood}, \citenamefont {Yoder}, \citenamefont {Merkel},
  \citenamefont {Gambetta}, \citenamefont {Temme},\ and\ \citenamefont
  {Kandala}}]{Kim2023}%
  \BibitemOpen
  \bibfield  {author} {\bibinfo {author} {\bibfnamefont {Y.}~\bibnamefont
  {Kim}}, \bibinfo {author} {\bibfnamefont {C.~J.}\ \bibnamefont {Wood}},
  \bibinfo {author} {\bibfnamefont {T.~J.}\ \bibnamefont {Yoder}}, \bibinfo
  {author} {\bibfnamefont {S.~T.}\ \bibnamefont {Merkel}}, \bibinfo {author}
  {\bibfnamefont {J.~M.}\ \bibnamefont {Gambetta}}, \bibinfo {author}
  {\bibfnamefont {K.}~\bibnamefont {Temme}},\ and\ \bibinfo {author}
  {\bibfnamefont {A.}~\bibnamefont {Kandala}},\ }\bibfield  {title} {\bibinfo
  {title} {{Scalable error mitigation for noisy quantum circuits produces
  competitive expectation values}},\ }\href
  {https://doi.org/10.1038/s41567-022-01914-3} {\bibfield  {journal} {\bibinfo
  {journal} {Nat. Phys.}\ }\textbf {\bibinfo {volume} {19}},\ \bibinfo {pages}
  {752} (\bibinfo {year} {2023})}\BibitemShut {NoStop}%
\bibitem [{\citenamefont {{Google Quantum AI and
  Collaborators}}(2024)}]{GoogleQuantAI2024}%
  \BibitemOpen
  \bibfield  {author} {\bibinfo {author} {\bibnamefont {{Google Quantum AI and
  Collaborators}}},\ }\bibfield  {title} {\bibinfo {title} {Quantum error
  correction below the surface code threshold},\ }\bibfield  {journal}
  {\bibinfo  {journal} {Nature}\ }\href
  {https://doi.org/10.1038/s41586-024-08449-y} {10.1038/s41586-024-08449-y}
  (\bibinfo {year} {2024})\BibitemShut {NoStop}%
\bibitem [{\citenamefont {Cardani}\ \emph {et~al.}(2021)\citenamefont
  {Cardani}, \citenamefont {Valenti}, \citenamefont {Casali}, \citenamefont
  {Catelani}, \citenamefont {Charpentier}, \citenamefont {Clemenza},
  \citenamefont {Colantoni}, \citenamefont {Cruciani}, \citenamefont
  {D{'}Imperio}, \citenamefont {Gironi}, \citenamefont
  {Gr{\ifmmode\ddot{u}\else\"{u}\fi}nhaupt}, \citenamefont {Gusenkova},
  \citenamefont {Henriques}, \citenamefont {Lagoin}, \citenamefont {Martinez},
  \citenamefont {Pettinari}, \citenamefont {Rusconi}, \citenamefont {Sander},
  \citenamefont {Tomei}, \citenamefont {Ustinov}, \citenamefont {Weber},
  \citenamefont {Wernsdorfer}, \citenamefont {Vignati}, \citenamefont {Pirro},\
  and\ \citenamefont {Pop}}]{Cardani2021}%
  \BibitemOpen
  \bibfield  {author} {\bibinfo {author} {\bibfnamefont {L.}~\bibnamefont
  {Cardani}}, \bibinfo {author} {\bibfnamefont {F.}~\bibnamefont {Valenti}},
  \bibinfo {author} {\bibfnamefont {N.}~\bibnamefont {Casali}}, \bibinfo
  {author} {\bibfnamefont {G.}~\bibnamefont {Catelani}}, \bibinfo {author}
  {\bibfnamefont {T.}~\bibnamefont {Charpentier}}, \bibinfo {author}
  {\bibfnamefont {M.}~\bibnamefont {Clemenza}}, \bibinfo {author}
  {\bibfnamefont {I.}~\bibnamefont {Colantoni}}, \bibinfo {author}
  {\bibfnamefont {A.}~\bibnamefont {Cruciani}}, \bibinfo {author}
  {\bibfnamefont {G.}~\bibnamefont {D{'}Imperio}}, \bibinfo {author}
  {\bibfnamefont {L.}~\bibnamefont {Gironi}}, \bibinfo {author} {\bibfnamefont
  {L.}~\bibnamefont {Gr{\ifmmode\ddot{u}\else\"{u}\fi}nhaupt}}, \bibinfo
  {author} {\bibfnamefont {D.}~\bibnamefont {Gusenkova}}, \bibinfo {author}
  {\bibfnamefont {F.}~\bibnamefont {Henriques}}, \bibinfo {author}
  {\bibfnamefont {M.}~\bibnamefont {Lagoin}}, \bibinfo {author} {\bibfnamefont
  {M.}~\bibnamefont {Martinez}}, \bibinfo {author} {\bibfnamefont
  {G.}~\bibnamefont {Pettinari}}, \bibinfo {author} {\bibfnamefont
  {C.}~\bibnamefont {Rusconi}}, \bibinfo {author} {\bibfnamefont
  {O.}~\bibnamefont {Sander}}, \bibinfo {author} {\bibfnamefont
  {C.}~\bibnamefont {Tomei}}, \bibinfo {author} {\bibfnamefont {A.~V.}\
  \bibnamefont {Ustinov}}, \bibinfo {author} {\bibfnamefont {M.}~\bibnamefont
  {Weber}}, \bibinfo {author} {\bibfnamefont {W.}~\bibnamefont {Wernsdorfer}},
  \bibinfo {author} {\bibfnamefont {M.}~\bibnamefont {Vignati}}, \bibinfo
  {author} {\bibfnamefont {S.}~\bibnamefont {Pirro}},\ and\ \bibinfo {author}
  {\bibfnamefont {I.~M.}\ \bibnamefont {Pop}},\ }\bibfield  {title} {\bibinfo
  {title} {{Reducing the impact of radioactivity on quantum circuits in a
  deep-underground facility}},\ }\href
  {https://doi.org/10.1038/s41467-021-23032-z} {\bibfield  {journal} {\bibinfo
  {journal} {Nat. Commun.}\ }\textbf {\bibinfo {volume} {12}},\ \bibinfo
  {pages} {1} (\bibinfo {year} {2021})}\BibitemShut {NoStop}%
\bibitem [{\citenamefont {McEwen}\ \emph {et~al.}(2021)\citenamefont {McEwen},
  \citenamefont {Faoro}, \citenamefont {Arya}, \citenamefont {Dunsworth},
  \citenamefont {Huang}, \citenamefont {Kim}, \citenamefont {Burkett},
  \citenamefont {Fowler}, \citenamefont {Arute}, \citenamefont {Bardin},
  \citenamefont {Bengtsson}, \citenamefont {Bilmes}, \citenamefont {Buckley},
  \citenamefont {Bushnell}, \citenamefont {Chen}, \citenamefont {Collins},
  \citenamefont {Demura}, \citenamefont {Derk}, \citenamefont {Erickson},
  \citenamefont {Giustina}, \citenamefont {Harrington}, \citenamefont {Hong},
  \citenamefont {Jeffrey}, \citenamefont {Kelly}, \citenamefont {Klimov},
  \citenamefont {Kostritsa}, \citenamefont {Laptev}, \citenamefont {Locharla},
  \citenamefont {Mi}, \citenamefont {Miao}, \citenamefont {Montazeri},
  \citenamefont {Mutus}, \citenamefont {Naaman}, \citenamefont {Neeley},
  \citenamefont {Neill}, \citenamefont {Opremcak}, \citenamefont {Quintana},
  \citenamefont {Redd}, \citenamefont {Roushan}, \citenamefont {Sank},
  \citenamefont {Satzinger}, \citenamefont {Shvarts}, \citenamefont {White},
  \citenamefont {Yao}, \citenamefont {Yeh}, \citenamefont {Yoo}, \citenamefont
  {Chen}, \citenamefont {Smelyanskiy}, \citenamefont {Martinis}, \citenamefont
  {Neven}, \citenamefont {Megrant}, \citenamefont {Ioffe},\ and\ \citenamefont
  {Barends}}]{McEwen__Resolving_catastrophic_error_bursts__2021}%
  \BibitemOpen
  \bibfield  {author} {\bibinfo {author} {\bibfnamefont {M.}~\bibnamefont
  {McEwen}}, \bibinfo {author} {\bibfnamefont {L.}~\bibnamefont {Faoro}},
  \bibinfo {author} {\bibfnamefont {K.}~\bibnamefont {Arya}}, \bibinfo {author}
  {\bibfnamefont {A.}~\bibnamefont {Dunsworth}}, \bibinfo {author}
  {\bibfnamefont {T.}~\bibnamefont {Huang}}, \bibinfo {author} {\bibfnamefont
  {S.}~\bibnamefont {Kim}}, \bibinfo {author} {\bibfnamefont {B.}~\bibnamefont
  {Burkett}}, \bibinfo {author} {\bibfnamefont {A.}~\bibnamefont {Fowler}},
  \bibinfo {author} {\bibfnamefont {F.}~\bibnamefont {Arute}}, \bibinfo
  {author} {\bibfnamefont {J.~C.}\ \bibnamefont {Bardin}}, \bibinfo {author}
  {\bibfnamefont {A.}~\bibnamefont {Bengtsson}}, \bibinfo {author}
  {\bibfnamefont {A.}~\bibnamefont {Bilmes}}, \bibinfo {author} {\bibfnamefont
  {B.~B.}\ \bibnamefont {Buckley}}, \bibinfo {author} {\bibfnamefont
  {N.}~\bibnamefont {Bushnell}}, \bibinfo {author} {\bibfnamefont
  {Z.}~\bibnamefont {Chen}}, \bibinfo {author} {\bibfnamefont {R.}~\bibnamefont
  {Collins}}, \bibinfo {author} {\bibfnamefont {S.}~\bibnamefont {Demura}},
  \bibinfo {author} {\bibfnamefont {A.~R.}\ \bibnamefont {Derk}}, \bibinfo
  {author} {\bibfnamefont {C.}~\bibnamefont {Erickson}}, \bibinfo {author}
  {\bibfnamefont {M.}~\bibnamefont {Giustina}}, \bibinfo {author}
  {\bibfnamefont {S.~D.}\ \bibnamefont {Harrington}}, \bibinfo {author}
  {\bibfnamefont {S.}~\bibnamefont {Hong}}, \bibinfo {author} {\bibfnamefont
  {E.}~\bibnamefont {Jeffrey}}, \bibinfo {author} {\bibfnamefont
  {J.}~\bibnamefont {Kelly}}, \bibinfo {author} {\bibfnamefont {P.~V.}\
  \bibnamefont {Klimov}}, \bibinfo {author} {\bibfnamefont {F.}~\bibnamefont
  {Kostritsa}}, \bibinfo {author} {\bibfnamefont {P.}~\bibnamefont {Laptev}},
  \bibinfo {author} {\bibfnamefont {A.}~\bibnamefont {Locharla}}, \bibinfo
  {author} {\bibfnamefont {X.}~\bibnamefont {Mi}}, \bibinfo {author}
  {\bibfnamefont {K.~C.}\ \bibnamefont {Miao}}, \bibinfo {author}
  {\bibfnamefont {S.}~\bibnamefont {Montazeri}}, \bibinfo {author}
  {\bibfnamefont {J.}~\bibnamefont {Mutus}}, \bibinfo {author} {\bibfnamefont
  {O.}~\bibnamefont {Naaman}}, \bibinfo {author} {\bibfnamefont
  {M.}~\bibnamefont {Neeley}}, \bibinfo {author} {\bibfnamefont
  {C.}~\bibnamefont {Neill}}, \bibinfo {author} {\bibfnamefont
  {A.}~\bibnamefont {Opremcak}}, \bibinfo {author} {\bibfnamefont
  {C.}~\bibnamefont {Quintana}}, \bibinfo {author} {\bibfnamefont
  {N.}~\bibnamefont {Redd}}, \bibinfo {author} {\bibfnamefont {P.}~\bibnamefont
  {Roushan}}, \bibinfo {author} {\bibfnamefont {D.}~\bibnamefont {Sank}},
  \bibinfo {author} {\bibfnamefont {K.~J.}\ \bibnamefont {Satzinger}}, \bibinfo
  {author} {\bibfnamefont {V.}~\bibnamefont {Shvarts}}, \bibinfo {author}
  {\bibfnamefont {T.}~\bibnamefont {White}}, \bibinfo {author} {\bibfnamefont
  {Z.~J.}\ \bibnamefont {Yao}}, \bibinfo {author} {\bibfnamefont
  {P.}~\bibnamefont {Yeh}}, \bibinfo {author} {\bibfnamefont {J.}~\bibnamefont
  {Yoo}}, \bibinfo {author} {\bibfnamefont {Y.}~\bibnamefont {Chen}}, \bibinfo
  {author} {\bibfnamefont {V.}~\bibnamefont {Smelyanskiy}}, \bibinfo {author}
  {\bibfnamefont {J.~M.}\ \bibnamefont {Martinis}}, \bibinfo {author}
  {\bibfnamefont {H.}~\bibnamefont {Neven}}, \bibinfo {author} {\bibfnamefont
  {A.}~\bibnamefont {Megrant}}, \bibinfo {author} {\bibfnamefont
  {L.}~\bibnamefont {Ioffe}},\ and\ \bibinfo {author} {\bibfnamefont
  {R.}~\bibnamefont {Barends}},\ }\bibfield  {title} {\bibinfo {title}
  {Resolving catastrophic error bursts from cosmic rays in large arrays of
  superconducting qubits},\ }\href {https://doi.org/10.1038/s41567-021-01432-8}
  {\bibfield  {journal} {\bibinfo  {journal} {Nature Physics}\ }\textbf
  {\bibinfo {volume} {18}},\ \bibinfo {pages} {107} (\bibinfo {year}
  {2021})}\BibitemShut {NoStop}%
\bibitem [{\citenamefont {Wilen}\ \emph {et~al.}(2021)\citenamefont {Wilen},
  \citenamefont {Abdullah}, \citenamefont {Kurinsky}, \citenamefont {Stanford},
  \citenamefont {Cardani}, \citenamefont {D{'}Imperio}, \citenamefont {Tomei},
  \citenamefont {Faoro}, \citenamefont {Ioffe}, \citenamefont {Liu},
  \citenamefont {Opremcak}, \citenamefont {Christensen}, \citenamefont
  {DuBois},\ and\ \citenamefont {McDermott}}]{Wilen2021}%
  \BibitemOpen
  \bibfield  {author} {\bibinfo {author} {\bibfnamefont {C.~D.}\ \bibnamefont
  {Wilen}}, \bibinfo {author} {\bibfnamefont {S.}~\bibnamefont {Abdullah}},
  \bibinfo {author} {\bibfnamefont {N.~A.}\ \bibnamefont {Kurinsky}}, \bibinfo
  {author} {\bibfnamefont {C.}~\bibnamefont {Stanford}}, \bibinfo {author}
  {\bibfnamefont {L.}~\bibnamefont {Cardani}}, \bibinfo {author} {\bibfnamefont
  {G.}~\bibnamefont {D{'}Imperio}}, \bibinfo {author} {\bibfnamefont
  {C.}~\bibnamefont {Tomei}}, \bibinfo {author} {\bibfnamefont
  {L.}~\bibnamefont {Faoro}}, \bibinfo {author} {\bibfnamefont {L.~B.}\
  \bibnamefont {Ioffe}}, \bibinfo {author} {\bibfnamefont {C.~H.}\ \bibnamefont
  {Liu}}, \bibinfo {author} {\bibfnamefont {A.}~\bibnamefont {Opremcak}},
  \bibinfo {author} {\bibfnamefont {B.~G.}\ \bibnamefont {Christensen}},
  \bibinfo {author} {\bibfnamefont {J.~L.}\ \bibnamefont {DuBois}},\ and\
  \bibinfo {author} {\bibfnamefont {R.}~\bibnamefont {McDermott}},\ }\bibfield
  {title} {\bibinfo {title} {{Correlated charge noise and relaxation errors in
  superconducting qubits}},\ }\href
  {https://doi.org/10.1038/s41586-021-03557-5} {\bibfield  {journal} {\bibinfo
  {journal} {Nature}\ }\textbf {\bibinfo {volume} {594}},\ \bibinfo {pages}
  {369} (\bibinfo {year} {2021})}\BibitemShut {NoStop}%
\bibitem [{\citenamefont {Thorbeck}\ \emph {et~al.}(2023)\citenamefont
  {Thorbeck}, \citenamefont {Eddins}, \citenamefont {Lauer}, \citenamefont
  {McClure},\ and\ \citenamefont {Carroll}}]{Thorbeck2023}%
  \BibitemOpen
  \bibfield  {author} {\bibinfo {author} {\bibfnamefont {T.}~\bibnamefont
  {Thorbeck}}, \bibinfo {author} {\bibfnamefont {A.}~\bibnamefont {Eddins}},
  \bibinfo {author} {\bibfnamefont {I.}~\bibnamefont {Lauer}}, \bibinfo
  {author} {\bibfnamefont {D.~T.}\ \bibnamefont {McClure}},\ and\ \bibinfo
  {author} {\bibfnamefont {M.}~\bibnamefont {Carroll}},\ }\bibfield  {title}
  {\bibinfo {title} {{Two-Level-System Dynamics in a Superconducting Qubit Due
  to Background Ionizing Radiation}},\ }\href
  {https://doi.org/10.1103/PRXQuantum.4.020356} {\bibfield  {journal} {\bibinfo
   {journal} {PRX Quantum}\ }\textbf {\bibinfo {volume} {4}},\ \bibinfo {pages}
  {020356} (\bibinfo {year} {2023})}\BibitemShut {NoStop}%
\bibitem [{\citenamefont {Yelton}\ \emph {et~al.}(2024)\citenamefont {Yelton},
  \citenamefont {Larson}, \citenamefont {Iaia}, \citenamefont {Dodge},
  \citenamefont {La~Magna}, \citenamefont {Baity}, \citenamefont
  {Pechenezhskiy}, \citenamefont {McDermott}, \citenamefont {Kurinsky},
  \citenamefont {Catelani},\ and\ \citenamefont {Plourde}}]{Yelton2024}%
  \BibitemOpen
  \bibfield  {author} {\bibinfo {author} {\bibfnamefont {E.}~\bibnamefont
  {Yelton}}, \bibinfo {author} {\bibfnamefont {C.~P.}\ \bibnamefont {Larson}},
  \bibinfo {author} {\bibfnamefont {V.}~\bibnamefont {Iaia}}, \bibinfo {author}
  {\bibfnamefont {K.}~\bibnamefont {Dodge}}, \bibinfo {author} {\bibfnamefont
  {G.}~\bibnamefont {La~Magna}}, \bibinfo {author} {\bibfnamefont {P.~G.}\
  \bibnamefont {Baity}}, \bibinfo {author} {\bibfnamefont {I.~V.}\ \bibnamefont
  {Pechenezhskiy}}, \bibinfo {author} {\bibfnamefont {R.}~\bibnamefont
  {McDermott}}, \bibinfo {author} {\bibfnamefont {N.~A.}\ \bibnamefont
  {Kurinsky}}, \bibinfo {author} {\bibfnamefont {G.}~\bibnamefont {Catelani}},\
  and\ \bibinfo {author} {\bibfnamefont {B.~L.~T.}\ \bibnamefont {Plourde}},\
  }\bibfield  {title} {\bibinfo {title} {{Modeling phonon-mediated
  quasiparticle poisoning in superconducting qubit arrays}},\ }\href
  {https://doi.org/10.1103/PhysRevB.110.024519} {\bibfield  {journal} {\bibinfo
   {journal} {Phys. Rev. B}\ }\textbf {\bibinfo {volume} {110}},\ \bibinfo
  {pages} {024519} (\bibinfo {year} {2024})}\BibitemShut {NoStop}%
\bibitem [{\citenamefont {Janzen}\ \emph {et~al.}(2022)\citenamefont {Janzen},
  \citenamefont {Kononenko}, \citenamefont {Ren},\ and\ \citenamefont
  {Lupascu}}]{Janzen__Aluminum_air_bridges__2022}%
  \BibitemOpen
  \bibfield  {author} {\bibinfo {author} {\bibfnamefont {N.}~\bibnamefont
  {Janzen}}, \bibinfo {author} {\bibfnamefont {M.}~\bibnamefont {Kononenko}},
  \bibinfo {author} {\bibfnamefont {S.}~\bibnamefont {Ren}},\ and\ \bibinfo
  {author} {\bibfnamefont {A.}~\bibnamefont {Lupascu}},\ }\bibfield  {title}
  {\bibinfo {title} {{Aluminum air bridges for superconducting quantum devices
  realized using a single-step electron-beam lithography process}},\ }\href
  {https://doi.org/10.1063/5.0103165} {\bibfield  {journal} {\bibinfo
  {journal} {Applied Physics Letters}\ }\textbf {\bibinfo {volume} {121}},\
  \bibinfo {pages} {094001} (\bibinfo {year} {2022})},\ \Eprint
  {https://arxiv.org/abs/https://pubs.aip.org/aip/apl/article-pdf/doi/10.1063/5.0103165/16640448/094001\_1\_online.pdf}
  {https://pubs.aip.org/aip/apl/article-pdf/doi/10.1063/5.0103165/16640448/094001\_1\_online.pdf}
  \BibitemShut {NoStop}%
\bibitem [{\citenamefont {Bronn}\ \emph {et~al.}(2018)\citenamefont {Bronn},
  \citenamefont {Adiga}, \citenamefont {Olivadese}, \citenamefont {Wu},
  \citenamefont {Chow},\ and\ \citenamefont
  {Pappas}}]{Bronn__High_coherence_plane_breaking_packaging_for_superconducting_qubits__2018}%
  \BibitemOpen
  \bibfield  {author} {\bibinfo {author} {\bibfnamefont {N.~T.}\ \bibnamefont
  {Bronn}}, \bibinfo {author} {\bibfnamefont {V.~P.}\ \bibnamefont {Adiga}},
  \bibinfo {author} {\bibfnamefont {S.~B.}\ \bibnamefont {Olivadese}}, \bibinfo
  {author} {\bibfnamefont {X.}~\bibnamefont {Wu}}, \bibinfo {author}
  {\bibfnamefont {J.~M.}\ \bibnamefont {Chow}},\ and\ \bibinfo {author}
  {\bibfnamefont {D.~P.}\ \bibnamefont {Pappas}},\ }\bibfield  {title}
  {\bibinfo {title} {High coherence plane breaking packaging for
  superconducting qubits},\ }\href {https://doi.org/10.1088/2058-9565/aaa645}
  {\bibfield  {journal} {\bibinfo  {journal} {Quantum Science and Technology}\
  }\textbf {\bibinfo {volume} {3}},\ \bibinfo {pages} {024007} (\bibinfo {year}
  {2018})}\BibitemShut {NoStop}%
\bibitem [{\citenamefont {Mallek}\ \emph {et~al.}(2021)\citenamefont {Mallek},
  \citenamefont {Yost}, \citenamefont {Rosenberg}, \citenamefont {Yoder},
  \citenamefont {Calusine}, \citenamefont {Cook}, \citenamefont {Das},
  \citenamefont {Day}, \citenamefont {Golden}, \citenamefont {Kim},
  \citenamefont {Knecht}, \citenamefont {Niedzielski}, \citenamefont
  {Schwartz}, \citenamefont {Sevi}, \citenamefont {Stull}, \citenamefont
  {Woods}, \citenamefont {Kerman},\ and\ \citenamefont
  {Oliver}}]{Mallek__Fabrication_of_superconducting_through_silicon_vias__2021}%
  \BibitemOpen
  \bibfield  {author} {\bibinfo {author} {\bibfnamefont {J.~L.}\ \bibnamefont
  {Mallek}}, \bibinfo {author} {\bibfnamefont {D.-R.~W.}\ \bibnamefont {Yost}},
  \bibinfo {author} {\bibfnamefont {D.}~\bibnamefont {Rosenberg}}, \bibinfo
  {author} {\bibfnamefont {J.~L.}\ \bibnamefont {Yoder}}, \bibinfo {author}
  {\bibfnamefont {G.}~\bibnamefont {Calusine}}, \bibinfo {author}
  {\bibfnamefont {M.}~\bibnamefont {Cook}}, \bibinfo {author} {\bibfnamefont
  {R.}~\bibnamefont {Das}}, \bibinfo {author} {\bibfnamefont {A.}~\bibnamefont
  {Day}}, \bibinfo {author} {\bibfnamefont {E.}~\bibnamefont {Golden}},
  \bibinfo {author} {\bibfnamefont {D.~K.}\ \bibnamefont {Kim}}, \bibinfo
  {author} {\bibfnamefont {J.}~\bibnamefont {Knecht}}, \bibinfo {author}
  {\bibfnamefont {B.~M.}\ \bibnamefont {Niedzielski}}, \bibinfo {author}
  {\bibfnamefont {M.}~\bibnamefont {Schwartz}}, \bibinfo {author}
  {\bibfnamefont {A.}~\bibnamefont {Sevi}}, \bibinfo {author} {\bibfnamefont
  {C.}~\bibnamefont {Stull}}, \bibinfo {author} {\bibfnamefont
  {W.}~\bibnamefont {Woods}}, \bibinfo {author} {\bibfnamefont {A.~J.}\
  \bibnamefont {Kerman}},\ and\ \bibinfo {author} {\bibfnamefont {W.~D.}\
  \bibnamefont {Oliver}},\ }\bibfield  {title} {\bibinfo {title} {{Fabrication
  of superconducting through-silicon vias}},\ }\bibfield  {journal} {\bibinfo
  {journal} {arXiv}\ }\href {https://doi.org/10.48550/arXiv.2103.08536}
  {10.48550/arXiv.2103.08536} (\bibinfo {year} {2021}),\ \Eprint
  {https://arxiv.org/abs/2103.08536} {2103.08536} \BibitemShut {NoStop}%
\bibitem [{\citenamefont {Mollenhauer}\ \emph {et~al.}(2024)\citenamefont
  {Mollenhauer}, \citenamefont {Irfan}, \citenamefont {Cao}, \citenamefont
  {Mandal},\ and\ \citenamefont
  {Pfaff}}]{Mollenhauer__high_efficiency_plug_and_play_superconducting_qubit__2024}%
  \BibitemOpen
  \bibfield  {author} {\bibinfo {author} {\bibfnamefont {M.}~\bibnamefont
  {Mollenhauer}}, \bibinfo {author} {\bibfnamefont {A.}~\bibnamefont {Irfan}},
  \bibinfo {author} {\bibfnamefont {X.}~\bibnamefont {Cao}}, \bibinfo {author}
  {\bibfnamefont {S.}~\bibnamefont {Mandal}},\ and\ \bibinfo {author}
  {\bibfnamefont {W.}~\bibnamefont {Pfaff}},\ }\href
  {https://arxiv.org/abs/2407.16743} {\bibinfo {title} {A high-efficiency
  plug-and-play superconducting qubit network}} (\bibinfo {year} {2024}),\
  \Eprint {https://arxiv.org/abs/2407.16743} {arXiv:2407.16743 [quant-ph]}
  \BibitemShut {NoStop}%
\bibitem [{\citenamefont {Chou}\ \emph {et~al.}(2024)\citenamefont {Chou},
  \citenamefont {Shemma}, \citenamefont {McCarrick}, \citenamefont {Chien},
  \citenamefont {Teoh}, \citenamefont {Winkel}, \citenamefont {Anderson},
  \citenamefont {Chen}, \citenamefont {Curtis}, \citenamefont {de~Graaf},
  \citenamefont {Garmon}, \citenamefont {Gudlewski}, \citenamefont {Kalfus},
  \citenamefont {Keen}, \citenamefont {Khedkar}, \citenamefont {Lei},
  \citenamefont {Liu}, \citenamefont {Lu}, \citenamefont {Lu}, \citenamefont
  {Maiti}, \citenamefont {Mastalli-Kelly}, \citenamefont {Mehta}, \citenamefont
  {Mundhada}, \citenamefont {Narla}, \citenamefont {Noh}, \citenamefont
  {Tsunoda}, \citenamefont {Xue}, \citenamefont {Yuan}, \citenamefont
  {Frunzio}, \citenamefont {Aumentado}, \citenamefont {Puri}, \citenamefont
  {Girvin}, \citenamefont {Moseley},\ and\ \citenamefont
  {Schoelkopf}}]{Chou__A_superconducting_dual_rail_cavity_qubit_with_erasure_detected_logical_measurements__2024}%
  \BibitemOpen
  \bibfield  {author} {\bibinfo {author} {\bibfnamefont {K.~S.}\ \bibnamefont
  {Chou}}, \bibinfo {author} {\bibfnamefont {T.}~\bibnamefont {Shemma}},
  \bibinfo {author} {\bibfnamefont {H.}~\bibnamefont {McCarrick}}, \bibinfo
  {author} {\bibfnamefont {T.-C.}\ \bibnamefont {Chien}}, \bibinfo {author}
  {\bibfnamefont {J.~D.}\ \bibnamefont {Teoh}}, \bibinfo {author}
  {\bibfnamefont {P.}~\bibnamefont {Winkel}}, \bibinfo {author} {\bibfnamefont
  {A.}~\bibnamefont {Anderson}}, \bibinfo {author} {\bibfnamefont
  {J.}~\bibnamefont {Chen}}, \bibinfo {author} {\bibfnamefont {J.~C.}\
  \bibnamefont {Curtis}}, \bibinfo {author} {\bibfnamefont {S.~J.}\
  \bibnamefont {de~Graaf}}, \bibinfo {author} {\bibfnamefont {J.~W.~O.}\
  \bibnamefont {Garmon}}, \bibinfo {author} {\bibfnamefont {B.}~\bibnamefont
  {Gudlewski}}, \bibinfo {author} {\bibfnamefont {W.~D.}\ \bibnamefont
  {Kalfus}}, \bibinfo {author} {\bibfnamefont {T.}~\bibnamefont {Keen}},
  \bibinfo {author} {\bibfnamefont {N.}~\bibnamefont {Khedkar}}, \bibinfo
  {author} {\bibfnamefont {C.~U.}\ \bibnamefont {Lei}}, \bibinfo {author}
  {\bibfnamefont {G.}~\bibnamefont {Liu}}, \bibinfo {author} {\bibfnamefont
  {P.}~\bibnamefont {Lu}}, \bibinfo {author} {\bibfnamefont {Y.}~\bibnamefont
  {Lu}}, \bibinfo {author} {\bibfnamefont {A.}~\bibnamefont {Maiti}}, \bibinfo
  {author} {\bibfnamefont {L.}~\bibnamefont {Mastalli-Kelly}}, \bibinfo
  {author} {\bibfnamefont {N.}~\bibnamefont {Mehta}}, \bibinfo {author}
  {\bibfnamefont {S.~O.}\ \bibnamefont {Mundhada}}, \bibinfo {author}
  {\bibfnamefont {A.}~\bibnamefont {Narla}}, \bibinfo {author} {\bibfnamefont
  {T.}~\bibnamefont {Noh}}, \bibinfo {author} {\bibfnamefont {T.}~\bibnamefont
  {Tsunoda}}, \bibinfo {author} {\bibfnamefont {S.~H.}\ \bibnamefont {Xue}},
  \bibinfo {author} {\bibfnamefont {J.~O.}\ \bibnamefont {Yuan}}, \bibinfo
  {author} {\bibfnamefont {L.}~\bibnamefont {Frunzio}}, \bibinfo {author}
  {\bibfnamefont {J.}~\bibnamefont {Aumentado}}, \bibinfo {author}
  {\bibfnamefont {S.}~\bibnamefont {Puri}}, \bibinfo {author} {\bibfnamefont
  {S.~M.}\ \bibnamefont {Girvin}}, \bibinfo {author} {\bibfnamefont {S.~H.}\
  \bibnamefont {Moseley}},\ and\ \bibinfo {author} {\bibfnamefont {R.~J.}\
  \bibnamefont {Schoelkopf}},\ }\bibfield  {title} {\bibinfo {title} {A
  superconducting dual-rail cavity qubit with erasure-detected logical
  measurements},\ }\bibfield  {journal} {\bibinfo  {journal} {Nature Physics}\
  }\href {https://doi.org/10.1038/s41567-024-02539-4}
  {10.1038/s41567-024-02539-4} (\bibinfo {year} {2024})\BibitemShut {NoStop}%
\bibitem [{\citenamefont {Copetudo}\ \emph {et~al.}(2024)\citenamefont
  {Copetudo}, \citenamefont {Fontaine}, \citenamefont {Valadares},\ and\
  \citenamefont {Gao}}]{Copetudo2024Feb}%
  \BibitemOpen
  \bibfield  {author} {\bibinfo {author} {\bibfnamefont {A.}~\bibnamefont
  {Copetudo}}, \bibinfo {author} {\bibfnamefont {C.~Y.}\ \bibnamefont
  {Fontaine}}, \bibinfo {author} {\bibfnamefont {F.}~\bibnamefont
  {Valadares}},\ and\ \bibinfo {author} {\bibfnamefont {Y.~Y.}\ \bibnamefont
  {Gao}},\ }\bibfield  {title} {\bibinfo {title} {{Shaping photons: Quantum
  information processing with bosonic cQED}},\ }\href
  {https://doi.org/10.1063/5.0183022} {\bibfield  {journal} {\bibinfo
  {journal} {Appl. Phys. Lett.}\ }\textbf {\bibinfo {volume} {124}},\ \bibinfo
  {pages} {080502} (\bibinfo {year} {2024})}\BibitemShut {NoStop}%
\bibitem [{\citenamefont {Spring}\ \emph {et~al.}(2022)\citenamefont {Spring},
  \citenamefont {Cao}, \citenamefont {Tsunoda}, \citenamefont {Campanaro},
  \citenamefont {Fasciati}, \citenamefont {Wills}, \citenamefont {Bakr},
  \citenamefont {Chidambaram}, \citenamefont {Shteynas}, \citenamefont
  {Carpenter}, \citenamefont {Gow}, \citenamefont {Gates}, \citenamefont
  {Vlastakis},\ and\ \citenamefont {Leek}}]{Spring2022}%
  \BibitemOpen
  \bibfield  {author} {\bibinfo {author} {\bibfnamefont {P.~A.}\ \bibnamefont
  {Spring}}, \bibinfo {author} {\bibfnamefont {S.}~\bibnamefont {Cao}},
  \bibinfo {author} {\bibfnamefont {T.}~\bibnamefont {Tsunoda}}, \bibinfo
  {author} {\bibfnamefont {G.}~\bibnamefont {Campanaro}}, \bibinfo {author}
  {\bibfnamefont {S.}~\bibnamefont {Fasciati}}, \bibinfo {author}
  {\bibfnamefont {J.}~\bibnamefont {Wills}}, \bibinfo {author} {\bibfnamefont
  {M.}~\bibnamefont {Bakr}}, \bibinfo {author} {\bibfnamefont {V.}~\bibnamefont
  {Chidambaram}}, \bibinfo {author} {\bibfnamefont {B.}~\bibnamefont
  {Shteynas}}, \bibinfo {author} {\bibfnamefont {L.}~\bibnamefont {Carpenter}},
  \bibinfo {author} {\bibfnamefont {P.}~\bibnamefont {Gow}}, \bibinfo {author}
  {\bibfnamefont {J.}~\bibnamefont {Gates}}, \bibinfo {author} {\bibfnamefont
  {B.}~\bibnamefont {Vlastakis}},\ and\ \bibinfo {author} {\bibfnamefont
  {P.~J.}\ \bibnamefont {Leek}},\ }\bibfield  {title} {\bibinfo {title} {High
  coherence and low cross-talk in a tileable 3d integrated superconducting
  circuit architecture},\ }\href {https://doi.org/10.1126/sciadv.abl6698}
  {\bibfield  {journal} {\bibinfo  {journal} {Science Advances}\ }\textbf
  {\bibinfo {volume} {8}},\ \bibinfo {pages} {eabl6698} (\bibinfo {year}
  {2022})},\ \Eprint
  {https://arxiv.org/abs/https://www.science.org/doi/pdf/10.1126/sciadv.abl6698}
  {https://www.science.org/doi/pdf/10.1126/sciadv.abl6698} \BibitemShut
  {NoStop}%
\bibitem [{\citenamefont {Kosen}\ \emph {et~al.}(2022)\citenamefont {Kosen},
  \citenamefont {Li}, \citenamefont {Rommel}, \citenamefont {Shiri},
  \citenamefont {Warren}, \citenamefont {Grönberg}, \citenamefont {Salonen},
  \citenamefont {Abad}, \citenamefont {Biznárová}, \citenamefont {Caputo},
  \citenamefont {Chen}, \citenamefont {Grigoras}, \citenamefont {Johansson},
  \citenamefont {Kockum}, \citenamefont {Križan}, \citenamefont {Lozano},
  \citenamefont {Norris}, \citenamefont {Osman}, \citenamefont
  {Fernández-Pendás}, \citenamefont {Ronzani}, \citenamefont {Roudsari},
  \citenamefont {Simbierowicz}, \citenamefont {Tancredi}, \citenamefont
  {Wallraff}, \citenamefont {Eichler}, \citenamefont {Govenius},\ and\
  \citenamefont {Bylander}}]{Kosen__Two_Tmon_FlipChip_Device__2022}%
  \BibitemOpen
  \bibfield  {author} {\bibinfo {author} {\bibfnamefont {S.}~\bibnamefont
  {Kosen}}, \bibinfo {author} {\bibfnamefont {H.-X.}\ \bibnamefont {Li}},
  \bibinfo {author} {\bibfnamefont {M.}~\bibnamefont {Rommel}}, \bibinfo
  {author} {\bibfnamefont {D.}~\bibnamefont {Shiri}}, \bibinfo {author}
  {\bibfnamefont {C.}~\bibnamefont {Warren}}, \bibinfo {author} {\bibfnamefont
  {L.}~\bibnamefont {Grönberg}}, \bibinfo {author} {\bibfnamefont
  {J.}~\bibnamefont {Salonen}}, \bibinfo {author} {\bibfnamefont
  {T.}~\bibnamefont {Abad}}, \bibinfo {author} {\bibfnamefont {J.}~\bibnamefont
  {Biznárová}}, \bibinfo {author} {\bibfnamefont {M.}~\bibnamefont {Caputo}},
  \bibinfo {author} {\bibfnamefont {L.}~\bibnamefont {Chen}}, \bibinfo {author}
  {\bibfnamefont {K.}~\bibnamefont {Grigoras}}, \bibinfo {author}
  {\bibfnamefont {G.}~\bibnamefont {Johansson}}, \bibinfo {author}
  {\bibfnamefont {A.~F.}\ \bibnamefont {Kockum}}, \bibinfo {author}
  {\bibfnamefont {C.}~\bibnamefont {Križan}}, \bibinfo {author} {\bibfnamefont
  {D.~P.}\ \bibnamefont {Lozano}}, \bibinfo {author} {\bibfnamefont {G.~J.}\
  \bibnamefont {Norris}}, \bibinfo {author} {\bibfnamefont {A.}~\bibnamefont
  {Osman}}, \bibinfo {author} {\bibfnamefont {J.}~\bibnamefont
  {Fernández-Pendás}}, \bibinfo {author} {\bibfnamefont {A.}~\bibnamefont
  {Ronzani}}, \bibinfo {author} {\bibfnamefont {A.~F.}\ \bibnamefont
  {Roudsari}}, \bibinfo {author} {\bibfnamefont {S.}~\bibnamefont
  {Simbierowicz}}, \bibinfo {author} {\bibfnamefont {G.}~\bibnamefont
  {Tancredi}}, \bibinfo {author} {\bibfnamefont {A.}~\bibnamefont {Wallraff}},
  \bibinfo {author} {\bibfnamefont {C.}~\bibnamefont {Eichler}}, \bibinfo
  {author} {\bibfnamefont {J.}~\bibnamefont {Govenius}},\ and\ \bibinfo
  {author} {\bibfnamefont {J.}~\bibnamefont {Bylander}},\ }\bibfield  {title}
  {\bibinfo {title} {Building blocks of a flip-chip integrated superconducting
  quantum processor},\ }\href {https://doi.org/10.1088/2058-9565/ac734b}
  {\bibfield  {journal} {\bibinfo  {journal} {Quantum Science and Technology}\
  }\textbf {\bibinfo {volume} {7}},\ \bibinfo {pages} {035018} (\bibinfo {year}
  {2022})}\BibitemShut {NoStop}%
\bibitem [{\citenamefont {Conner}\ \emph {et~al.}(2021)\citenamefont {Conner},
  \citenamefont {Bienfait}, \citenamefont {Chang}, \citenamefont {Chou},
  \citenamefont {Dumur}, \citenamefont {Grebel}, \citenamefont {Peairs},
  \citenamefont {Povey}, \citenamefont {Yan}, \citenamefont {Zhong},\ and\
  \citenamefont
  {Cleland}}]{Conner__Superconducting_qubits_in_a_flip_chip_architecture__2021}%
  \BibitemOpen
  \bibfield  {author} {\bibinfo {author} {\bibfnamefont {C.~R.}\ \bibnamefont
  {Conner}}, \bibinfo {author} {\bibfnamefont {A.}~\bibnamefont {Bienfait}},
  \bibinfo {author} {\bibfnamefont {H.-S.}\ \bibnamefont {Chang}}, \bibinfo
  {author} {\bibfnamefont {M.-H.}\ \bibnamefont {Chou}}, \bibinfo {author}
  {\bibfnamefont {Ã.}~\bibnamefont {Dumur}}, \bibinfo {author} {\bibfnamefont
  {J.}~\bibnamefont {Grebel}}, \bibinfo {author} {\bibfnamefont {G.~A.}\
  \bibnamefont {Peairs}}, \bibinfo {author} {\bibfnamefont {R.~G.}\
  \bibnamefont {Povey}}, \bibinfo {author} {\bibfnamefont {H.}~\bibnamefont
  {Yan}}, \bibinfo {author} {\bibfnamefont {Y.~P.}\ \bibnamefont {Zhong}},\
  and\ \bibinfo {author} {\bibfnamefont {A.~N.}\ \bibnamefont {Cleland}},\
  }\bibfield  {title} {\bibinfo {title} {Superconducting qubits in a flip-chip
  architecture},\ }\bibfield  {journal} {\bibinfo  {journal} {Applied Physics
  Letters}\ }\textbf {\bibinfo {volume} {118}},\ \href
  {https://doi.org/10.1063/5.0050173} {10.1063/5.0050173} (\bibinfo {year}
  {2021})\BibitemShut {NoStop}%
\bibitem [{\citenamefont {Field}\ \emph {et~al.}(2024)\citenamefont {Field},
  \citenamefont {Chen}, \citenamefont {Scharmann}, \citenamefont {Sete},
  \citenamefont {Oruc}, \citenamefont {Vu}, \citenamefont {Kosenko},
  \citenamefont {Mutus}, \citenamefont {Poletto},\ and\ \citenamefont
  {Bestwick}}]{Field__Modular_Superconducting_Qubit_Architecture_with_a_Multi_chip_Tunable_Coupler__2024}%
  \BibitemOpen
  \bibfield  {author} {\bibinfo {author} {\bibfnamefont {M.}~\bibnamefont
  {Field}}, \bibinfo {author} {\bibfnamefont {A.~Q.}\ \bibnamefont {Chen}},
  \bibinfo {author} {\bibfnamefont {B.}~\bibnamefont {Scharmann}}, \bibinfo
  {author} {\bibfnamefont {E.~A.}\ \bibnamefont {Sete}}, \bibinfo {author}
  {\bibfnamefont {F.}~\bibnamefont {Oruc}}, \bibinfo {author} {\bibfnamefont
  {K.}~\bibnamefont {Vu}}, \bibinfo {author} {\bibfnamefont {V.}~\bibnamefont
  {Kosenko}}, \bibinfo {author} {\bibfnamefont {J.~Y.}\ \bibnamefont {Mutus}},
  \bibinfo {author} {\bibfnamefont {S.}~\bibnamefont {Poletto}},\ and\ \bibinfo
  {author} {\bibfnamefont {A.}~\bibnamefont {Bestwick}},\ }\bibfield  {title}
  {\bibinfo {title} {{Modular superconducting-qubit architecture with a
  multichip tunable coupler}},\ }\href
  {https://doi.org/10.1103/PhysRevApplied.21.054063} {\bibfield  {journal}
  {\bibinfo  {journal} {Phys. Rev. Appl.}\ }\textbf {\bibinfo {volume} {21}},\
  \bibinfo {pages} {054063} (\bibinfo {year} {2024})}\BibitemShut {NoStop}%
\bibitem [{\citenamefont {Bild}\ \emph {et~al.}(2023)\citenamefont {Bild},
  \citenamefont {Fadel}, \citenamefont {Yang}, \citenamefont {von
  L{\ifmmode\ddot{u}\else\"{u}\fi}pke}, \citenamefont {Martin}, \citenamefont
  {Bruno},\ and\ \citenamefont {Chu}}]{Bild2023Apr}%
  \BibitemOpen
  \bibfield  {author} {\bibinfo {author} {\bibfnamefont {M.}~\bibnamefont
  {Bild}}, \bibinfo {author} {\bibfnamefont {M.}~\bibnamefont {Fadel}},
  \bibinfo {author} {\bibfnamefont {Y.}~\bibnamefont {Yang}}, \bibinfo {author}
  {\bibfnamefont {U.}~\bibnamefont {von L{\ifmmode\ddot{u}\else\"{u}\fi}pke}},
  \bibinfo {author} {\bibfnamefont {P.}~\bibnamefont {Martin}}, \bibinfo
  {author} {\bibfnamefont {A.}~\bibnamefont {Bruno}},\ and\ \bibinfo {author}
  {\bibfnamefont {Y.}~\bibnamefont {Chu}},\ }\bibfield  {title} {\bibinfo
  {title} {{Schr{\ifmmode\ddot{o}\else\"{o}\fi}dinger cat states of a
  16-microgram mechanical oscillator}},\ }\href
  {https://doi.org/10.1126/science.adf7553} {\bibfield  {journal} {\bibinfo
  {journal} {Science}\ }\textbf {\bibinfo {volume} {380}},\ \bibinfo {pages}
  {274} (\bibinfo {year} {2023})}\BibitemShut {NoStop}%
\bibitem [{\citenamefont {Yost}\ \emph {et~al.}(2020)\citenamefont {Yost},
  \citenamefont {Schwartz}, \citenamefont {Mallek}, \citenamefont {Rosenberg},
  \citenamefont {Stull}, \citenamefont {Yoder}, \citenamefont {Calusine},
  \citenamefont {Cook}, \citenamefont {Das}, \citenamefont {Day}, \citenamefont
  {Golden}, \citenamefont {Kim}, \citenamefont {Melville}, \citenamefont
  {Niedzielski}, \citenamefont {Woods}, \citenamefont {Kerman},\ and\
  \citenamefont
  {Oliver}}]{Yost__Solid_state_qubits_integrated_with_superconducting_through_silicon_vias__2020}%
  \BibitemOpen
  \bibfield  {author} {\bibinfo {author} {\bibfnamefont {D.~R.~W.}\
  \bibnamefont {Yost}}, \bibinfo {author} {\bibfnamefont {M.~E.}\ \bibnamefont
  {Schwartz}}, \bibinfo {author} {\bibfnamefont {J.}~\bibnamefont {Mallek}},
  \bibinfo {author} {\bibfnamefont {D.}~\bibnamefont {Rosenberg}}, \bibinfo
  {author} {\bibfnamefont {C.}~\bibnamefont {Stull}}, \bibinfo {author}
  {\bibfnamefont {J.~L.}\ \bibnamefont {Yoder}}, \bibinfo {author}
  {\bibfnamefont {G.}~\bibnamefont {Calusine}}, \bibinfo {author}
  {\bibfnamefont {M.}~\bibnamefont {Cook}}, \bibinfo {author} {\bibfnamefont
  {R.}~\bibnamefont {Das}}, \bibinfo {author} {\bibfnamefont {A.~L.}\
  \bibnamefont {Day}}, \bibinfo {author} {\bibfnamefont {E.~B.}\ \bibnamefont
  {Golden}}, \bibinfo {author} {\bibfnamefont {D.~K.}\ \bibnamefont {Kim}},
  \bibinfo {author} {\bibfnamefont {A.}~\bibnamefont {Melville}}, \bibinfo
  {author} {\bibfnamefont {B.~M.}\ \bibnamefont {Niedzielski}}, \bibinfo
  {author} {\bibfnamefont {W.}~\bibnamefont {Woods}}, \bibinfo {author}
  {\bibfnamefont {A.~J.}\ \bibnamefont {Kerman}},\ and\ \bibinfo {author}
  {\bibfnamefont {W.~D.}\ \bibnamefont {Oliver}},\ }\bibfield  {title}
  {\bibinfo {title} {Solid-state qubits integrated with superconducting
  through-silicon vias},\ }\bibfield  {journal} {\bibinfo  {journal} {npj
  Quantum Information}\ }\textbf {\bibinfo {volume} {6}},\ \href
  {https://doi.org/10.1038/s41534-020-00289-8} {10.1038/s41534-020-00289-8}
  (\bibinfo {year} {2020})\BibitemShut {NoStop}%
\bibitem [{\citenamefont {Geisert}\ \emph {et~al.}(2024)\citenamefont
  {Geisert}, \citenamefont {Ihssen}, \citenamefont {Winkel}, \citenamefont
  {Spiecker}, \citenamefont {Fechant}, \citenamefont {Paluch}, \citenamefont
  {Gosling}, \citenamefont {Zapata}, \citenamefont
  {G{\ifmmode\ddot{u}\else\"{u}\fi}nzler}, \citenamefont {Rieger},
  \citenamefont
  {B{\ifmmode\acute{e}\else\'{e}\fi}n{\ifmmode\hat{a}\else\^{a}\fi}tre},
  \citenamefont {Reisinger}, \citenamefont {Wernsdorfer},\ and\ \citenamefont
  {Pop}}]{Geisert__GFQ__2024}%
  \BibitemOpen
  \bibfield  {author} {\bibinfo {author} {\bibfnamefont {S.}~\bibnamefont
  {Geisert}}, \bibinfo {author} {\bibfnamefont {S.}~\bibnamefont {Ihssen}},
  \bibinfo {author} {\bibfnamefont {P.}~\bibnamefont {Winkel}}, \bibinfo
  {author} {\bibfnamefont {M.}~\bibnamefont {Spiecker}}, \bibinfo {author}
  {\bibfnamefont {M.}~\bibnamefont {Fechant}}, \bibinfo {author} {\bibfnamefont
  {P.}~\bibnamefont {Paluch}}, \bibinfo {author} {\bibfnamefont
  {N.}~\bibnamefont {Gosling}}, \bibinfo {author} {\bibfnamefont
  {N.}~\bibnamefont {Zapata}}, \bibinfo {author} {\bibfnamefont
  {S.}~\bibnamefont {G{\ifmmode\ddot{u}\else\"{u}\fi}nzler}}, \bibinfo {author}
  {\bibfnamefont {D.}~\bibnamefont {Rieger}}, \bibinfo {author} {\bibfnamefont
  {D.}~\bibnamefont
  {B{\ifmmode\acute{e}\else\'{e}\fi}n{\ifmmode\hat{a}\else\^{a}\fi}tre}},
  \bibinfo {author} {\bibfnamefont {T.}~\bibnamefont {Reisinger}}, \bibinfo
  {author} {\bibfnamefont {W.}~\bibnamefont {Wernsdorfer}},\ and\ \bibinfo
  {author} {\bibfnamefont {I.~M.}\ \bibnamefont {Pop}},\ }\bibfield  {title}
  {\bibinfo {title} {{Pure kinetic inductance coupling for cQED with flux
  qubits}},\ }\href {https://doi.org/10.1063/5.0218361} {\bibfield  {journal}
  {\bibinfo  {journal} {Appl. Phys. Lett.}\ }\textbf {\bibinfo {volume}
  {125}},\ \bibinfo {pages} {064002} (\bibinfo {year} {2024})}\BibitemShut
  {NoStop}%
\bibitem [{\citenamefont {Winkel}\ \emph {et~al.}(2020)\citenamefont {Winkel},
  \citenamefont {Takmakov}, \citenamefont {Rieger}, \citenamefont {Planat},
  \citenamefont {Hasch-Guichard}, \citenamefont {Gr\"unhaupt}, \citenamefont
  {Maleeva}, \citenamefont {Foroughi}, \citenamefont {Henriques}, \citenamefont
  {Borisov}, \citenamefont {Ferrero}, \citenamefont {Ustinov}, \citenamefont
  {Wernsdorfer}, \citenamefont {Roch},\ and\ \citenamefont
  {Pop}}]{Winkel__Amplifier__2020}%
  \BibitemOpen
  \bibfield  {author} {\bibinfo {author} {\bibfnamefont {P.}~\bibnamefont
  {Winkel}}, \bibinfo {author} {\bibfnamefont {I.}~\bibnamefont {Takmakov}},
  \bibinfo {author} {\bibfnamefont {D.}~\bibnamefont {Rieger}}, \bibinfo
  {author} {\bibfnamefont {L.}~\bibnamefont {Planat}}, \bibinfo {author}
  {\bibfnamefont {W.}~\bibnamefont {Hasch-Guichard}}, \bibinfo {author}
  {\bibfnamefont {L.}~\bibnamefont {Gr\"unhaupt}}, \bibinfo {author}
  {\bibfnamefont {N.}~\bibnamefont {Maleeva}}, \bibinfo {author} {\bibfnamefont
  {F.}~\bibnamefont {Foroughi}}, \bibinfo {author} {\bibfnamefont
  {F.}~\bibnamefont {Henriques}}, \bibinfo {author} {\bibfnamefont
  {K.}~\bibnamefont {Borisov}}, \bibinfo {author} {\bibfnamefont
  {J.}~\bibnamefont {Ferrero}}, \bibinfo {author} {\bibfnamefont {A.~V.}\
  \bibnamefont {Ustinov}}, \bibinfo {author} {\bibfnamefont {W.}~\bibnamefont
  {Wernsdorfer}}, \bibinfo {author} {\bibfnamefont {N.}~\bibnamefont {Roch}},\
  and\ \bibinfo {author} {\bibfnamefont {I.~M.}\ \bibnamefont {Pop}},\
  }\bibfield  {title} {\bibinfo {title} {Nondegenerate parametric amplifiers
  based on dispersion-engineered josephson-junction arrays},\ }\href
  {https://doi.org/10.1103/PhysRevApplied.13.024015} {\bibfield  {journal}
  {\bibinfo  {journal} {Phys. Rev. Appl.}\ }\textbf {\bibinfo {volume} {13}},\
  \bibinfo {pages} {024015} (\bibinfo {year} {2020})}\BibitemShut {NoStop}%
\bibitem [{qde()}]{qdevil_shielding}%
  \BibitemOpen
  \href@noop {} {\bibinfo {title} {Magnetic shielding}},\ \bibinfo
  {howpublished} {\url{https://www.quantum-machines.co/products/qcage/}},\
  \bibinfo {note} {online; accessed 15-July-2024}\BibitemShut {NoStop}%
\bibitem [{\citenamefont {Reed}\ \emph {et~al.}(2010)\citenamefont {Reed},
  \citenamefont {Johnson}, \citenamefont {Houck}, \citenamefont {DiCarlo},
  \citenamefont {Chow}, \citenamefont {Schuster}, \citenamefont {Frunzio},\
  and\ \citenamefont {Schoelkopf}}]{Reed2010}%
  \BibitemOpen
  \bibfield  {author} {\bibinfo {author} {\bibfnamefont {M.~D.}\ \bibnamefont
  {Reed}}, \bibinfo {author} {\bibfnamefont {B.~R.}\ \bibnamefont {Johnson}},
  \bibinfo {author} {\bibfnamefont {A.~A.}\ \bibnamefont {Houck}}, \bibinfo
  {author} {\bibfnamefont {L.}~\bibnamefont {DiCarlo}}, \bibinfo {author}
  {\bibfnamefont {J.~M.}\ \bibnamefont {Chow}}, \bibinfo {author}
  {\bibfnamefont {D.~I.}\ \bibnamefont {Schuster}}, \bibinfo {author}
  {\bibfnamefont {L.}~\bibnamefont {Frunzio}},\ and\ \bibinfo {author}
  {\bibfnamefont {R.~J.}\ \bibnamefont {Schoelkopf}},\ }\bibfield  {title}
  {\bibinfo {title} {{Fast reset and suppressing spontaneous emission of a
  superconducting qubit}},\ }\href {https://doi.org/10.1063/1.3435463}
  {\bibfield  {journal} {\bibinfo  {journal} {Appl. Phys. Lett.}\ }\textbf
  {\bibinfo {volume} {96}},\ \bibinfo {pages} {203110} (\bibinfo {year}
  {2010})}\BibitemShut {NoStop}%
\bibitem [{\citenamefont {Sete}\ \emph {et~al.}(2015)\citenamefont {Sete},
  \citenamefont {Martinis},\ and\ \citenamefont {Korotkov}}]{Sete2015}%
  \BibitemOpen
  \bibfield  {author} {\bibinfo {author} {\bibfnamefont {E.~A.}\ \bibnamefont
  {Sete}}, \bibinfo {author} {\bibfnamefont {J.~M.}\ \bibnamefont {Martinis}},\
  and\ \bibinfo {author} {\bibfnamefont {A.~N.}\ \bibnamefont {Korotkov}},\
  }\bibfield  {title} {\bibinfo {title} {Quantum theory of a bandpass purcell
  filter for qubit readout},\ }\href
  {https://doi.org/10.1103/PhysRevA.92.012325} {\bibfield  {journal} {\bibinfo
  {journal} {Phys. Rev. A}\ }\textbf {\bibinfo {volume} {92}},\ \bibinfo
  {pages} {012325} (\bibinfo {year} {2015})}\BibitemShut {NoStop}%
\bibitem [{\citenamefont {Gambetta}\ \emph {et~al.}(2006)\citenamefont
  {Gambetta}, \citenamefont {Blais}, \citenamefont {Schuster}, \citenamefont
  {Wallraff}, \citenamefont {Frunzio}, \citenamefont {Majer}, \citenamefont
  {Devoret}, \citenamefont {Girvin},\ and\ \citenamefont
  {Schoelkopf}}]{Gambetta2006}%
  \BibitemOpen
  \bibfield  {author} {\bibinfo {author} {\bibfnamefont {J.}~\bibnamefont
  {Gambetta}}, \bibinfo {author} {\bibfnamefont {A.}~\bibnamefont {Blais}},
  \bibinfo {author} {\bibfnamefont {D.~I.}\ \bibnamefont {Schuster}}, \bibinfo
  {author} {\bibfnamefont {A.}~\bibnamefont {Wallraff}}, \bibinfo {author}
  {\bibfnamefont {L.}~\bibnamefont {Frunzio}}, \bibinfo {author} {\bibfnamefont
  {J.}~\bibnamefont {Majer}}, \bibinfo {author} {\bibfnamefont {M.~H.}\
  \bibnamefont {Devoret}}, \bibinfo {author} {\bibfnamefont {S.~M.}\
  \bibnamefont {Girvin}},\ and\ \bibinfo {author} {\bibfnamefont {R.~J.}\
  \bibnamefont {Schoelkopf}},\ }\bibfield  {title} {\bibinfo {title}
  {Qubit-photon interactions in a cavity: Measurement-induced dephasing and
  number splitting},\ }\href {https://doi.org/10.1103/PhysRevA.74.042318}
  {\bibfield  {journal} {\bibinfo  {journal} {Phys. Rev. A}\ }\textbf {\bibinfo
  {volume} {74}},\ \bibinfo {pages} {042318} (\bibinfo {year}
  {2006})}\BibitemShut {NoStop}%
\bibitem [{\citenamefont {Kosen}\ \emph {et~al.}(2024)\citenamefont {Kosen},
  \citenamefont {Li}, \citenamefont {Rommel}, \citenamefont {Rehammar},
  \citenamefont {Caputo}, \citenamefont {Gr\"onberg}, \citenamefont
  {Fern\'andez-Pend\'as}, \citenamefont {Kockum}, \citenamefont
  {Bizn\'arov\'a}, \citenamefont {Chen}, \citenamefont
  {Kri\ifmmode~\check{z}\else \v{z}\fi{}an}, \citenamefont {Nylander},
  \citenamefont {Osman}, \citenamefont {Roudsari}, \citenamefont {Shiri},
  \citenamefont {Tancredi}, \citenamefont {Govenius},\ and\ \citenamefont
  {Bylander}}]{Kosen2024}%
  \BibitemOpen
  \bibfield  {author} {\bibinfo {author} {\bibfnamefont {S.}~\bibnamefont
  {Kosen}}, \bibinfo {author} {\bibfnamefont {H.-X.}\ \bibnamefont {Li}},
  \bibinfo {author} {\bibfnamefont {M.}~\bibnamefont {Rommel}}, \bibinfo
  {author} {\bibfnamefont {R.}~\bibnamefont {Rehammar}}, \bibinfo {author}
  {\bibfnamefont {M.}~\bibnamefont {Caputo}}, \bibinfo {author} {\bibfnamefont
  {L.}~\bibnamefont {Gr\"onberg}}, \bibinfo {author} {\bibfnamefont
  {J.}~\bibnamefont {Fern\'andez-Pend\'as}}, \bibinfo {author} {\bibfnamefont
  {A.~F.}\ \bibnamefont {Kockum}}, \bibinfo {author} {\bibfnamefont
  {J.}~\bibnamefont {Bizn\'arov\'a}}, \bibinfo {author} {\bibfnamefont
  {L.}~\bibnamefont {Chen}}, \bibinfo {author} {\bibfnamefont {C.}~\bibnamefont
  {Kri\ifmmode~\check{z}\else \v{z}\fi{}an}}, \bibinfo {author} {\bibfnamefont
  {A.}~\bibnamefont {Nylander}}, \bibinfo {author} {\bibfnamefont
  {A.}~\bibnamefont {Osman}}, \bibinfo {author} {\bibfnamefont {A.~F.}\
  \bibnamefont {Roudsari}}, \bibinfo {author} {\bibfnamefont {D.}~\bibnamefont
  {Shiri}}, \bibinfo {author} {\bibfnamefont {G.}~\bibnamefont {Tancredi}},
  \bibinfo {author} {\bibfnamefont {J.}~\bibnamefont {Govenius}},\ and\
  \bibinfo {author} {\bibfnamefont {J.}~\bibnamefont {Bylander}},\ }\bibfield
  {title} {\bibinfo {title} {Signal crosstalk in a flip-chip quantum
  processor},\ }\href {https://doi.org/10.1103/PRXQuantum.5.030350} {\bibfield
  {journal} {\bibinfo  {journal} {PRX Quantum}\ }\textbf {\bibinfo {volume}
  {5}},\ \bibinfo {pages} {030350} (\bibinfo {year} {2024})}\BibitemShut
  {NoStop}%
\bibitem [{\citenamefont {Hita-P{\ifmmode\acute{e}\else\'{e}\fi}rez}\ \emph
  {et~al.}(2021)\citenamefont {Hita-P{\ifmmode\acute{e}\else\'{e}\fi}rez},
  \citenamefont {Jaum{\ifmmode\grave{a}\else\`{a}\fi}}, \citenamefont {Pino},\
  and\ \citenamefont
  {Garc{\ifmmode\acute{\imath}\else\'{\i}\fi}a-Ripoll}}]{hita2021three}%
  \BibitemOpen
  \bibfield  {author} {\bibinfo {author} {\bibfnamefont {M.}~\bibnamefont
  {Hita-P{\ifmmode\acute{e}\else\'{e}\fi}rez}}, \bibinfo {author}
  {\bibfnamefont {G.}~\bibnamefont {Jaum{\ifmmode\grave{a}\else\`{a}\fi}}},
  \bibinfo {author} {\bibfnamefont {M.}~\bibnamefont {Pino}},\ and\ \bibinfo
  {author} {\bibfnamefont {J.~J.}\ \bibnamefont
  {Garc{\ifmmode\acute{\imath}\else\'{\i}\fi}a-Ripoll}},\ }\bibfield  {title}
  {\bibinfo {title} {{Three-Josephson junctions flux qubit couplings}},\ }\href
  {https://doi.org/10.1063/5.0069530} {\bibfield  {journal} {\bibinfo
  {journal} {Appl. Phys. Lett.}\ }\textbf {\bibinfo {volume} {119}},\ \bibinfo
  {pages} {222601} (\bibinfo {year} {2021})}\BibitemShut {NoStop}%
\bibitem [{\citenamefont {Consani}\ and\ \citenamefont
  {Warburton}(2020)}]{consani2020effective}%
  \BibitemOpen
  \bibfield  {author} {\bibinfo {author} {\bibfnamefont {G.}~\bibnamefont
  {Consani}}\ and\ \bibinfo {author} {\bibfnamefont {P.~A.}\ \bibnamefont
  {Warburton}},\ }\bibfield  {title} {\bibinfo {title} {{Effective Hamiltonians
  for interacting superconducting qubits: local basis reduction and the
  Schrieffer{\textendash}Wolff transformation}},\ }\href
  {https://doi.org/10.1088/1367-2630/ab83d1} {\bibfield  {journal} {\bibinfo
  {journal} {New J. Phys.}\ }\textbf {\bibinfo {volume} {22}},\ \bibinfo
  {pages} {053040} (\bibinfo {year} {2020})}\BibitemShut {NoStop}%
\bibitem [{\citenamefont {Sung}\ \emph {et~al.}(2021)\citenamefont {Sung},
  \citenamefont {Ding}, \citenamefont
  {Braum{\ifmmode\ddot{u}\else\"{u}\fi}ller}, \citenamefont
  {Veps{\ifmmode\ddot{a}\else\"{a}\fi}l{\ifmmode\ddot{a}\else\"{a}\fi}inen},
  \citenamefont {Kannan}, \citenamefont {Kjaergaard}, \citenamefont {Greene},
  \citenamefont {Samach}, \citenamefont {McNally}, \citenamefont {Kim},
  \citenamefont {Melville}, \citenamefont {Niedzielski}, \citenamefont
  {Schwartz}, \citenamefont {Yoder}, \citenamefont {Orlando}, \citenamefont
  {Gustavsson},\ and\ \citenamefont {Oliver}}]{Sung2021}%
  \BibitemOpen
  \bibfield  {author} {\bibinfo {author} {\bibfnamefont {Y.}~\bibnamefont
  {Sung}}, \bibinfo {author} {\bibfnamefont {L.}~\bibnamefont {Ding}}, \bibinfo
  {author} {\bibfnamefont {J.}~\bibnamefont
  {Braum{\ifmmode\ddot{u}\else\"{u}\fi}ller}}, \bibinfo {author} {\bibfnamefont
  {A.}~\bibnamefont
  {Veps{\ifmmode\ddot{a}\else\"{a}\fi}l{\ifmmode\ddot{a}\else\"{a}\fi}inen}},
  \bibinfo {author} {\bibfnamefont {B.}~\bibnamefont {Kannan}}, \bibinfo
  {author} {\bibfnamefont {M.}~\bibnamefont {Kjaergaard}}, \bibinfo {author}
  {\bibfnamefont {A.}~\bibnamefont {Greene}}, \bibinfo {author} {\bibfnamefont
  {G.~O.}\ \bibnamefont {Samach}}, \bibinfo {author} {\bibfnamefont
  {C.}~\bibnamefont {McNally}}, \bibinfo {author} {\bibfnamefont
  {D.}~\bibnamefont {Kim}}, \bibinfo {author} {\bibfnamefont {A.}~\bibnamefont
  {Melville}}, \bibinfo {author} {\bibfnamefont {B.~M.}\ \bibnamefont
  {Niedzielski}}, \bibinfo {author} {\bibfnamefont {M.~E.}\ \bibnamefont
  {Schwartz}}, \bibinfo {author} {\bibfnamefont {J.~L.}\ \bibnamefont {Yoder}},
  \bibinfo {author} {\bibfnamefont {T.~P.}\ \bibnamefont {Orlando}}, \bibinfo
  {author} {\bibfnamefont {S.}~\bibnamefont {Gustavsson}},\ and\ \bibinfo
  {author} {\bibfnamefont {W.~D.}\ \bibnamefont {Oliver}},\ }\bibfield  {title}
  {\bibinfo {title} {{Realization of High-Fidelity CZ and $ZZ$-Free iSWAP Gates
  with a Tunable Coupler}},\ }\href
  {https://doi.org/10.1103/PhysRevX.11.021058} {\bibfield  {journal} {\bibinfo
  {journal} {Phys. Rev. X}\ }\textbf {\bibinfo {volume} {11}},\ \bibinfo
  {pages} {021058} (\bibinfo {year} {2021})}\BibitemShut {NoStop}%
\bibitem [{\citenamefont {Chen}\ \emph {et~al.}(2014)\citenamefont {Chen},
  \citenamefont {Neill}, \citenamefont {Roushan}, \citenamefont {Leung},
  \citenamefont {Fang}, \citenamefont {Barends}, \citenamefont {Kelly},
  \citenamefont {Campbell}, \citenamefont {Chen}, \citenamefont {Chiaro},
  \citenamefont {Dunsworth}, \citenamefont {Jeffrey}, \citenamefont {Megrant},
  \citenamefont {Mutus}, \citenamefont {O{'}Malley}, \citenamefont {Quintana},
  \citenamefont {Sank}, \citenamefont {Vainsencher}, \citenamefont {Wenner},
  \citenamefont {White}, \citenamefont {Geller}, \citenamefont {Cleland},\ and\
  \citenamefont {Martinis}}]{Chen2014}%
  \BibitemOpen
  \bibfield  {author} {\bibinfo {author} {\bibfnamefont {Y.}~\bibnamefont
  {Chen}}, \bibinfo {author} {\bibfnamefont {C.}~\bibnamefont {Neill}},
  \bibinfo {author} {\bibfnamefont {P.}~\bibnamefont {Roushan}}, \bibinfo
  {author} {\bibfnamefont {N.}~\bibnamefont {Leung}}, \bibinfo {author}
  {\bibfnamefont {M.}~\bibnamefont {Fang}}, \bibinfo {author} {\bibfnamefont
  {R.}~\bibnamefont {Barends}}, \bibinfo {author} {\bibfnamefont
  {J.}~\bibnamefont {Kelly}}, \bibinfo {author} {\bibfnamefont
  {B.}~\bibnamefont {Campbell}}, \bibinfo {author} {\bibfnamefont
  {Z.}~\bibnamefont {Chen}}, \bibinfo {author} {\bibfnamefont {B.}~\bibnamefont
  {Chiaro}}, \bibinfo {author} {\bibfnamefont {A.}~\bibnamefont {Dunsworth}},
  \bibinfo {author} {\bibfnamefont {E.}~\bibnamefont {Jeffrey}}, \bibinfo
  {author} {\bibfnamefont {A.}~\bibnamefont {Megrant}}, \bibinfo {author}
  {\bibfnamefont {J.~Y.}\ \bibnamefont {Mutus}}, \bibinfo {author}
  {\bibfnamefont {P.~J.~J.}\ \bibnamefont {O{'}Malley}}, \bibinfo {author}
  {\bibfnamefont {C.~M.}\ \bibnamefont {Quintana}}, \bibinfo {author}
  {\bibfnamefont {D.}~\bibnamefont {Sank}}, \bibinfo {author} {\bibfnamefont
  {A.}~\bibnamefont {Vainsencher}}, \bibinfo {author} {\bibfnamefont
  {J.}~\bibnamefont {Wenner}}, \bibinfo {author} {\bibfnamefont {T.~C.}\
  \bibnamefont {White}}, \bibinfo {author} {\bibfnamefont {M.~R.}\ \bibnamefont
  {Geller}}, \bibinfo {author} {\bibfnamefont {A.~N.}\ \bibnamefont
  {Cleland}},\ and\ \bibinfo {author} {\bibfnamefont {J.~M.}\ \bibnamefont
  {Martinis}},\ }\bibfield  {title} {\bibinfo {title} {{Qubit Architecture with
  High Coherence and Fast Tunable Coupling}},\ }\href
  {https://doi.org/10.1103/PhysRevLett.113.220502} {\bibfield  {journal}
  {\bibinfo  {journal} {Phys. Rev. Lett.}\ }\textbf {\bibinfo {volume} {113}},\
  \bibinfo {pages} {220502} (\bibinfo {year} {2014})}\BibitemShut {NoStop}%
\bibitem [{\citenamefont {McKay}\ \emph {et~al.}(2016)\citenamefont {McKay},
  \citenamefont {Filipp}, \citenamefont {Mezzacapo}, \citenamefont {Magesan},
  \citenamefont {Chow},\ and\ \citenamefont {Gambetta}}]{McKay2016}%
  \BibitemOpen
  \bibfield  {author} {\bibinfo {author} {\bibfnamefont {D.~C.}\ \bibnamefont
  {McKay}}, \bibinfo {author} {\bibfnamefont {S.}~\bibnamefont {Filipp}},
  \bibinfo {author} {\bibfnamefont {A.}~\bibnamefont {Mezzacapo}}, \bibinfo
  {author} {\bibfnamefont {E.}~\bibnamefont {Magesan}}, \bibinfo {author}
  {\bibfnamefont {J.~M.}\ \bibnamefont {Chow}},\ and\ \bibinfo {author}
  {\bibfnamefont {J.~M.}\ \bibnamefont {Gambetta}},\ }\bibfield  {title}
  {\bibinfo {title} {{Universal Gate for Fixed-Frequency Qubits via a Tunable
  Bus}},\ }\href {https://doi.org/10.1103/PhysRevApplied.6.064007} {\bibfield
  {journal} {\bibinfo  {journal} {Phys. Rev. Appl.}\ }\textbf {\bibinfo
  {volume} {6}},\ \bibinfo {pages} {064007} (\bibinfo {year}
  {2016})}\BibitemShut {NoStop}%
\bibitem [{\citenamefont {Weber}\ \emph {et~al.}(2017)\citenamefont {Weber},
  \citenamefont {Samach}, \citenamefont {Hover}, \citenamefont {Gustavsson},
  \citenamefont {Kim}, \citenamefont {Melville}, \citenamefont {Rosenberg},
  \citenamefont {Sears}, \citenamefont {Yan}, \citenamefont {Yoder},
  \citenamefont {Oliver},\ and\ \citenamefont {Kerman}}]{Weber2017}%
  \BibitemOpen
  \bibfield  {author} {\bibinfo {author} {\bibfnamefont {S.~J.}\ \bibnamefont
  {Weber}}, \bibinfo {author} {\bibfnamefont {G.~O.}\ \bibnamefont {Samach}},
  \bibinfo {author} {\bibfnamefont {D.}~\bibnamefont {Hover}}, \bibinfo
  {author} {\bibfnamefont {S.}~\bibnamefont {Gustavsson}}, \bibinfo {author}
  {\bibfnamefont {D.~K.}\ \bibnamefont {Kim}}, \bibinfo {author} {\bibfnamefont
  {A.}~\bibnamefont {Melville}}, \bibinfo {author} {\bibfnamefont
  {D.}~\bibnamefont {Rosenberg}}, \bibinfo {author} {\bibfnamefont {A.~P.}\
  \bibnamefont {Sears}}, \bibinfo {author} {\bibfnamefont {F.}~\bibnamefont
  {Yan}}, \bibinfo {author} {\bibfnamefont {J.~L.}\ \bibnamefont {Yoder}},
  \bibinfo {author} {\bibfnamefont {W.~D.}\ \bibnamefont {Oliver}},\ and\
  \bibinfo {author} {\bibfnamefont {A.~J.}\ \bibnamefont {Kerman}},\ }\bibfield
   {title} {\bibinfo {title} {{Coherent Coupled Qubits for Quantum
  Annealing}},\ }\href {https://doi.org/10.1103/PhysRevApplied.8.014004}
  {\bibfield  {journal} {\bibinfo  {journal} {Phys. Rev. Appl.}\ }\textbf
  {\bibinfo {volume} {8}},\ \bibinfo {pages} {014004} (\bibinfo {year}
  {2017})}\BibitemShut {NoStop}%
\bibitem [{\citenamefont {Moskalenko}\ \emph {et~al.}(2022)\citenamefont
  {Moskalenko}, \citenamefont {Simakov}, \citenamefont {Abramov}, \citenamefont
  {Grigorev}, \citenamefont {Moskalev}, \citenamefont {Pishchimova},
  \citenamefont {Smirnov}, \citenamefont {Zikiy}, \citenamefont {Rodionov},\
  and\ \citenamefont {Besedin}}]{Moskalenko2022}%
  \BibitemOpen
  \bibfield  {author} {\bibinfo {author} {\bibfnamefont {I.~N.}\ \bibnamefont
  {Moskalenko}}, \bibinfo {author} {\bibfnamefont {I.~A.}\ \bibnamefont
  {Simakov}}, \bibinfo {author} {\bibfnamefont {N.~N.}\ \bibnamefont
  {Abramov}}, \bibinfo {author} {\bibfnamefont {A.~A.}\ \bibnamefont
  {Grigorev}}, \bibinfo {author} {\bibfnamefont {D.~O.}\ \bibnamefont
  {Moskalev}}, \bibinfo {author} {\bibfnamefont {A.~A.}\ \bibnamefont
  {Pishchimova}}, \bibinfo {author} {\bibfnamefont {N.~S.}\ \bibnamefont
  {Smirnov}}, \bibinfo {author} {\bibfnamefont {E.~V.}\ \bibnamefont {Zikiy}},
  \bibinfo {author} {\bibfnamefont {I.~A.}\ \bibnamefont {Rodionov}},\ and\
  \bibinfo {author} {\bibfnamefont {I.~S.}\ \bibnamefont {Besedin}},\
  }\bibfield  {title} {\bibinfo {title} {{High fidelity two-qubit gates on
  fluxoniums using a tunable coupler}},\ }\href
  {https://doi.org/10.1038/s41534-022-00644-x} {\bibfield  {journal} {\bibinfo
  {journal} {npj Quantum Inf.}\ }\textbf {\bibinfo {volume} {8}},\ \bibinfo
  {pages} {1} (\bibinfo {year} {2022})}\BibitemShut {NoStop}%
\bibitem [{\citenamefont {Brecht}\ \emph {et~al.}(2017)\citenamefont {Brecht},
  \citenamefont {Chu}, \citenamefont {Axline}, \citenamefont {Pfaff},
  \citenamefont {Blumoff}, \citenamefont {Chou}, \citenamefont {Krayzman},
  \citenamefont {Frunzio},\ and\ \citenamefont {Schoelkopf}}]{Brecht2017}%
  \BibitemOpen
  \bibfield  {author} {\bibinfo {author} {\bibfnamefont {T.}~\bibnamefont
  {Brecht}}, \bibinfo {author} {\bibfnamefont {Y.}~\bibnamefont {Chu}},
  \bibinfo {author} {\bibfnamefont {C.}~\bibnamefont {Axline}}, \bibinfo
  {author} {\bibfnamefont {W.}~\bibnamefont {Pfaff}}, \bibinfo {author}
  {\bibfnamefont {J.~Z.}\ \bibnamefont {Blumoff}}, \bibinfo {author}
  {\bibfnamefont {K.}~\bibnamefont {Chou}}, \bibinfo {author} {\bibfnamefont
  {L.}~\bibnamefont {Krayzman}}, \bibinfo {author} {\bibfnamefont
  {L.}~\bibnamefont {Frunzio}},\ and\ \bibinfo {author} {\bibfnamefont {R.~J.}\
  \bibnamefont {Schoelkopf}},\ }\bibfield  {title} {\bibinfo {title}
  {{Micromachined Integrated Quantum Circuit Containing a Superconducting
  Qubit}},\ }\href {https://doi.org/10.1103/PhysRevApplied.7.044018} {\bibfield
   {journal} {\bibinfo  {journal} {Phys. Rev. Appl.}\ }\textbf {\bibinfo
  {volume} {7}},\ \bibinfo {pages} {044018} (\bibinfo {year}
  {2017})}\BibitemShut {NoStop}%
\bibitem [{\citenamefont {Rosenberg}\ \emph {et~al.}(2020)\citenamefont
  {Rosenberg}, \citenamefont {Weber}, \citenamefont {Conway}, \citenamefont
  {Yost}, \citenamefont {Mallek},\ and\ \citenamefont
  {Calusine}}]{Rosenberg2020}%
  \BibitemOpen
  \bibfield  {author} {\bibinfo {author} {\bibfnamefont {D.}~\bibnamefont
  {Rosenberg}}, \bibinfo {author} {\bibfnamefont {S.~J.}\ \bibnamefont
  {Weber}}, \bibinfo {author} {\bibfnamefont {D.}~\bibnamefont {Conway}},
  \bibinfo {author} {\bibfnamefont {D.-R.~W.}\ \bibnamefont {Yost}}, \bibinfo
  {author} {\bibfnamefont {J.}~\bibnamefont {Mallek}},\ and\ \bibinfo {author}
  {\bibfnamefont {G.}~\bibnamefont {Calusine}},\ }\bibfield  {title} {\bibinfo
  {title} {{Solid-State Qubits: 3D Integration and Packaging}},\ }\href
  {https://doi.org/10.1109/MMM.2020.2993478} {\bibfield  {journal} {\bibinfo
  {journal} {IEEE Microwave Mag.}\ }\textbf {\bibinfo {volume} {21}},\ \bibinfo
  {pages} {72} (\bibinfo {year} {2020})}\BibitemShut {NoStop}%
\bibitem [{\citenamefont {Bravyi}\ \emph {et~al.}(2011)\citenamefont {Bravyi},
  \citenamefont {DiVincenzo},\ and\ \citenamefont
  {Loss}}]{bravyi2011schrieffer}%
  \BibitemOpen
  \bibfield  {author} {\bibinfo {author} {\bibfnamefont {S.}~\bibnamefont
  {Bravyi}}, \bibinfo {author} {\bibfnamefont {D.~P.}\ \bibnamefont
  {DiVincenzo}},\ and\ \bibinfo {author} {\bibfnamefont {D.}~\bibnamefont
  {Loss}},\ }\bibfield  {title} {\bibinfo {title}
  {{Schrieffer{\textendash}Wolff transformation for quantum many-body
  systems}},\ }\href {https://doi.org/10.1016/j.aop.2011.06.004} {\bibfield
  {journal} {\bibinfo  {journal} {Ann. Phys.}\ }\textbf {\bibinfo {volume}
  {326}},\ \bibinfo {pages} {2793} (\bibinfo {year} {2011})}\BibitemShut
  {NoStop}%
\bibitem [{\citenamefont {Hita-P{\ifmmode\acute{e}\else\'{e}\fi}rez}\ \emph
  {et~al.}(2022)\citenamefont {Hita-P{\ifmmode\acute{e}\else\'{e}\fi}rez},
  \citenamefont {Jaum{\ifmmode\grave{a}\else\`{a}\fi}}, \citenamefont {Pino},\
  and\ \citenamefont
  {Garc{\ifmmode\acute{\imath}\else\'{\i}\fi}a-Ripoll}}]{hita2022ultrastrong}%
  \BibitemOpen
  \bibfield  {author} {\bibinfo {author} {\bibfnamefont {M.}~\bibnamefont
  {Hita-P{\ifmmode\acute{e}\else\'{e}\fi}rez}}, \bibinfo {author}
  {\bibfnamefont {G.}~\bibnamefont {Jaum{\ifmmode\grave{a}\else\`{a}\fi}}},
  \bibinfo {author} {\bibfnamefont {M.}~\bibnamefont {Pino}},\ and\ \bibinfo
  {author} {\bibfnamefont {J.~J.}\ \bibnamefont
  {Garc{\ifmmode\acute{\imath}\else\'{\i}\fi}a-Ripoll}},\ }\bibfield  {title}
  {\bibinfo {title} {{Ultrastrong Capacitive Coupling of Flux Qubits}},\ }\href
  {https://doi.org/10.1103/PhysRevApplied.17.014028} {\bibfield  {journal}
  {\bibinfo  {journal} {Phys. Rev. Appl.}\ }\textbf {\bibinfo {volume} {17}},\
  \bibinfo {pages} {014028} (\bibinfo {year} {2022})}\BibitemShut {NoStop}%
\bibitem [{\citenamefont {Yan}\ \emph {et~al.}(2018)\citenamefont {Yan},
  \citenamefont {Krantz}, \citenamefont {Sung}, \citenamefont {Kjaergaard},
  \citenamefont {Campbell}, \citenamefont {Orlando}, \citenamefont
  {Gustavsson},\ and\ \citenamefont {Oliver}}]{yan2018tunable}%
  \BibitemOpen
  \bibfield  {author} {\bibinfo {author} {\bibfnamefont {F.}~\bibnamefont
  {Yan}}, \bibinfo {author} {\bibfnamefont {P.}~\bibnamefont {Krantz}},
  \bibinfo {author} {\bibfnamefont {Y.}~\bibnamefont {Sung}}, \bibinfo {author}
  {\bibfnamefont {M.}~\bibnamefont {Kjaergaard}}, \bibinfo {author}
  {\bibfnamefont {D.~L.}\ \bibnamefont {Campbell}}, \bibinfo {author}
  {\bibfnamefont {T.~P.}\ \bibnamefont {Orlando}}, \bibinfo {author}
  {\bibfnamefont {S.}~\bibnamefont {Gustavsson}},\ and\ \bibinfo {author}
  {\bibfnamefont {W.~D.}\ \bibnamefont {Oliver}},\ }\bibfield  {title}
  {\bibinfo {title} {{Tunable Coupling Scheme for Implementing High-Fidelity
  Two-Qubit Gates}},\ }\href {https://doi.org/10.1103/PhysRevApplied.10.054062}
  {\bibfield  {journal} {\bibinfo  {journal} {Phys. Rev. Appl.}\ }\textbf
  {\bibinfo {volume} {10}},\ \bibinfo {pages} {054062} (\bibinfo {year}
  {2018})}\BibitemShut {NoStop}%
\bibitem [{\citenamefont {Wu}\ \emph {et~al.}(2024)\citenamefont {Wu},
  \citenamefont {Yan}, \citenamefont {Andersson}, \citenamefont {Anferov},
  \citenamefont {Chou}, \citenamefont {Conner}, \citenamefont {Grebel},
  \citenamefont {Joshi}, \citenamefont {Li}, \citenamefont {Miller},
  \citenamefont {Povey}, \citenamefont {Qiao},\ and\ \citenamefont
  {Cleland}}]{wu2024modularquantumprocessoralltoall}%
  \BibitemOpen
  \bibfield  {author} {\bibinfo {author} {\bibfnamefont {X.}~\bibnamefont
  {Wu}}, \bibinfo {author} {\bibfnamefont {H.}~\bibnamefont {Yan}}, \bibinfo
  {author} {\bibfnamefont {G.}~\bibnamefont {Andersson}}, \bibinfo {author}
  {\bibfnamefont {A.}~\bibnamefont {Anferov}}, \bibinfo {author} {\bibfnamefont
  {M.-H.}\ \bibnamefont {Chou}}, \bibinfo {author} {\bibfnamefont {C.~R.}\
  \bibnamefont {Conner}}, \bibinfo {author} {\bibfnamefont {J.}~\bibnamefont
  {Grebel}}, \bibinfo {author} {\bibfnamefont {Y.~J.}\ \bibnamefont {Joshi}},
  \bibinfo {author} {\bibfnamefont {S.}~\bibnamefont {Li}}, \bibinfo {author}
  {\bibfnamefont {J.~M.}\ \bibnamefont {Miller}}, \bibinfo {author}
  {\bibfnamefont {R.~G.}\ \bibnamefont {Povey}}, \bibinfo {author}
  {\bibfnamefont {H.}~\bibnamefont {Qiao}},\ and\ \bibinfo {author}
  {\bibfnamefont {A.~N.}\ \bibnamefont {Cleland}},\ }\href
  {https://arxiv.org/abs/2407.20134} {\bibinfo {title} {Modular quantum
  processor with an all-to-all reconfigurable router}} (\bibinfo {year}
  {2024}),\ \Eprint {https://arxiv.org/abs/2407.20134} {arXiv:2407.20134
  [quant-ph]} \BibitemShut {NoStop}%
\end{thebibliography}%

\newpage
\onecolumngrid
\appendix

\setcounter{equation}{0}
\setcounter{figure}{0}
\setcounter{table}{0}
\makeatletter
\renewcommand{\theequation}{A\arabic{equation}}
\renewcommand{\thefigure}{A\arabic{figure}}
\renewcommand{\thetable}{A\arabic{table}}

\newpage

\section{\label{sec:ANSYS_simulations} Finite element simulations}

\begin{figure*}[hbt!]
        \includegraphics[width=1.0\textwidth]{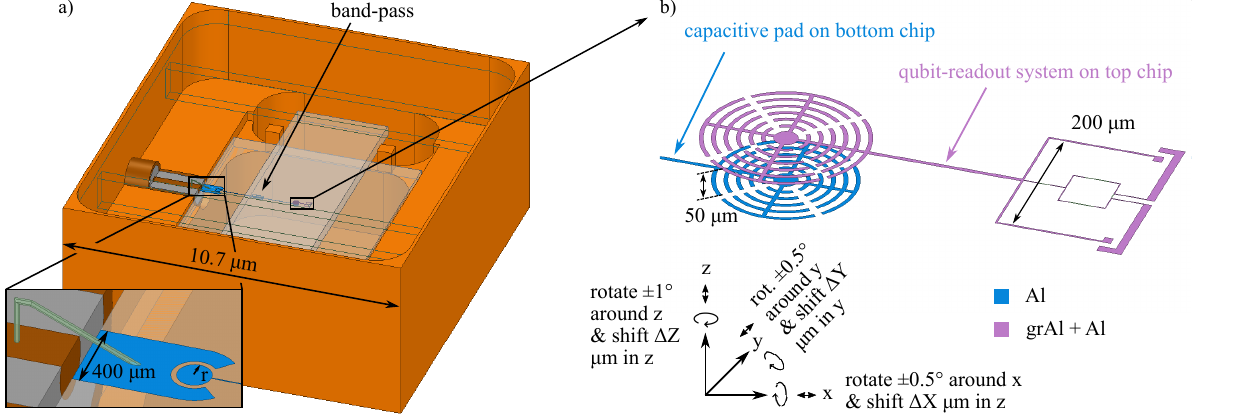}
        \caption{ \textbf{Samplebox for ANSYS simulations. (a)} Simplified copper sample box in which we perform the ANSYS finite-element simulations with transparent lid for better visibility. The band-pass filter is capacitively coupled to the bondpad through a circular capacitor with variable radius $r$. To save computational resources, we omit the capacitive extenders and replace the finger capacitance of the band-pass filter with a variable lumped capacitance. \textbf{(b)}\ Zoom-in towards the overlapping capacitive pad on the control chip and the QR system on the top chip, which is based on QR system 1. Both chips are spaced by $50\,\upmu$m. As indicated by the coordinate system in the bottom left corner, the qubit chip is shifted up to $\pm40\,\upmu$m in X, Y \& Z direction and rotated by $\pm0.5^\circ$ in X\&Y / by $\pm1.0^\circ$ in Z direction. The color legend indicates the material used for each circuit element: blue for aluminum (Al) and purple for Al covered with granular aluminum (grAl).
        \label{fig:App_SampleBox}
        }
\end{figure*}

To investigate the influence of qubit chip misalignment, mismatch between readout and band-pass filter frequency and the contribution of the copper box to energy relaxation, we perform eigenmode simulations with the finite element solver Ansys HFSS. We run the simulations in a detailed 3D model of a single enclosure that is shown in \figref{fig:App_SampleBox}. The simulations are terminated at a convergence error of less than 1\,\% and are based on the parameters of qubit 1. We model the pure grAl parts in the QR system with lumped inductors, i.e. $L_\text{r}=15\,$nH, $L_\text{q}=25\,$fF and $\Delta = 0.5\,$ nH. The qubit mode is simulated as a harmonic mode which is why we replace the JJ with a lumped capacitor $C_\text{j}=4\,$fF resulting in a resonance frequency of $f_\text{sim}\,\approx 7.7\,$GHz.

\begin{figure*}[hbt!]
        \includegraphics[width=1.0\textwidth]{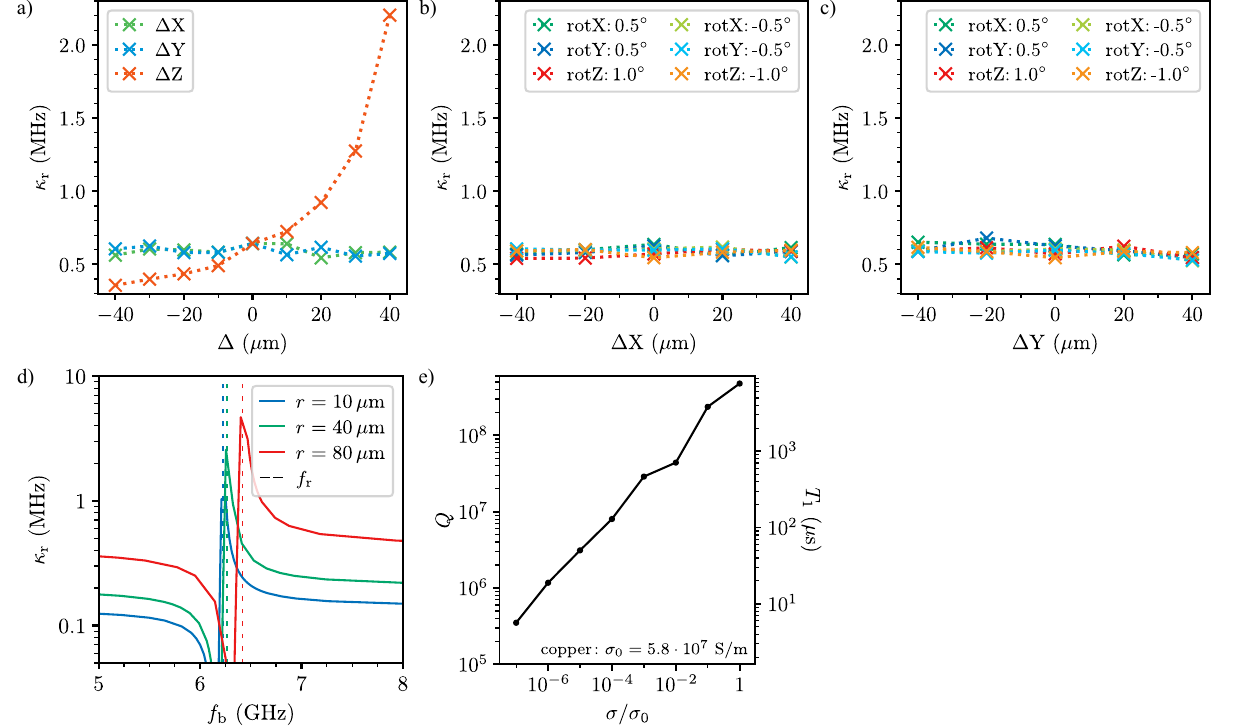}
        \caption{ \textbf{ANSYS simulations.} Bandwidth of the readout resonator $\kappa_\text{r}$ vs. \textbf{(a)} shift in X, Y \& Z direction and \textbf{(b/c)} shift in X/Y direction for rotations of $\pm0.5^\circ$ around the X\&Y axis and $\pm1.0^\circ$ around the Z axis of the qubit chip. \textbf{(d)} $\kappa_\text{r}$ for different radii $r$ of the circular capacitor vs. the frequency of the band-pass filter $f_\text{b}$. The frequencies of the readout resonators are indicated by the dashed lines. \textbf{(e)} Simulated quality factor $Q$ of a harmonic mode including the JJ, oscillating at a frequency of $f_\text{sim}\approx 7.7\,$GHz versus the conductivity of the bulk sample holder material. The predicted qubit energy relaxation time $T_\mathrm{1}$ for inductive loss is calculated via \eqref{eq:Fermi}. 
        \label{fig:App_Displacment}
        }
\end{figure*}

To analyze the architecture susceptibility to misalignment, we simulate the bandwidth $\kappa_\text{r}=f_\text{r}/Q_r$ of the readout mode for a qubit chip that is shifted and rotated in X, Y \& Z directions. The results are shown in \figref{fig:App_Displacment}.a,b,c. In \figref{fig:App_Displacment}.d, we show the dependence of $\kappa_\text{r}$ on the band-pass frequency $f_\text{b}$ and the capacitive coupling strength to the rf-port (expressed by the radius $r$ of the circular capacitance). To assess the contribution of the box to the qubit energy relaxation time, we simulate the quality factor of a mode that includes the JJ for increasing losses in the copper box, i.e. we reduce the conductivity $\sigma_0=5.8\cdot 10^7$ of the simulated sample box material. To calculate the qubit $T_1$ from the simulated $Q$, we use Fermi's Golden Rule
\begin{equation}
    \frac{1}{T_\mathrm{1}} = \frac{8\pi^3E_{\text{L}}}{hQ} |\braket{0|\hat{\varphi}|1}|^2 \left( 1 + \coth{\frac{hf_{\text{q}}}{2k_{\text{B}}T}}\right),
    \label{eq:Fermi}
\end{equation}
where $E_L$ is the inductive energy, $\hat{\varphi}$ the flux operator in units of $\Phi_0$, and $f_\text{q}$ the qubit frequency. The simulated $Q$ and calculated $T_\mathrm{1}$ values are shown in \figref{fig:App_Displacment}.e and exceed $T_\mathrm{1}>10^3\,\upmu$s for copper indicating that at the moment these losses are not a limiting factor.

\ \\
\ \\ 

\ \\
\ \\

\newpage

\ \\

\newpage

\section{\label{sec:fabrication} Sample fabrication}

The qubit chips are fabricated using e-beam lithography and shadow evaporation. The Josephson junction and the capacitive structures are a result of an aluminum double-angle evaporation with an intermediate static oxidation process for the junction insulating barrier. The inductive parts are implemented with granular aluminum wires, deposited with a zero-angle aluminum evaporation in combination with a dynamical oxygen flow in the evaporation chamber. The patterning and deposition is conducted in a single lithographic and evaporation step and follows the recipe from Ref.~\cite{Geisert__GFQ__2024}. In summary the deposition process is started at a pressure below $5\cdot10^{-7}$\,mbar with a $0^{\circ}$ plasma cleaning step using a Kaufman source (200\,V, 10\,mA, 10\,sccm $\text{O}_2$, 5 sccm Ar) that is followed by a titanium evaporation with a closed shutter (10\,s at 0.2\,nm/s). The fist aluminum layer is evaporated at a $-30^{\circ}$ angle (20\,nm at 1\,nm/s) and then statically oxidized at 50\,mbar $\text{O}_2$ for 4\,minutes. The second aluminum layer is deposited at a $+30^{\circ}$ angle (30 nm at 1 nm/s), followed by argon milling step at $0^{\circ}$ (400 V, 15 mA, 4 sccm Ar) to ensure contact to the final 70\,nm thick granular aluminum layer (evaporated with 2\,nm/s at a 0° angle with a $\text{O}_2$-flow of 9.4 sccm and planetary rotation at 5 rpm).

\begin{figure*}[!htb]
    \resizebox{1\textwidth}{!}{ 
        \def\svgwidth{\textwidth}
        %% Creator: Inkscape 1.1.1 (3bf5ae0d25, 2021-09-20), www.inkscape.org
%% Accompanies image file '2024-10-09__Exploded_Junction.pdf' (pdf, eps, ps)
\begingroup%
  \makeatletter%
  \providecommand\color[2][]{%
    \errmessage{(Inkscape) Color is used for the text in Inkscape, but the package 'color.sty' is not loaded}%
    \renewcommand\color[2][]{}%
  }%
  \providecommand\transparent[1]{%
    \errmessage{(Inkscape) Transparency is used (non-zero) for the text in Inkscape, but the package 'transparent.sty' is not loaded}%
    \renewcommand\transparent[1]{}%
  }%
  \providecommand\rotatebox[2]{#2}%
  \newcommand*\fsize{\dimexpr\f@size pt\relax}%
  \newcommand*\lineheight[1]{\fontsize{\fsize}{#1\fsize}\selectfont}%
  \ifx\svgwidth\undefined%
    \setlength{\unitlength}{566.92913386bp}%
    \ifx\svgscale\undefined%
      \relax%
    \else%
      \setlength{\unitlength}{\unitlength * \real{\svgscale}}%
    \fi%
  \else%
    \setlength{\unitlength}{\svgwidth}%
  \fi%
  \global\let\svgwidth\undefined%
  \global\let\svgscale\undefined%
  \makeatother%
  \begin{picture}(1,0.385)%
    \lineheight{1}%
    \setlength\tabcolsep{0pt}%
    \put(0,0){\includegraphics[width=\unitlength,page=1]{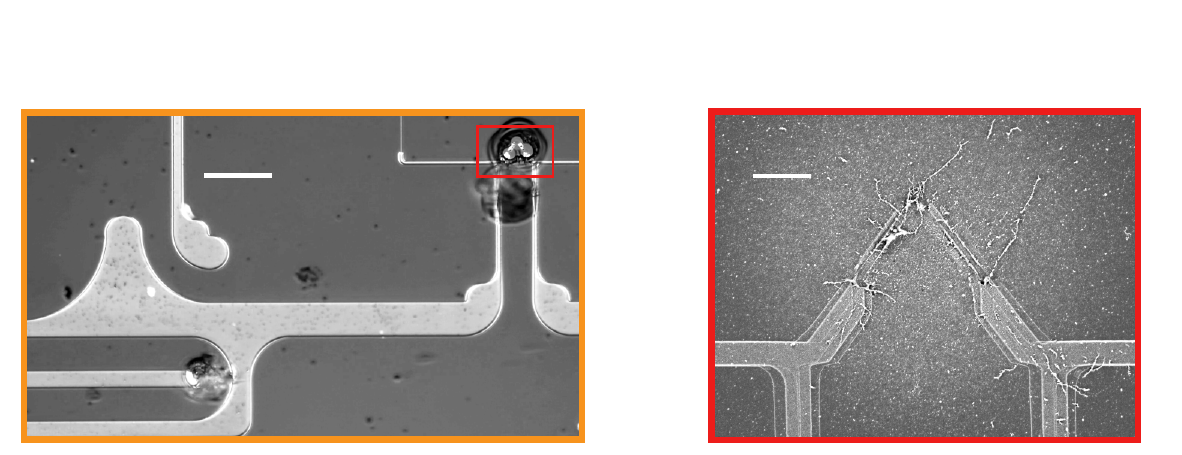}}%
    \put(0.64308731,0.24672343){\makebox(0,0)[lt]{\lineheight{1.25}\smash{\begin{tabular}[t]{l}\color{white} 2\,$\upmu$m\end{tabular}}}}%
    \put(0.17430634,0.24702321){\makebox(0,0)[lt]{\lineheight{1.25}\smash{\begin{tabular}[t]{l}\color{white}10\,$\upmu$m\end{tabular}}}}%
    \put(0,0){\includegraphics[width=\unitlength,page=2]{2024-10-09__Exploded_Junction.pdf}}%
    \put(0.11509314,0.31803753){\makebox(0,0)[lt]{\lineheight{1.25}\smash{\begin{tabular}[t]{l}8500\,$\upmu$m\end{tabular}}}}%
    \put(0.02783397,0.31706068){\makebox(0,0)[lt]{\lineheight{1.25}\smash{\begin{tabular}[t]{l}a)\end{tabular}}}}%
    \put(0.03024645,0.26602289){\color[rgb]{1,1,1}\transparent{0.941176}\makebox(0,0)[lt]{\lineheight{1.25}\smash{\begin{tabular}[t]{l}b)\end{tabular}}}}%
    \put(0.60873959,0.26650635){\color[rgb]{1,1,1}\makebox(0,0)[lt]{\lineheight{1.25}\smash{\begin{tabular}[t]{l}c)\end{tabular}}}}%
  \end{picture}%
\endgroup%
 
    }
    \caption{\textbf{Exploded junctions due to electric discharge. a)} The layout of the qubit chip shows the capacitive extenders reaching close to the edges of the sapphire chip with dimensions $\approx 9$\,mm. The QR system in the middle of the chip couples capacitively to the extenders via the Josephson junction electrodes. \textbf{b)} The inset shows an optical image of the center part of the chip after the final dicing process of the 2" sapphire wafer into $2.85\times10\,\text{mm}^2$ qubit chips. Residuals of explosions are visible in the protective resist (Microposit S1808) used for dicing at the junction position as well as the coupling capacitor. \textbf{c)} A scanning-electron micrograph of an exploded junction reveals that the explosion happened exactly at the junction position, leaving intact the bigger capacitive pads, but destroying the junction and its connecting films.
    \label{fig:App_exploded_JJ}
    }
\end{figure*}

The capacitive extenders ($\approx 4$\,mm) that each span half the qubit chip length, pose an additional fabrication challenge, as they increase the risk of electric discharge during resist spinning or chip dicing processes. As visible in \figref{fig:App_exploded_JJ}, destroyed junctions can be directly related to electric discharge induced by the extenders that capacitively couple the junction electrodes to the chip edges. This problem is remedied by shunting the junction with a low impedance aluminum film, which is etched away after the final dicing process. 
\ \\
\ \\

\newpage

\section{\label{sec:single_qubit}Readout and single qubit gate fidelity}

\begin{figure*}[!htb]
        \includegraphics[width=1.0\textwidth]{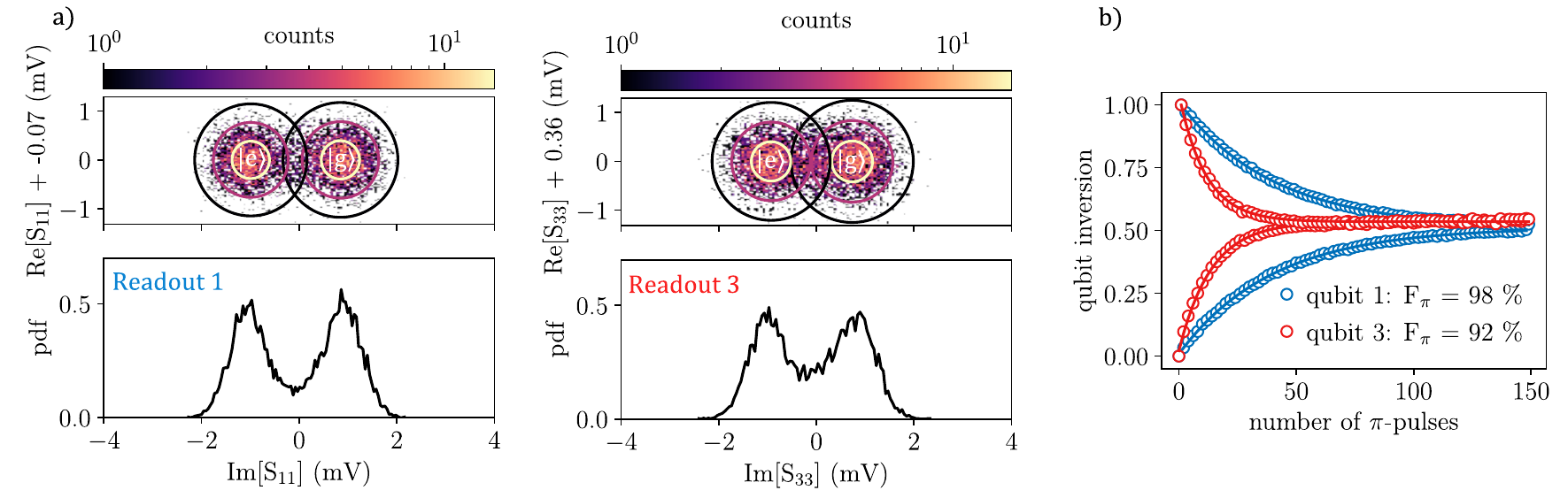}
        \caption{ 
        \label{fig:single_qubit}
        \textbf{Readout and $\pi$-pulse fidelity} \textbf{a)} Pointer state distributions for readout 1 and readout 3 for continuously driven qubits. The measurement outcomes of contiguously played readout pulses are visualized as histograms in the complex planes of the reflection coefficients $S_{11}$ and $S_{33}$. \textbf{b)} Simultaneous $\pi$-pulse calibration for qubits 1 (blue) and 3 (red).}
\end{figure*}

We show simultaneous readout and single qubit $\pi$-pulse calibration measurements in \figref{fig:single_qubit}. The readout pulse lengths are equal to the integration time $t_{\mathrm{int}} = 1 \ \upmu$s, which is on the order of the qubit relaxation time. As visible from the overlapping pointer state distributions and absence of additional states in \figref{fig:single_qubit}a, qubit relaxation during readout is the main limitation of the single shot readout separation fidelity. The $\pi$-pulse is calibrated as a Gaussian envelope pulse with $t_{\mathrm{\pi}} = 16$ ns pulse length. We show sequences of subsequently played $\pi$-pulses in \figref{fig:single_qubit}b. The abscence of beating patterns proves an optimal power calibration. We define the $\pi$-pulse fidelity $F_{\pi}$ via the exponential decay versus the pulse sequence length and find $F_{\pi} = 98 \%$ and $F_{\pi} = 92 \%$ for qubit 1 and qubit 3, respectively, which is limited by qubit relaxation.

\newpage

\section{\label{sec:power_calibration}Power calibration}

Following Ref.~\cite{Gambetta2006}, the dephasing rate $\gamma_{\mathrm{m},i}$ added to qubit $i$ when driving the microwave line connected to port $j$ at frequency $f_{\mathrm{d}}$ with amplitude $A_{\mathrm{port},j}$ is given by
\begin{equation}
    \gamma_{\mathrm{m},i} = \epsilon_i^2 \left(\frac{1}{\kappa_i^2/4 + (f_{\mathrm{d}} - f_{\mathrm{r},i} + \chi_i/2)^2} + \frac{1}{\kappa_i^2/4 + (f_{\mathrm{d}} - f_{\mathrm{r},i} - \chi_i/2)^2} \right)
    \frac{\kappa_i \chi_i^2/4}{\kappa_i^2/4 + \chi_i^2/4 + (f_{\mathrm{d}} - f_{\mathrm{r},i})^2},
    \label{eq:power_cal}
\end{equation}
where $f_{\mathrm{r},i}$, $\kappa_i$ are the resonance frequency and the linewidth of the resonator coupled to qubit $i$ via the dispersive shift $\chi_i$, which is defined as the frequency difference of the readout resonator frequency when the qubit is in the ground or the excited state, respectively. The two terms in the parenthesis correspond to the circulating photon number in the resonator for the qubit in the ground and excited state. The microwave field amplitude $\epsilon_i$ seen by resonator $i$ is related to $A_{\mathrm{port},j}$ via the power transfer coefficient $\eta_{ij}$: $\epsilon_i^2 = \eta_{ij} A_{\mathrm{port},j}^2$. Accordingly, we define the port-to-resonator isolation on a logarithmic power scale as $d = 10\log_{10}(\eta_{13}/\eta_{33})$.

\begin{figure*}[!htb]
        \includegraphics[width=1.0\textwidth]{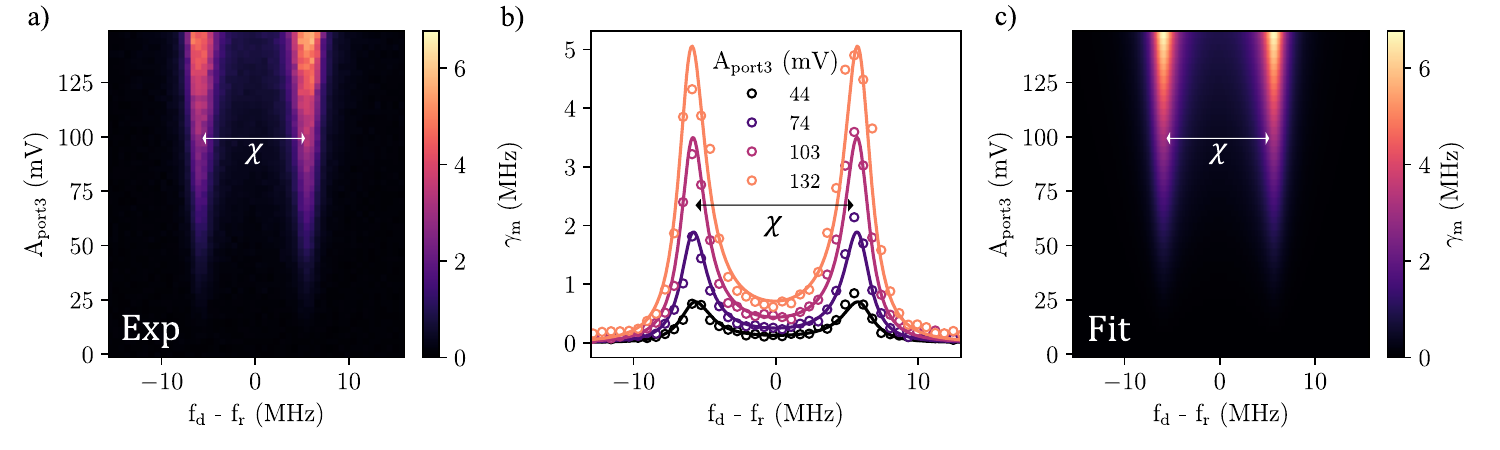}
        \caption{ 
        \label{fig:power_calibration}
        \textbf{Power calibration via measurement-induced dephasing. }\textbf{a)} Additional dephasing $\gamma_m$ of qubit 1 due to resonator photons measured from exponentially damped Ramsey fringes for different drive frequencies and drive amplitudes $A_{\text{port3}}$ when driven through port 3. The features correspond to driving at $f_{\mathrm{d}} = f_{\mathrm{r},i} \pm \frac{\chi}{2}$. \textbf{b)} The plot shows the data points and individual fits for different drive amplitudes, yielding a set of possible transfer coefficients for different drive amplitudes. \textbf{c)} The plot shows the theory prediction calculated with the mean values of the fit parameters obtained in \textbf{b}.}
\end{figure*}

We measure $\gamma_{\mathrm{m},i}$ for various $f_{\mathrm{d}}$ and $A_{\mathrm{port},j}$ via Ramsey interferometry. As an example, the data for $\gamma_{\mathrm{m},1}$ when driving through port 3 is shown in \figref{fig:power_calibration}a. In \figref{fig:isolation_experiments}.b in the main text, we report values for $\gamma_{\mathrm{m},1}$ and $\gamma_{\mathrm{m},3}$ for $f_{\mathrm{d}} - f_{\mathrm{r},1,3} = 2.4 \ \mathrm{MHz}$. For each $A_{\mathrm{port},j}$, we fit the data to \eqref{eq:power_cal} using the fitting parameters $\eta_{ij}$, $f_{\mathrm{r},i}$, $\chi_i$ and $\kappa_i$, as shown in \figref{fig:power_calibration}b. This yields different values of $\eta_{ij}$ for different amplitudes, giving a measure of the uncertainty in the extraction of $d$. In \figref{fig:power_calibration}c, we show the calculated dephasing using the mean values of the fitted parameters. By comparing the dephasing of qubit 1 and qubit 3 while driving through port 3, we extract $d = 64\ \pm 0.5\mathrm{dB}$.

\newpage

\section{\label{sec:avoided_crossings} Measurements of qubit-qubit avoided crossings }

\begin{figure*}[!htb]
        \includegraphics[width=1.0\textwidth]{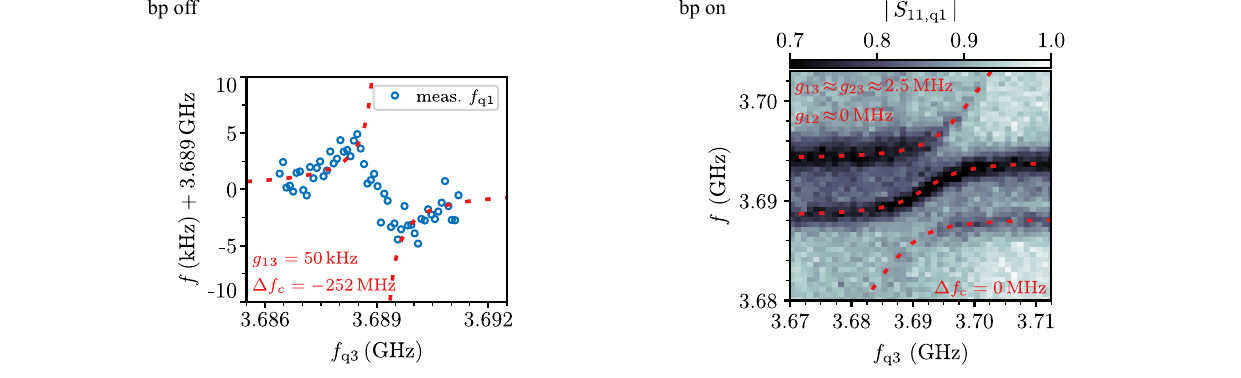}
        \caption{ \textbf{Avoided Level Crossings between q1 and q3 for bp off ($\Delta f_{\text{c}}=-252\,$MHz) and bp on ($\Delta f_{\text{c}}=0\,$MHz).} At the bp off ($\Delta f_\text{c}=-252\,$MHz) we use Ramsey fringes to measure $f_\text{q1}$, otherwise we perform spectroscopy on q1. The colormap shows the amplitude of the reflected measurement signal on q1, $|S_{11,\text{q1}}|$. The extracted $g_{13}$ values from the fits (dashed red lines) are listed in red. 
        \label{fig:App_AC_extrema}
        }
\end{figure*}

\begin{figure*}[!htb]
        \includegraphics[width=1.0\textwidth]{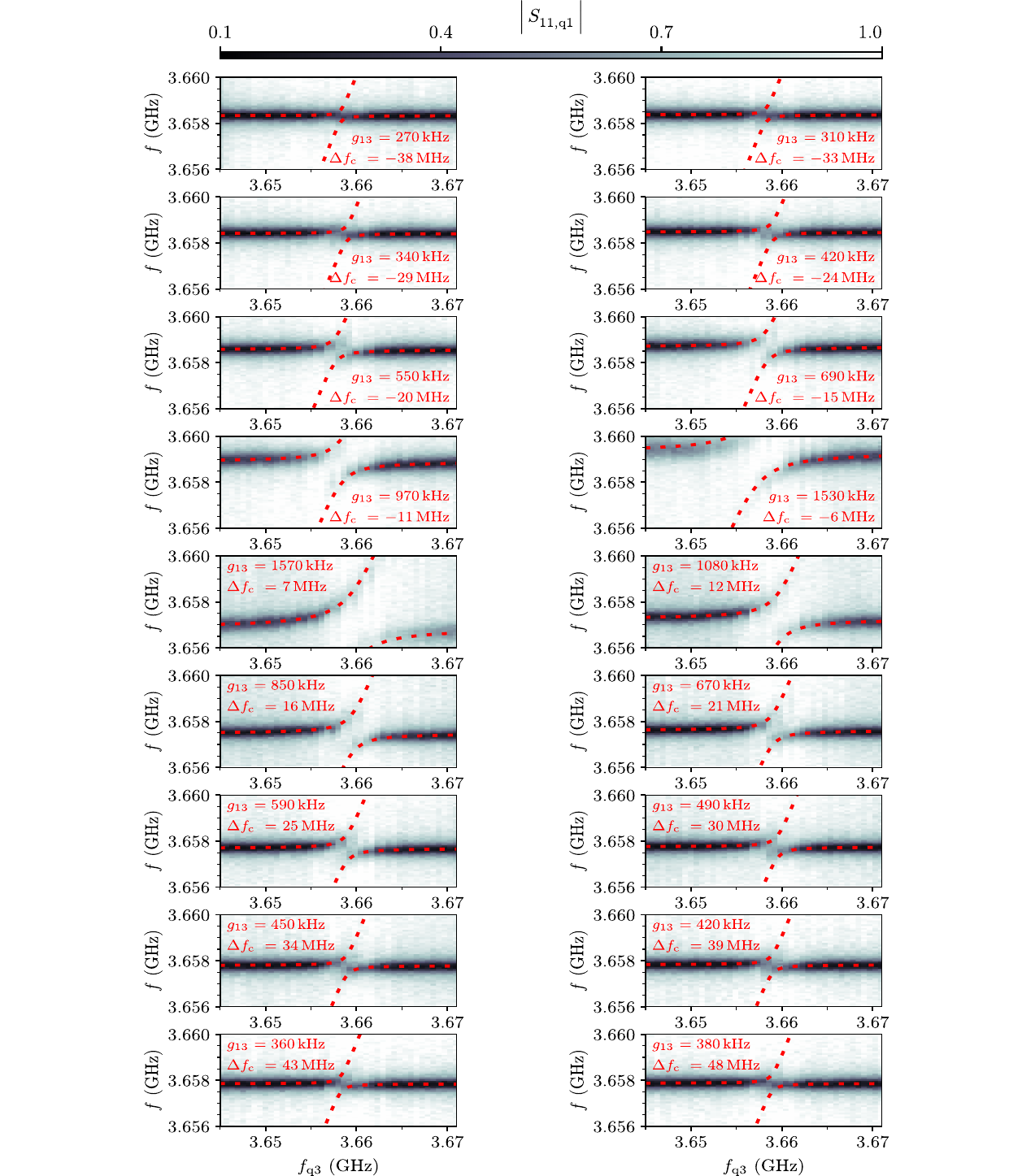}
        \caption{ \textbf{Avoided Level Crossings between q1 and q3 for different coupler detunings $\Delta f_{\text{c}}$,} which we measured by performing spectroscopy on q1. The colormap shows the amplitude of the reflected measurement signal on q1, $|S_{11,\text{q1}}|$. The extracted $g_{13}$ values from the fits (dashed red lines) are listed in red. 
        \label{fig:App_AC}
        }
\end{figure*}

In \figref{fig:App_AC_extrema} and \figref{fig:App_AC} we show the measured avoided level splittings between qubits 1 and 3 for for different coupler detunings $\Delta f_\text{c}$, which are used in ~\figref{fig:static_coupling_experiments} in the main text. Measurement data for $\Delta f_\text{c}>-250\,$MHz is obtained by performing two-tone spectroscopy on qubit 1. At the idle point ($\Delta f_\text{c}=-252\,$MHz) we measure Ramsey fringes on q1 to characterize the avoided level splitting.

\newpage

\newpage

\section{\label{sec:SWT}Schrieffer-Wolff transformation}

In this appendix, we provide a detailed explanation of the calculations that result in the theoretical prediction for the effective qubit-qubit coupling $g_{13}^{\text{eff}}$ shown in Fig.~\ref{fig:static_coupling_experiments} of the main text. These calculations are based on two different applications of the Schrieffer-Wolff transformation \cite{consani2020effective, bravyi2011schrieffer} (SWT) to the superconducting circuit, composed of three unit cells, shown in \figref{fig:sample_box} g). Both SWTs yield an effective two-qubit model that captures how the coupling between qubits in e1 and e3 depends on the detuning of the qubit in e2 (coupler). The two SWTs complement each other: the first is exact and fully relies on the numerical methods developed in Ref. \cite{hita2021three, hita2022ultrastrong} (continuous line in~\figref{fig:static_coupling_experiments}), while the second uses a semi-analytical approach that gives us a qualitative understanding of the mediated coupling (dashed line in~\figref{fig:static_coupling_experiments}).

The circuit under study can be decomposed into three sub-circuits, or unit cells, that are capacitively coupled. As shown in \figref{fig:sample_box}g, this involves a direct coupling between sub-circuits 1 and 2 through capacitor $C_{12}$, between sub-circuits 2 and 3 through capacitor $C_{23}$, and an indirect coupling between sub-circuits 1 and 3 through the sample box, $C_{13}$. As discussed in Ref. \cite{Geisert__GFQ__2024}, each of the sub-circuits under study can be modeled as a flux qubit and a resonator inductively coupled, however, this inductive coupling can be neglected in the treatment below, because during the experiment the qubits are far detuned from their respective resonators. We use a Hamiltonian of three capacitively coupled flux qubits:
\begin{equation}
\begin{aligned}
H =& \sum_{i=1}^3 H_{\text{FQ},i} + \frac{1}{2} \sum_{i,j=1}^3 C_{ij}^{-1} Q_i Q_j,\\
H_{\text{FQ},i} = &  \frac{C^{-1}_{ii}}{2} Q_{i}^2 + \frac{1}{2L_i} \Phi_{i}^2 - E_{\text{J},i}\cos\left[\frac{2\pi}{\Phi_0}\left(\Phi_{i} - \Phi_i^\text{ext}\right)\right],
\label{eq:H_coupled_qubits}
\end{aligned}
\end{equation}
where $H_{\text{FQ},i}$, $L_i$ and $E_{\text{J},i}$ are the Hamiltonian, the linear inductance and the Josephson energies of qubit $i$ \cite{Geisert__GFQ__2024}. The symbols $C_{ij}^{-1}$ are the elements $ij$ of the inverse of the capacitance matrix $C$. The non-diagonal elements of $C^{-1}$ couple different flux qubits. All single qubit parameters in \eqref{eq:H_coupled_qubits} are determined by fitting the flux-dependent spectra obtained for the qubits in the experiments, as shown in \suppref{sec:qubit_parameters}.

In the following, we explain how to derive an effective model that describes the coupling between qubits 1 and 3 as a function of the frequency of the coupler qubit. That is, we map the Hamiltonian of the full circuit shown in \eqref{eq:H_coupled_qubits} to an effective one of the form
\begin{align}
        H_\text{eff} = \frac{\omega_1}{2}{\sigma}^z_1 + \frac{{\omega}_3}{2}{\sigma}^z_3  +  \sum_{i,j=\{x,y,z\}}J_{ij} \sigma^i_1{\sigma}^j_3\,,
\label{eq:target}
\end{align}
where $\omega_i$ is the renormalized frequency of qubit $i$, slightly shifted from the frequency of the bare qubits, and $J_{ij}$ are the coupling strengths in the different directions. 

Let us assume a Hamiltonian that can be decomposed in an unperturbed part plus a perturbation, $H=H_0+V$. The SWT is a unitary transformation $U$ of the Hamiltonian that decouples an eigensubspace of interest (the two-qubit subspace) of the interacting model $H$ from the rest of the Hilbert space through a unitary map to the equivalent eigensubspace of the non-interacting problem $H_0$. If we call the projectors onto the qubit eigensubspace with and without interactions $P$ and $P_0$, and the complementary eigensubspace projectors $Q=\mathbbm{1}-P$ and $Q_0=\mathbbm{1}-P_0$,  the SWT maps those projectors, $UPU^\dagger=P_0$, and $UQU^\dagger=Q_0$. It has been shown \cite{bravyi2011schrieffer} that
\begin{equation}
    U = \sqrt{(P_0-Q_0)(P-Q)}\,.
    \label{eq:SWT}
\end{equation}
With the SWT we can obtain an effective Hamiltonian $H_\text{eff}$ that describes the dynamics of our coupled two-qubit subspace as
\begin{equation}
H_\text{eff} = P_0 U P H P U^\dagger P_0.
\label{eq:H_eff}
\end{equation}
Note that this procedure is valid whenever the subspaces $P$ and $P_0$ are gapped from the rest of the spectrum regardless of whether they involve low energy states or not.

In this manuscript we use the SWT in two strategies to derive an effective Hamiltonian in the low-energy subspace of the first and third qubits. In the first and more complete strategy our starting point is the full circuit mode Hamiltonian \eqref{eq:H_coupled_qubits}, while in the second one we start from an effective model of three coupled qubits and apply the SWT in a perturbative fashion.

In the first approach, we rely on a numerically exact procedure \cite{hita2021three} which allows us to work directly with the circuit Hamiltonian \eqref{eq:H_coupled_qubits}. Here, our unperturbed Hamiltonian is the sum of the bare Hamiltonians of the flux qubits that appear in \eqref{eq:H_coupled_qubits}, that is, $H_0= \sum_{i=1}^3 H_{\text{FQ},i}$, and the perturbation is the capacitive coupling between them,  $V= \frac{1}{2} \sum_{i,j=1}^3 C_{ij}^{-1} Q_i Q_j$. Since we are interested in the effective coupling between qubits 1 and 3, our subspace of interest will include the eigenstates that have up to one excitation in either qubit 1 or 3, with zero excitations in the coupler qubit, that is, $P_0=\{\{\ket{0}, \ket{1}\}_1 \otimes \{{\ket{0}}\}_2 \otimes \{{\ket{0}, \ket{1}}\}_3\}$. The projector $P$ is composed of the numerically estimated eigenstates of $H$ that correspond to the same qubit subspace. From $P_0$ and $P$, using the procedure introduced in Ref. \cite{hita2021three} we arrive at the effective Hamiltonian
 \begin{align}
        H_\text{eff}^{\text{SWT}_1} = \frac{\omega_1}{2}{\sigma}^z_1 + \frac{{\omega}_3}{2}{\sigma}^z_3 + J_{xx} \sigma^x_1{\sigma}^x_3+ J_{yy} \sigma^y_1{\sigma}^y_3+ J_{zz} \sigma^z_1{\sigma}^z_3.
\label{eq:SWTnumeric}
\end{align}

Experimentally, we characterize the effective coupling strength by half the energy splitting between the one excitation eigenstates, solid points shown in~\figref{fig:static_coupling_experiments}. To obtain this energy difference we diagonalize the Hamiltonian above, which can be written as a $4\times4$ matrix
\begin{equation}
H_\text{eff}^{\text{SWT}_1} = 
    \begin{pmatrix}
    \frac{\omega_1+\omega_3}{2} + J_{zz} & 0 &0 & J_{xx}-J_{yy}  \\
    0 & \frac{\omega_1-\omega_3}{2} - J_{zz} & J_{xx}+J_{yy}  & 0 \\
    0 & J_{xx}+J_{yy}  & \frac{-\omega_1+\omega_3}{2} - J_{zz}  & 0 \\
     J_{xx}-J_{yy}   & 0 & 0 & \frac{-\omega_1-\omega_3}{2} + J_{zz}  \\
    \end{pmatrix}\,.
\end{equation}
Assuming that qubits 1 and 3 are on resonance, the energy splitting is directly given by $2g^{\text{eff}}_{13} = 2|J_{xx}+J_{yy}|$, corresponding to the solid line in~\figref{fig:static_coupling_experiments}. Furthermore, as evidenced by our numerical simulations, $J_{zz}\sim(10^{-3} ~\text{to}~ 10^{-2})\times|J_{xx} + J_{yy}|$, a quantity below our experimental resolution, see \figref{fig:isolation_experiments}c.
 
In the second approach, we derive an analytical approximation for the energy splitting $g^{\text{eff}}_{13}$. Following Ref.\ \cite{hita2022ultrastrong} we start from a model of three capacitively coupled qubits in the limit in which the coupler qubit is highly detuned
\begin{equation}
    H_\text{low} =  \frac{\tilde{\omega}_{1}}{2}\boldsymbol{\sigma}^z_1  + \frac{\tilde{\omega}_{2}}{2}\boldsymbol{\sigma}^z_2  + \frac{\tilde{\omega}_{3}}{2}\boldsymbol{\sigma}^z_3 + g_{12} \boldsymbol{\sigma}^y_1\boldsymbol{\sigma}^y_2 + g_{23} \boldsymbol{\sigma}^y_2\boldsymbol{\sigma}^y_3 .
\label{eq:H_low}
\end{equation}
This model is derived by projecting the coupled flux qubit Hamiltonian \eqref{eq:H_coupled_qubits} onto the non-interacting subspace of the three qubits, $\{\{\ket{0}, \ket{1}\}_1 \otimes \{{\ket{0}, \ket{1}}\}_2 \otimes \{{\ket{0}, \ket{1}}\}_3\}$. Here $\tilde{\omega}_i$ are the qubit gaps of the bare model.  The interaction is derived from the projection of the capacitive coupling  $g_{ij}=C_{ij}^{-1}\braket{0_i|Q_i|1_i}\braket{0_j|Q_j|1_j}$.

Similar to Ref.~\cite{yan2018tunable}, we implement an analytical SWT using perturbative series up to second order with $H_\text{0} = \frac{\tilde{\omega}_{1}}{2}\boldsymbol{\sigma}^z_1  + \frac{\tilde{\omega}_{2}}{2}\boldsymbol{\sigma}^z_2  + \frac{\tilde{\omega}_{3}}{2}\boldsymbol{\sigma}^z_3 $, and, $V=g_{12} \boldsymbol{\sigma}^y_1\boldsymbol{\sigma}^y_2 + g_{23} \boldsymbol{\sigma}^y_2\boldsymbol{\sigma}^y_3 $. The second-order expansion of the effective Hamiltonian includes the projection of the perturbed Hamiltonian onto our subspace of interest $ H_\text{eff,1} = P_0 H P_0$ and the virtual transitions mediated by excited states $k\in\{\ket{010},\ket{110},\ket{011},\ket{111}\}$
\begin{equation}
    H_\text{eff,2} = H_\text{eff,1} + \frac{1}{2}\sum_{ijk}\left( \frac{1}{E_i-E_k} + \frac{1}{E_j-E_k} \right) \braket{i|V|k}\braket{k|V|j}\ket{i}\bra{j}\;.
\end{equation}
As qubits 1 and 3 are on resonance in the experiment, $\omega_1=\omega_3=\omega$, we recover the effective Hamiltonian
\begin{equation}
    H_\text{eff,2} = \frac{\tilde{\omega}_1}{2}\boldsymbol{\sigma}^z_1 + \frac{\tilde{\omega}_3}{2}\boldsymbol{\sigma}^z_3  + \tilde{g} \boldsymbol{\sigma}^y_1\boldsymbol{\sigma}^y_3 , 
\end{equation}
with the renormalized qubit gaps
\begin{equation}
        \tilde{\omega}_1 = \omega + g_{12}^2\left(\frac{1}{\Delta}-\frac{1}{\Sigma}\right) , \; 
        \tilde{\omega}_3 = \omega + g_{23}^2\left(\frac{1}{\Delta}-\frac{1}{\Sigma}\right) ,
\end{equation}
and the interaction 
\begin{equation}
    g^{\text{eff}}_{13} = g_{12}g_{23}\left(\frac{1}{\Delta}-\frac{1}{\Sigma}\right) .
    \label{eq:gper}
\end{equation}
Since in this case there is no term proportional to $\sigma_1^x\sigma_3^x$ the coupling $g^{\text{eff}}_{13}$ corresponds directly to the measured energy splitting, and is shown as a dotted line in Fig.~\ref{fig:static_coupling_experiments}.  In the equations above we have defined $\Delta = \omega-\omega_2$ and $\Sigma = \omega+\omega_2.$  Note that this second order expansion presents divergences that are absent in the complete SWT \cite{bravyi2011schrieffer}.

Let us emphasize that the SWT on the full quantum circuit is numerically exact and describes the three interacting qubits without approximations (except for possible crosstalks or couplings that are not considered in the circuit). In contrast, the analytical SWT is based on an effective pairwise model that only captures the dominant contributions to the final coupling, exhibits non-existent divergences and cannot reflect phenomena such as interaction mediated by excited states or $\sigma_z\sigma_z$ perturbative couplings between the qubits.

\newpage

\section{\label{sec:qubit_parameters} Qubit parameters}

\begin{figure*}[!htb]
        \includegraphics[width=1.0\textwidth]{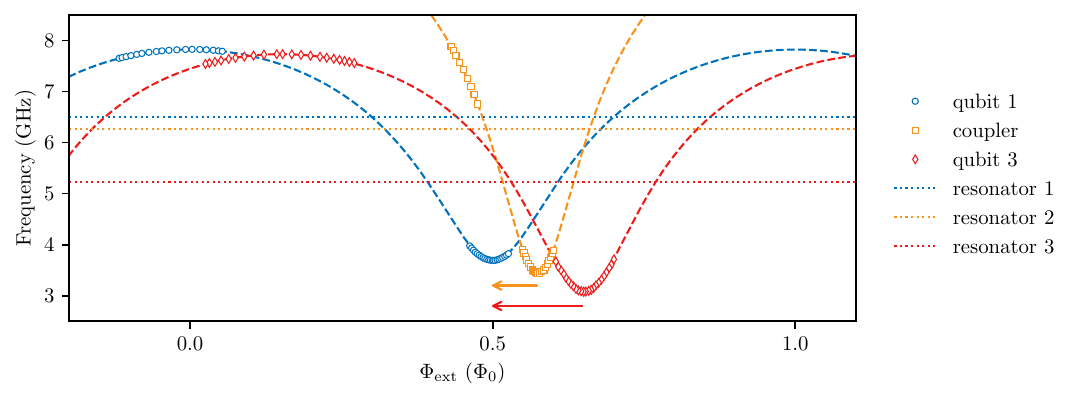}
        \caption{ 
        \label{fig:qubit_spectra}
        \textbf{Qubit spectra vs. external flux} The flux qubit spectra are shown with respect to the applied external flux modulo integer numbers of flux quanta $\Phi_{\text{ext}}$. The markers correspond to measured data. The dashed lines are fits to the flux qubit Hamiltonian yielding the parameters given in Tab.\ref{supp:qubit_parameters}. The dotted lines represent the readout resonators. For clarity, the spectra of the coupler (orange) and qubit 3 (red) have been shifted with respect to qubit 1 (blue) as indicated by the arrows.}
\end{figure*}

We model the single qubit spectra using the flux qubit Hamiltonian and neglecting the resonator: 

\begin{equation}
    \hat{H} = \frac{\hat{\text{Q}}^2}{2C_{\Sigma}} + \frac{\hat{\Phi}^2}{2L_\text{Q}} - E_\text{J} \cos(\hat{\Phi} - \Phi_{\text{ext}}),
\end{equation}
where $\hat{\Phi}$ and $\hat{Q}$ are the qubit flux and charge operators, $\Phi_{\text{ext}}$ is the external flux applied to the qubit loop and $C_{\Sigma}$, $L_\text{Q}$ and $E_\text{J}$ are the qubit capacitance, inductance and the junction Josephson energy. These three parameters are fitted to spectroscopic measurements. The fits are plotted in \figref{fig:qubit_spectra} and the fit results are summarized in \tabref{supp:qubit_parameters}.

The flux qubit coherence is typically best at half flux bias, aka sweet spot, due to the first order insensitivity to flux noise. To operate the qubits on resonance, however, qubit 3 is detuned from its respective sweet spot to match the frequency of qubit 1, at the expense of coherence. The energy relaxation times $T_1$ and Ramsey coherence times $T_2^{*}$ for the qubits and the coupler are summarized in \tabref{supp:qubit_parameters}. The frequency $f_\text{r}$, bandwidth $\kappa_\text{r}$ and loaded quality factor $Q_\text{L}$ of the devices readout resonators are listed in \tabref{supp:rout_parameters}.

\begin{table*}[!htb]\centering   
    \begin{tabular}{|c||c|c|c|c|c|c|c|c|}
        \hline
        device & $E_{\text{J}}/h$\,[GHz] & $L_{\text{Q}}$\,[nH] & $C_{\Sigma}$\,[fF] & $f_\text{q}$ @ $\Phi_0/2$\,[GHz] & $T_1$ @ $\Phi_0 /2$\,[$\upmu$s] & $T_2^{*}$ @ $\Phi_0 /2$\,[$\upmu$s] & $T_1$ @ bp\,[$\upmu$s] & $T_2^{*}$ @ bp\,[$\upmu$s] \\
        \hline
        \hline
 
        qubit 1 & 6.2 & 22.1 & 32.2 & 3.689  & 2.1 & 1.7 & -- & -- \\
        \hline
        coupler & 9.6 & 20.2 & 22.7 & 3.437 & -- & -- & -- & --\\ 
        \hline
        qubit 3 & 5.6 & 31.6 & 25.2 & 3.084 & 1.7 & 1.1 & 0.8 & 0.17  \\
        \hline
\end{tabular}
    \caption{\label{supp:qubit_parameters} 
    \textbf{Qubit parameters.} The fitted lumped element parameters, frequencies $f_\text{q}$ and measured coherence times for the qubits at the sweet spot or the bias point, respectively.}
\end{table*}

\begin{table*}[!htb]\centering   
    \begin{tabular}{|c||c|c|c|}
        \hline
        device & $f_\text{r}$ @ $\Phi_0/2$\,[GHz] & $\kappa_\text{r}$\,[MHz] & $Q_\text{L}$ \\
        \hline
        \hline
        readout 1 & 6.508 &  2.2 & 3000 \\
        \hline
        rout-coupler & 6.274 &  4.2 & 1500 \\ 
        \hline
        readout 3 & 5.226 &  1.3 & 3900 \\
        \hline
\end{tabular}
    \caption{\label{supp:rout_parameters} 
    \textbf{Readout parameters.} The frequencies $f_\text{r}$, bandwidth $\kappa_\text{r}$ and total quality factors $Q_\text{L}$ of the devices readout resonators.}
\end{table*}

\newpage

\section{\label{sec:pop_transfer} Qubit-qubit population transfer}
\begin{figure*}[!htb]
        \includegraphics[width=1.0\textwidth]{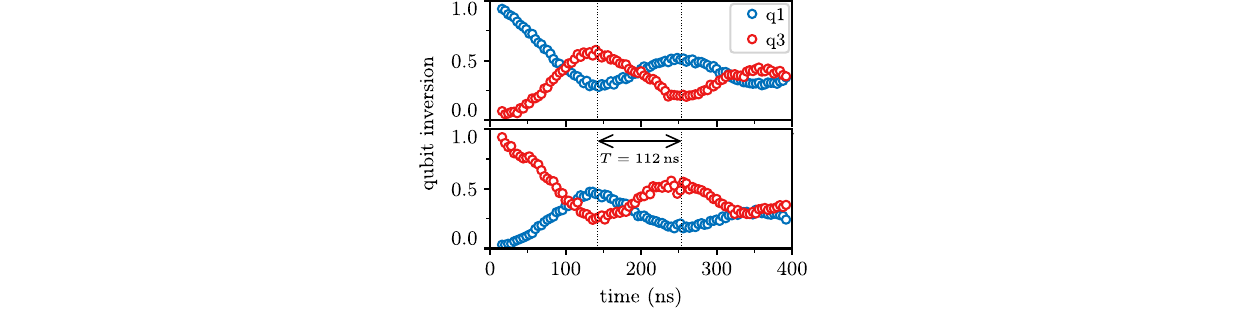}
        \caption{ \textbf{The qubit inversion between q1 (blue circles) and q3 (red circles) vs. time.} We excite q1 (top plot)/ q3 (bottom plot) with a $\pi$ pulse and then use the FBL to tune the coupler from the idle point to the bp. The qubit inversion is simultaneously measured for varying pulse lengths. We observe a population swapping in $T=112$\,ns.
        \label{fig:pop_swapping}
        }
\end{figure*}

To demonstrate population transfer via the tunable coupler, we use the fast flux bias line to tune the coupler from the idle point to the bias point with a baseband rectangular pulse of variable length and subsequently read out both qubits simultaneously. Preparing either qubit 1 or 3 in the excited state with a $\pi$ pulse yields oscillations in the qubit populations as shown in \figref{fig:pop_swapping}. The extracted swap time is $T \approx 112$\,ns which is consistent with the effective qubit-qubit coupling strength of 2.5\,MHz obtained via spectroscopy. The transfer fidelity is likely limited by the coherence time of qubit 2 and 3, which both operate off the sweet spot when the coupling is switched on.

\newpage

\section{\label{sec:5erBox} Scaling in 2 dimensions}

\begin{figure*}[hbt!]
        \includegraphics[width=1.0\textwidth]{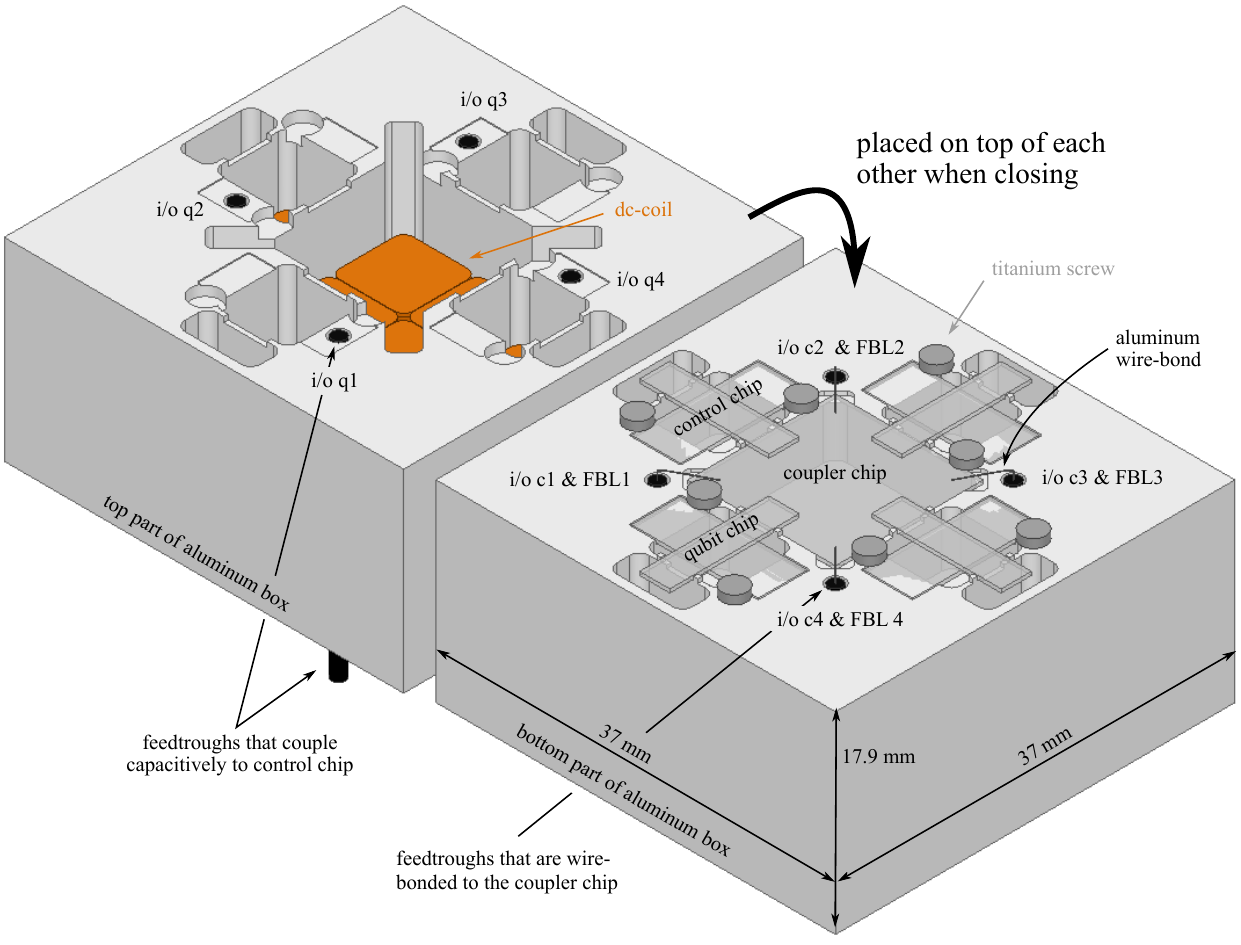}
        \caption{ \textbf{Concept for a scalable two dimensional qubit array in a modular architecture.} The aluminum box hosts individual enclosures for four qubits on the sides and a single center enclosure containing the coupling elements. Each qubit enclosure contains one control and one qubit chip, is accessed via a feed-through through the lid that couples capacitively to the control chip (i.e. the pad for bonding in \figref{fig:sample_box}.e), and has an individual dc-coil. The dc-coils are attached on the inside of the lid. The middle enclosure holds the coupler chip ($13.7\times 13.7\,\upmu$m$^2$) that are wire-bonded to four feed-throughs embedded in the bottom part of the box.  
        \label{fig:App_5erBox}
        }
\end{figure*}

To upgrade the one-dimensional array in two dimensions, the rf- and dc-ports of the enclosures need to be repositioned. As illustrated in \figref{fig:App_5erBox}, access to the enclosures is gained through out-of-plane feedthroughs which couple capacitively to the control chips and are wire-bonded to the coupler chips. As coupling element, an all-to-all reconfigurable router similar to~\cite{wu2024modularquantumprocessoralltoall} could be used.
\ \\
\ \\

\end{document}